%% file: main.tex
\documentclass[journal]{IEEEtran}

\usepackage{graphicx}
\usepackage{glossaries}
\usepackage{todonotes}
\usepackage[binary-units]{siunitx}
\usepackage[hidelinks]{hyperref}
\usepackage[capitalise]{cleveref}
\usepackage{nicefrac}
\usepackage{amsmath}
\usepackage{amssymb}
\usepackage{xcolor}
\usepackage{tabularx}
\usepackage{booktabs}
\usepackage{changes}
\usepackage{multirow}
\usepackage{bbm}
\usepackage[ngerman, english]{babel}
\usepackage{mathtools}
\usepackage{commath}
\usepackage[nocompress]{cite}
\usepackage{tikz}
\usetikzlibrary{shapes.geometric}

\DeclareMathOperator*{\argmax}{\arg\!\max}
\DeclareMathOperator*{\argmin}{\arg\!\min}
\DeclareSIUnit{\au}{a.u.}

\definecolor{red}{HTML}{C92682}
\definecolor{blue}{HTML}{296CD5}
\definecolor{green}{HTML}{12940C}
\definecolor{yellow}{HTML}{F2AC1D}

\input{tex/acronyms}

\begin{document}
\title{Spiking Neural Network Nonlinear Demapping\\on Neuromorphic Hardware for IM/DD Optical Communication}

\author{Elias~Arnold, Georg~B\"ocherer, Florian~Strasser, Eric~M\"uller, Philipp~Spilger, Sebastian~Billaudelle, Johannes~Weis, Johannes~Schemmel, Stefano~Calabrò, Maxim~Kuschnerov
\thanks{E. Arnold, E. M\"uller, P. Spilger, S. Billaudelle, J. Weis, and J. Schemmel are with the Electronics Vision(s) Group at the Kirchhoff-Institute for Physics, Heidelberg University, Germany, e-mail: elias.arnold@kip.uni-heidelberg.de}
\thanks{G. B\"ocherer, F. Strasser, S. Calabrò, and M. Kuschnerov are with Huawei Technologies D\"usseldorf GmbH, Munich Research Center, Germany, e-mail: georg.bocherer@huawei.com}}

\IEEEoverridecommandlockouts
\IEEEpubid{\begin{minipage}[t]{\textwidth}\ \\[15pt] \centering \fbox{\begin{minipage}[t]{.9\textwidth}
    \footnotesize{\copyright 2023 IEEE. Personal use of this material is permitted. Permission from IEEE must be obtained for all other uses, in any current or future media, including reprinting/republishing this material for advertising or promotional purposes, creating new collective works, for resale or redistribution to servers or lists, or reuse of any copyrighted component of this work in other works.}
\end{minipage}} \end{minipage} }

\maketitle
\IEEEpubidadjcol

\input{tex/abstract}

\IEEEpeerreviewmaketitle

\input{tex/introduction}
\input{tex/imdd_refs}
\input{tex/spiking}
\input{tex/bss2}
\input{tex/training}
\input{tex/results}
\input{tex/conclusion}
\input{tex/acknowledgment}

\ifCLASSOPTIONcaptionsoff
  \newpage
\fi

\bibliographystyle{IEEEtranDOI}
\bibliography{IEEEabrv, bib}

\end{document}

%% file: tex/acronyms.tex
\newacronym{adc}{ADC}{analog-to-digital converter}
\newacronym{adex}{AdEx}{adaptive exponential integrate-and-fire}
\newacronym{ai}{AI}{artificial intelligence}
\newacronym{anncore}{ANNCORE}{analog neural network core}
\newacronym{ann}{ANN}{artificial neural network}
\newacronym{awgn}{AWGN}{additive white Gaussian noise}
\newacronym{asic}{ASIC}{application-specific integrated circuit}
\newacronym{bss-2}{\mbox{BSS-2}}{\mbox{BrainScaleS-2}}
\newacronym{ber}{BER}{bit error rate}
\newacronym{bptt}{BPTT}{backpropagation through time}
\newacronym{cadc}{CADC}{columnar \acrshort{adc}}
\newacronym{cd}{CD}{chromatic dispersion}
\newacronym{ctle}{CTLE}{continuous-time linear equalizer}
\newacronym{ce}{CE}{cross-entropy}
\newacronym{cmos}{CMOS}{complementary metal–oxide–semiconductor}
\newacronym{dc}{DC}{direct current}
\newacronym{dfe}{DFE}{decision feedback equalization}
\newacronym{dsp}{DSP}{digital signal processing}
\newacronym{dram}{DRAM}{dynamic random access memory}
\newacronym{fec}{FEC}{forward error correction}
\newacronym{ffe}{FFE}{feed forward equalizer}
\newacronym{fpga}{FPGA}{Field Programmable Gate Array}
\newacronym{hd}{HD}{hard decision}
\newacronym{imc}{IMC}{in-memory-computing}
\newacronym{li}{LI}{leaky integrator}
\newacronym{itl}{ITL}{in-the-loop}
\newacronym{imdd}{IM/DD}{intensity-modulation direct-detection}
\newacronym{lif}{LIF}{leaky integrate-and-fire}
\newacronym{lmmse}{LMMSE}{linear minimum mean square error}
\newacronym{le}{LE}{linear equalization}
\newacronym{pd}{PD}{photodiode}
\newacronym{pam4}{PAM-4}{pulse-amplitude-modulation 4-level}
\newacronym{mse}{MSE}{mean squared error}
\newacronym{nn}{NN}{neural network}
\newacronym{nlll}{NLLL}{negative log likelihood loss}
\newacronym{nrz}{NRZ}{non-return-to-zero}
\newacronym{snn}{SNN}{spiking neural network}
\newacronym{rnn}{RNN}{recurrent neural network}
\newacronym{rrc}{RRC}{root-raised-cosine}
\newacronym{sd}{SD}{soft decision}
\newacronym{simd}{SIMD}{single instruction, multiple data}
\newacronym{snr}{SNR}{signal-to-noise ratio}
\newacronym{ode}{ODE}{ordinary differential equation}
\newacronym{vnle}{VNLE}{Volterra nonlinear equalizer}
\newacronym{isi}{ISI}{inter-symbol interference}
\newacronym[\glslongpluralkey={orders of magnitude},
]{oom}{OOM}{order of magnitude}

%% file: tex/abstract.tex
\begin{abstract}
Neuromorphic computing implementing spiking neural networks (SNN) is a promising technology for reducing the footprint of optical transceivers, as required by the fast-paced growth of data center traffic. In this work, an SNN nonlinear demapper is designed and evaluated on a simulated intensity-modulation direct-detection link with chromatic dispersion.
The SNN demapper is implemented in software and on the analog neuromorphic hardware system BrainScaleS-2 (BSS-2).
For comparison, linear equalization (LE), Volterra nonlinear equalization (VNLE), and nonlinear demapping by an artifical neural network (ANN) implemented in software are considered.
At a pre-forward error correction bit error rate of \num[exponent-product=\cdot]{2e-3}, the software SNN outperforms LE by \SI{1.5}{dB}, VNLE by \SI{0.3}{dB} and the ANN by \SI{0.5}{dB}.
The hardware penalty of the SNN on BSS-2 is only \SI{0.2}{dB}, i.e., also on hardware, the SNN performs better than all software implementations of the reference approaches.
Hence, this work demonstrates that SNN demappers implemented on electrical analog hardware can realize powerful and accurate signal processing fulfilling the strict requirements of optical communications.
\end{abstract}

%% file: tex/introduction.tex
\section{Introduction}

\IEEEPARstart{T}{he} fast-paced growth of data center traffic is the driver behind the increase in bit rate and, at the same time, the footprint reduction of the optical transceivers. This trend results in an urgent need to decrease the power consumption per bit. Whereas evolutionary steps can mitigate the problem, the exponential traffic growth asks for a paradigm shift. To resolve this dilemma, recent research envisions moving parts of \gls{dsp} to analog frontends with lower power consumption. 

One approach is photonic neuromorphic computing~\cite{shastri2021photonics}, which has been proposed, e.g., for \gls{cd} compensation and nonlinear equalization in short-reach optical transmission~\cite{li2021micro,ranzini2021experimental,sozos2021photonic}. However, although photonics can operate faster than electronic hardware, the latter scales better in terms of footprint and power consumption.

The return to analog electrical adaptive equalizers is also gaining traction, e.g., in \cite{caruso2022real}, the transmitter \gls{dsp} feeds two electrical \gls{nrz} signals to an analog \gls{pam4} encoder, whose output is filtered by a \gls{ctle} and a 3-tap \gls{ffe}. 

At the same time, the research community is striving to implement more powerful nonlinear algorithms, e.g. based on \gls{ai} techniques, on analog electronics.
An important subfield is \gls{imc}~\cite{burr2017neuromorphic}, which aims for efficient calculation of vector-matrix multiplications.
Research on \gls{imc} is mainly driven by the urgent need for \gls{ai} accelerators for \glspl{ann}.
Eventually, \gls{imc} may enable the use of \glspl{ann} for signal processing in the data path of communication systems, see, e.g.,~\cite{eldebiky2022power}. 

Analog electronic neuromorphic computing offers an alternative path towards \gls{ai}-based signal processing. \Glspl{snn}~\cite{gerstner2014neuronal} in analog hardware~\cite{pehle2022brainscales2}, adopt the brain’s unique power efficiency by imitating the basic functioning of the human brain. They combine the sparse representation of information by event-based spiking signals with power efficient \gls{imc}. 
In \cite{arnold2022spiking}, we have shown  that \gls{snn} \glspl{ffe} simulated in software can compensate nonlinear impairments in \gls{imdd} links.
In~\cite{bansbach2022spiking}, \gls{snn} \gls{dfe} is considered for compensating severe linear \gls{isi}.

Recently, \gls{itl} training of \glspl{snn} on analog hardware~\cite{schmitt2017neuromorphic} has shown promising results by achieving state-of-the-art performance in inference tasks~\cite{cramer2022surrogate}.
In \cite{arnold2022spikingneuro}, we presented preliminary results on the design and evaluation of an \gls{snn} demapper on the analog neuromorphic \gls{bss-2} system~\cite{pehle2022brainscales2}.
Specifically, we considered the detection of a \gls{pam4} signal in a simulated \gls{imdd} link, which is impaired by \gls{cd} and \gls{awgn}, as displayed in \cref{fig:imdd_link}.
Our results in \cite{arnold2022spikingneuro} show that \glspl{snn} emulated on the neuromorphic 
\gls{bss-2} hardware outperform linear equalization in software, while the gap between software and hardware \gls{snn} is slightly below \SI{1}{dB}.

In this work, we detail and extend our previous work on \gls{snn}-based neuromorphic demapping \cite{arnold2022spikingneuro}.
For the same \gls{imdd} link model as in \cite{arnold2022spikingneuro} (see~\cref{fig:imdd_link}), we reduce the \gls{snn} software-hardware penalty to below \SI{.2}{dB}. We achieve this by optimizing the hardware operation point, tuning the training procedure, and adjusting the input-spike encoding.
We compare the proposed solution with software implementations of a linear equalizer, a $5$-th order \gls{vnle}, and a nonlinear \gls{ann} demapper.
Despite the nonzero hardware penalty, our hardware \gls{snn} demapper performs better than the considered simulated reference algorithms. At the assumed \gls{fec} \gls{ber} threshold of \num[exponent-product=\cdot]{2e-3}, the gain over a linear equalizer is approximately \SI{1.5}{dB}.

The remainder of this work is organized as follows.
In \cref{sec:imdd_refs}, we outline the \gls{imdd} link and explain the implementation of the reference demappers.
\cref{sec:ssn_eq} details the \gls{snn} demapper and the input encoding scheme.
Subsequently, we provide an overview of the \gls{bss-2} platform in \cref{sec:bss-2}.
The training procedure is explained in \cref{sec:training}.
In \cref{sec:results}, we show our results and in \cref{sec:conclusion}, we present our conclusions.

%% file: tex/imdd_refs.tex
\section{\Acrshort{imdd} Model and Reference Demappers}
\label{sec:imdd_refs}

\begin{figure}[t]
    \centering
    \tikzset{
        panel/.style={
            inner sep=0pt,
            outer sep=0pt,
            execute at begin node={\tikzset{anchor=center, inner sep=.33333em}}},
        label/.style={
            anchor=north west,
            inner sep=0,
            outer sep=0}}

    \begin{tikzpicture}
        \node[panel, anchor=north west] (a) at (0,  7) {
            \input{./figures/imdd_model.tex}};
        \node[label] at (a.north west) {\textbf{A}};
    \end{tikzpicture}
        
    \caption{The simulated \acrshort{imdd} link schematics. A bit sequence is mapped at the transmitter (Tx) to a \acrshort{pam4} signal and is impaired by \acrshort{cd} in the fiber. At the receiver (Rx), after square-law detection, \acrshort{awgn} is added. An equalizer/demapper recovers the transmitted bits.}
    \label{fig:imdd_link}
\end{figure}
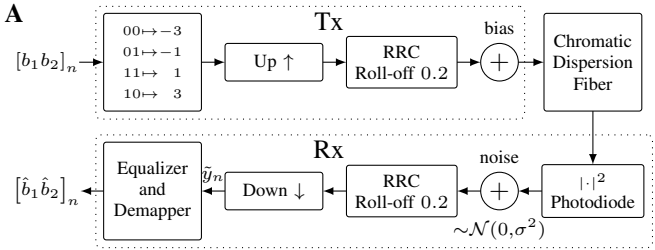

In this section, we detail our \gls{imdd} link model and specify the reference algorithms, i.e., \gls{le} and \gls{vnle} followed by \gls{hd} demapping, and \gls{ann} nonlinear demapping. All reference demappers are simulated in double-precision floating-point arithmetic, except for the \gls{ann}, which uses single-precision floating-point arithmetic. The considered \gls{ann} and \gls{vnle} architectures are rather complex, i.e., the \gls{ann} has two nonlinear hidden layers and the \gls{vnle} uses the full filter length also for the higher order terms. The purpose of considering complex \gls{ann} and \gls{vnle} processing is to benchmark what performance we can achieve by nonlinear processing without considering resource usage, and then to compare the \gls{snn} performance to such benchmark.

\newcommand{\vecy}{\boldsymbol{y}}

\subsection{Simulated \acrshort{imdd} Link}\label{sec:simulated imdd link}
We simulate the transmission of \gls{pam4} symbols in the O-band at a baudrate of \SI{112}{GBd}. Assuming an \gls{fec} overhead of \SI{12}{\percent} with a \gls{ber} threshold of \num[exponent-product=\cdot]{2e-3}, we target a corresponding net bit rate of \SI{200}{\giga\bit\per\second}.

We display the simulated link in Fig.~\ref{fig:imdd_link} and the corresponding parameters in Table~\ref{tab:imdd_link}.
At the transmitter, a bit sequence $[b_1b_2]^N$ is mapped to a length $N$ \gls{pam4} signal $\vecy=y^N$ according to a Gray-labelled alphabet $\mathcal{A}=\{-3, -1, 1, 3\}$.
This signal is upsampled, \gls{rrc} filtered, and offset by a bias.
The resulting sequence is impaired by \gls{cd}, modelled linearly following, e.g., \cite[Sec.~3.2]{mello2021digital}, to simulate the effect of the fibre on the propagating optical signal.
We assume that the power dissipated into the fiber is low and we ignore fiber non-linearities in our simulated link.
At the receiver, the signal goes through a \gls{pd}, which is modeled as a square-law device, and \gls{awgn} is added.
The resulting signal is \gls{rrc} filtered and downsampled, resulting in the received sequence $\tilde{\vecy}=\Tilde{y}^N$.
Finally, bit decisions $[\hat{b}_1\hat{b}_2]^N$ are output by the respective device. We index the bit sequence and signal elements with $n$, $0\leq n < N$. Note that a constellation with non-equidistant signal points to precompensate the squaring of the PD may be beneficial, however, this is beyond the scope of this work.

\begin{table}[!t]
    \begin{center}
        \caption{Parameters of the Simulated \Acrshort{imdd} Link}
        \vspace{-5px}
        \label{tab:imdd_link}
        \begin{tabular}{cc}
            \hline \hline
            Parameter & Value \\ \hline
            Net bit rate & \SI{200}{\giga\bit\per\second} \\
            \Acrshort{fec} threshold & \SI{12}{\percent} \\
            Baudrate & \SI{112}{GBd} \\
            Wavelength $\lambda$ & \SI{1270}{\nano\meter} \\
            Dispersion $D_\text{CD}$ & -\SI{5}{\pico\second\per\nano\meter\per\kilo\meter} \\
            Fiber length $l$ & \SI{4}{\kilo\meter} \\ \hline
            Alphabet $\mathcal{A}$ & $\lbrace -3, -1, 1, 3\rbrace$ \\
            Sequence length $N$ & 10000 \\
            Bias $b$ & $2.25$ \\
            \Acrshort{rrc} roll-off $\alpha$ & $0.2$ \\
            Upsampling $n_\text{up}$ & $3$ \\
            Downsampling $n_\text{down}$ & $3$ \\
            \hline \hline
        \end{tabular}
    \end{center}
    Remarks on the simulated IM/DD link parameters:
    \begin{enumerate}
        \item Wavelength and dispersion are in the range specified in \cite[Table~9.6]{itut202150gpon}.
        \item For the considered baudrate and fiber length, the dispersion in terms of delay spread between the frequency components at $\pm$ Nyquist frequency is 1.35 symbols. 
        \item The bias results in a carrier-to-signal-power-ratio (CSPR) of \SI{9.6}{dB}.
        \item The combination of CD and PD results in a band limitation, despite the fact that CD alone acts as an allpass filter. Consider
        \begin{align*}
        |\text{signal}+\text{carrier}|^2=|\text{signal}|^2+2\text{Re}(\text{signal}\cdot\text{carrier})+|\text{carrier}|^2.
        \end{align*}
        For the considered parameters, CD and PD cause for the linear term $2\text{Re}(\text{signal}\cdot\text{carrier})$ an attenuation of \SI{6.2}{dB} at the Nyquist frequency, compared to frequency 0.
    \end{enumerate}
\end{table}
\newcommand{\ty}{\tilde{y}}

\subsection{Linear Minimum Mean Squared Error (LMMSE) Equalization}
\newcommand{\ntap}{n_\textnormal{tap}}
Our first reference detector consists of \gls{le} followed by \gls{hd} demapping. To simplify the notation in the following, we specify the samples considered for equalizing the $n$-th sample via double-indexing,
\begin{align}
\tilde{\vecy}_n &= \left[ \ty_{n,0}, \ty_{n,1}, \dotsc, \ty_{n,\ntap-1} \right], \label{eq:chunk} \\ 
&:=\left[ \tilde{y}_{n-\lfloor\ntap/2\rfloor}, \dotsc,\tilde{y}_n, \dotsc, \tilde{y}_{n+\lfloor\ntap/2\rfloor} \right].
\end{align}
Specifically, the \gls{le} calculates
\begin{align}
    \hat{y}_n = c + \sum_{j=0}^{\ntap-1} \tilde{y}_{n,j} h_j, 
\end{align}
where the bias $c$ accounts for residual \gls{dc} and $\boldsymbol{h}$ are the filter coefficients.
The number $\ntap$ of taps is the filter width and is assumed to be odd. 
We use data-aided training to calculate $\boldsymbol{h}$ and $c$ so as to minimize the \gls{mse}, $\nicefrac{1}{N}\sum_n(\hat{y}_n - y_n)^2$. To this end, we form the feature matrix
\begin{align}
\boldsymbol{A}=
    \left[\begin{matrix}
    \dotsb & 1 & 1 & 1 & \dotsb\\
    \dotsb & \tilde{\vecy}_{n-1}^\top & \tilde{\vecy}_{n}^\top  & \tilde{\vecy}_{n+1}^\top & \dotsb 
    \end{matrix}\right]^\top
\end{align}
and we then solve
\begin{align}
    [c^*,\boldsymbol{h}^*]=\argmin_{c, \boldsymbol{h}}\lVert\boldsymbol{A}[c,\boldsymbol{h}]^\top-\boldsymbol{y}^\top\rVert_2^2.\label{eq:le  training}
\end{align}
The demapper calculates an \gls{hd} $[\hat{b}_{1}\hat{b}_{2}]_n$ from the equalized sample $\hat{y}_n$ via three decision boundaries, which are chosen such that the \gls{ber} is minimized. Note that at the transmitter, the signal points in the \gls{pam4} constellation are equidistant, while the received signal points are not equidistant anymore, because of the nonlinear transfer function of the \gls{pd}.
The \gls{le} cannot compensate nonlinear distortions, so the received signal points remain non-equidistant after \gls{le}.
This is compensated in part by the demapper, as the decision boundaries are optimized with respect to the received and equalized signal points $\hat{y}_n$, not the transmitted signal points $y_n$.
In the following, we refer to the combination of a \gls{le} and a memoryless demapper by \gls{lmmse} equalization.

\subsection{Volterra Nonlinear Equalizer (VNLE)}
\label{sec:vnle}
\newcommand{\Avnle}{\boldsymbol{A}_\text{VNLE}}
\newcommand{\Ale}{\boldsymbol{A}_\text{LE}}
\newcommand{\setf}{\mathcal{F}}
\newcommand{\setx}{\mathcal{X}}
\newcommand{\sety}{\mathcal{Y}}
\newcommand{\vecf}{\boldsymbol{f}}
As nonlinear reference, we consider a $5$-th order \gls{vnle}, see, e.g., \cite[Chap.~14]{benedetto1999principles}, \cite{schadler2021soft}. The \gls{vnle} is very similar to the \gls{le}, with the difference that the \gls{vnle} feature matrix contains additional columns for the monomials of order higher than $1$.
The $0$-th and $1$-st order feature vectors considered by the \gls{le} are
\begin{align}
    \vecf_{n,0}&=[1],\\
    \vecf_{n,1}&=\Bigl[\ty_{n,j} \Big| 0\leq j < \ntap\Bigr]=\tilde{\vecy}_n.
\end{align}
Accordingly, the feature vectors of order two and higher are defined by
\begin{align}
    \vecf_{n,2}&=\Bigl[ \ty_{n,j} \cdot \ty_{n,k} \Big| 0\leq j\leq k < \ntap \Bigr],\\
    \vecf_{n,3}&=\Bigl[ \ty_{n,j} \cdot \ty_{n,k} \cdot \ty_{n,\ell} \Big| 0\leq j\leq k\leq \ell < \ntap\Bigr],\\
    &\vdots\nonumber
\end{align}
The $n$-th row of the feature matrix then consists of the concatenation of the feature vectors $\vecf_{n,m}$, $m=0,\dotsc, 5$. The number of features of order $m$ is given by
\begin{align}
    \text{length}(\vecf_{n,m}) = {m + \ntap - 1\choose m}
\end{align}
and accordingly, for $\ntap=7$, the total number of coefficients of the \gls{vnle} is
\begin{align}
\sum_{m=0}^5\text{length}(\vecf_{n,m})=792.    
\end{align}
The optimization of the coefficients and the specification of the demapper is data-aided and follows exactly the \gls{le} procedure described above, where the key step is solving \eqref{eq:le  training}.

\subsection{Nonlinear \Acrshort{ann} Demapper}
\label{sec:ann}
We consider an \gls{ann} with $\ntap=7$ input units, a first hidden layer with 40 neurons, followed by a second hidden layer with 20 neurons, both activated by the $\tanh$ function, and a linear output layer with 4 neurons.
The output values are interpreted as $\log$-probabilities providing a \gls{sd} on the \gls{pam4} symbols.
A symbol-wise \gls{hd} is obtained by choosing the symbol of highest probability and the bitwise \gls{hd} is obtained from the symbol decisions via the Gray label.

%% file: figures/imdd_model.tex
\begin{tikzpicture}
    \tikzset{
    device/.style={
        draw,
        rounded corners=1pt,
        font=\scriptsize},
    outer/.style={
        draw,
        dotted,
        rounded corners=2pt,
        anchor=south west},
    tx/.style={
        minimum width = 5.7cm,
        minimum height = 1.5cm},
    rx/.style={
        minimum width = 7.4cm,
        minimum height = 1.5cm}}
    \matrix[column sep=0.3cm, row sep=0.4cm, minimum height=0.5cm]{
        \node[outer sep=0, inner sep=0](src) {$\scriptstyle \left[b_1b_2\right]_n$};
        & \node[device,align=center, minimum width=1.3cm, minimum height=1.3cm](map){
            $\scriptstyle 00\mapsto -3$\\
            $\scriptstyle 01\mapsto -1$\\
            $\scriptstyle 11\mapsto \;\;\,1$\\
            $\scriptstyle 10\mapsto \;\;\,3$};
        & \node[device, minimum width=1.3cm](up){Up $\uparrow$};
        & \node[device, minimum width=1.3cm, align=center](rrc){RRC\\Roll-off $0.2$};
        & \node[circle, draw, inner sep=0cm](bias){$+$};
        & \node[device, align=center, minimum width=1.3cm, minimum height=1.3cm](cd){Chromatic\\Dispersion\\Fiber};
        \\
        \node[outer sep=0, inner sep=0](sink) {$\scriptstyle\left[\hat{b}_1\hat{b}_2\right]_n$};
        & \node[device, align=center, minimum width=1.3cm, minimum height=1.3cm](demap){Equalizer\\and\\Demapper};
        & \node[device, minimum width=1.3cm](down){Down $\downarrow$};
        & \node[device, minimum width=1.3cm, align=center](rrc_rx){RRC\\Roll-off $0.2$};
        & \node[circle, draw, inner sep=0cm](noise){$+$};
        & \node[device, minimum width=1.3cm, align=center](pd){$\scriptstyle |\cdot|^2$\\Photodiode};
        \\
    };

    \draw[-latex](src)--(map);
    \draw[-latex](map)--(up);
    \draw[-latex](up)--(rrc);
    \draw[-latex](rrc)--(bias);
    \draw[-latex](bias)--(cd);
    \draw[-latex](cd)--(pd);
    \draw[-latex](pd)--(noise);
    \draw[-latex](noise)--(rrc_rx);
    \draw[-latex](rrc_rx)--(down);
    \draw[-latex](down)--node[above]{$\scriptstyle \Tilde{y}_n$}(demap);
    \draw[-latex](demap)--(sink);

    \node[outer, tx](outer_tx) at (-3.10, 0.1) {};
    \node[outer, rx](outer_rx) at (-3.10, -1.61) {};

    \node[above=0.0cm of bias, font=\scriptsize]{bias};
    \node[above=0.0cm of noise, font=\scriptsize]{noise};
    \node[below=0.0cm of noise](noise_val){$\scriptstyle \sim\mathcal{N}(0,\sigma^2)$};

    \node (tx) at (0, 1.4) {Tx};
    \node (tx) at (0, -0.3) {Rx};
\end{tikzpicture}

%% file: tex/spiking.tex
\section{Spiking Neural Networks for Equalization}
\label{sec:ssn_eq}

\begin{figure*}[!t]
    \centering
    \tikzset{
        panel/.style={
            inner sep=0pt,
            outer sep=0pt,
            execute at begin node={\tikzset{anchor=center, inner sep=.33333em}}},
        label/.style={
            anchor=north west,
            inner sep=0,
            outer sep=0}}

    \begin{tikzpicture}
        \node[panel, anchor=north west] (a) at (1.5,  2.8) {
            \input{figures/lif_dynamic.pgf}};
        \node[label] at (0., 2.8) {\textbf{A}};

        \node[panel, anchor=north west] (b) at (0,  0) {
            \input{figures/network_setup.tex}};
        \node[label] at (0., -0.2) {\textbf{B}};

        \node[panel, anchor=north west] (c) at (12,  3.1) {
            \input{figures/linear_encoding.pgf}};
        \node[label] at (c.north west) {\textbf{C}};
    \end{tikzpicture}
    \vspace{-10px}
    \caption{
        \textbf{(A)} \Acrshort{lif} dynamic: a \acrshort{lif} neuron receives input spikes (yellow arrows), causing a synaptic current (purple) onto the neuron's membrane, thereby deflecting its membrane potential (blue). As the potential crosses the threshold (gray), the neuron sends out a spike (blue arrow) and is reset.
        \textbf{(B)} The considered \acrshort{snn} setup for joint equalization and demapping. A chunk $\Tilde{\boldsymbol{y}}_n$ of samples is translated to spike times of input neurons, projected onto one hidden \acrshort{lif} layer. A readout layer of \acrshort{li} neurons adjust their membrane voltage by integrating these hidden spike events. Symbol-level $\log$-probabilities are calculated by taking the maximum membrane value over time of the readout neurons, allowing to infer bit-decisions.
        \textbf{(C)} Schematic drawing of the input spike-encoding: a sample $\Tilde{y}_{n,\ell}$ (purple dotted) is translated into spike times according to the distance to reference points assigned to $10$ neurons (gray lines). Spikes occurring after a cutoff time $t_\text{c}$ are omitted.}
    \label{fig:demapper_setup}
\end{figure*}
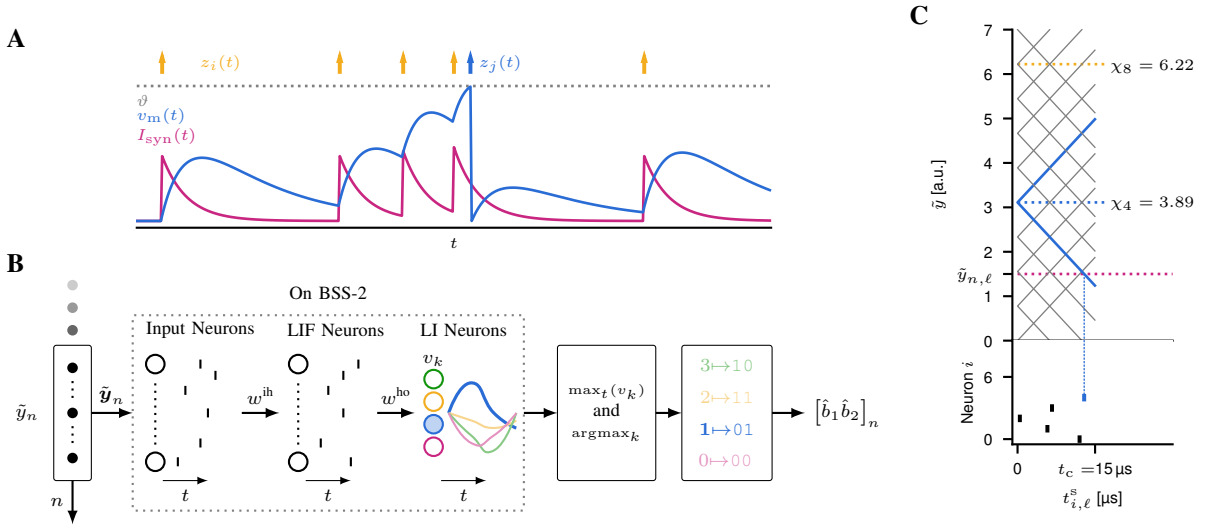

This section outlines the \gls{snn} demappers. We detail their emulation on \gls{bss-2} in Sec.~\ref{sec:bss-2}.

\Glspl{snn} consist of neurons, evolving in time $t$, and communicating via binary spike events.
The \gls{lif} spiking neuron model \cite[Sec.~1.3]{gerstner2014neuronal} captures some of the core dynamics observed in biological neurons while at the same time maintaining a tractable complexity.
\Gls{lif} neurons integrate synaptic input current $I(t)$ onto their internal membrane voltage state $v_\mathrm{m}(t)$ according to the dynamics described by the \gls{ode}
\begin{equation}
    \tau_\mathrm{m}\frac{\text{d}v_\mathrm{m}(t)}{\text{d}t} = \left[v_\text{l} - v_\mathrm{m}(t)\right] + R_\text{l} \cdot I(t).
    \label{eq:lifdyn}
\end{equation}
Here, $\tau_\mathrm{m}$ is the membrane time constant, $R_\text{l}$ is the leakage resistance, and $v_\text{l}$ is the leakage potential.
When the membrane potential reaches a threshold potential $\vartheta$ at time $t^\text{s}$, the neuron emits a spike $z(t) = \delta\left( t - t^\text{s} \right)$, with $\delta$ being the Dirac delta distribution, and $v_\mathrm{m}$ is reset to a potential $v_\mathrm{m}(t^\text{s}) = v_\mathrm{r}$.
The synaptic current $I$ is induced by \emph{presynaptic} neurons $\lbrace n_i \rbrace$, projecting spike events $z_i(t) = \delta \left(t -t_i^\text{s}\right)$ at times $\lbrace t_i^\text{s} \rbrace$ onto the \emph{postsynaptic} neuron through synapses with weights $w_i$, thereby causing an exponentially decaying current described by the \gls{ode}
\begin{equation}
    \frac{\text{d}I(t)}{\text{d}t} = -\frac{I(t)}{\tau_\mathrm{s}} + \sum_i w_i z_i(t).
\end{equation}
$\tau_\text{s}$ denotes the synaptic time constant.
The \gls{lif} dynamics are exemplified in \cref{fig:demapper_setup}A.
Neurons with a disabled spiking mechanism and membrane dynamics according to \eqref{eq:lifdyn}, are referred to as \gls{li} neurons \cite[Sec.~1.3]{gerstner2014neuronal}.

In the following, we consider an \gls{snn} with the structure outlined in \cite{arnold2022spikingneuro} and depicted in \cref{fig:demapper_setup}B.
It consists of one hidden layer constituted by $N^\text{h}$ \gls{lif} neurons $\lbrace n^\text{h}_j\rbrace$, projecting its spike events onto one output layer with $N^\text{o}=4$ non-spiking \gls{li} readout neurons $\lbrace n^\text{o}_k \rbrace$.
The hidden layer receives spike events from the input layer, encoding a set of input samples $\Tilde{\boldsymbol{y}}_n$.
The readout layer's outputs are translated to symbol-level $\log$-probabilities.
Spike-input encoding and output decoding are explained in the following.

\paragraph{Input Spike-Encoding}
To demap a sample $\Tilde{y}_n$, we consider the chunk $\Tilde{\vecy}_n$ defined in \eqref{eq:chunk} and assign to each sample $\Tilde{y}_{n,\ell}$ a set of input neurons $\lbrace n^\text{i}_{i, \ell} \rbrace _{i=0}^{<\bar{N}^\text{i}_\ell}$, encoding the sample value in their spike times $\lbrace t^\text{s}_{i, \ell} \rbrace$.
Here, $\ell$ indexes the samples within $\Tilde{\vecy}_n$ and $\bar{N}^\text{i}_\ell \in \mathbb{N}$ is the number of neurons associated to sample $\Tilde{y}_{n,\ell}$, such that $N^\text{i} = \sum_{\ell=0}^{<n_\text{tap}} \bar{N}^\text{i}_\ell$ is the size of the input layer.
Further, we assign each input neuron $n_{i, \ell}^\text{i}$ a \emph{reference point} $\chi_{i,\ell}$, which we choose together with $\bar{N}^\text{i}_\ell$ to be independent of $\ell$, $\chi_{i,\ell}=\chi_{i}$ and $\bar{N}^\text{i}_\ell = \bar{N}^\text{i}$.
Finally, we compute the spike time $t_{i,\ell}^\text{s}$ by scaling the distance of $\Tilde{y}_{n,\ell}$ to $\chi_{i}$,
\begin{equation}
    t^\text{s}_{i,\ell} = \alpha \big| \Tilde{y}_{n,\ell} -  \chi_{i} \big| + o,
\end{equation}
where $\alpha$ is a scaling factor and $o$ is an offset.
This spike-encoding preserves all information and encodes the value $\Tilde{y}_{n,\ell}$ redundantly in $\bar{N}^\text{i}$ spike times in order to increase the network's activity and enrich information in time.
The values $\chi_i$, $\bar{N}^\text{i}$ and $\alpha$ are subject to tuning and are chosen to augment the network's activity by the right amount to achieve optimal performance.
Here, the $\chi_i$s are equidistantly spaced in the domain of $\Tilde{y}_{n,\ell}$ and $\alpha$ is selected to obtain spike times comparable to the membrane time constants.
Note, while larger $\bar{N}^\text{i}$ increases the network's complexity, it potentially stabilizes the network's performance on a noisy analog substrate like \gls{bss-2}, see \cref{sec:bss-2}.
We further introduce a cutoff time $t_\text{c}$ after which input neurons are not allowed to emit spike events and we do not expect the \gls{snn} to gain information afterwards.
The spike encoding is illustrated in \cref{fig:demapper_setup}C.
A sample $\Tilde{y}_{n,\ell}$ (purple, dotted) is translated into spike times according to its distance to the reference points, e.g., the distance to $\chi_4$ (blue, solid) results in a spike from input neuron $n_{4,\ell}^\text{i}$ depicted in blue.
The input neuron $n_{8,\ell}^\text{i}$, corresponding to $\chi_8$ (yellow, dotted), remains silent.

\paragraph{Output Membrane-Decoding}
Each of the 4 neurons in the readout layer is assigned to one element in the \gls{pam4} alphabet $\mathcal{A}$.
We take the maximum membrane voltages produced over time, i.e., $\hat{s}_k = \max_t v_k(t)$, which are interpreted as $\log$-probabilities providing an \gls{sd} on the \gls{pam4} symbols.
Then, the symbol-wise \gls{hd} is obtained by choosing the symbol of highest probability and the bitwise \gls{hd} is obtained from the symbol decisions via the Gray label.
Hence, the network learns to place its hidden layer spike events in time, such that the membrane trace of the correct output neuron is deflected upwards while the traces of the others are suppressed.

\subsection{Training}
Time-discretized \glspl{snn} are mathematically \glspl{rnn} \cite{neftci2019superspike} and can be trained with the gradient-based \gls{bptt}.
For this, the derivative of the spiking output of the \gls{lif} neurons with respect to their membrane potential has to be known.
This derivative is ill-defined due to the threshold activation function.
Often surrogate gradients, smoothing out the neurons' activation functions, are used to bypass this issue and allow backpropagating the gradient.
Here, we rely on the SuperSpike \cite{neftci2019superspike} surrogate gradient.
The model parameters are optimized by the Adam optimizer \cite{kingma2012adam}.

In the simulation, the \gls{snn} is integrated with a step size $\Delta t = \SI{0.5}{\micro\second}$ for $T=\SI{30}{\micro\second}$, suitable for \gls{bss-2} (see \cref{sec:bss-2}).
Our simulated \gls{snn} demappers are implemented using the \texttt{PyTorch}-based Norse~\cite{norse2021} framework.
To estimate the hardware gradient for the \gls{snn} demappers emulated on \gls{bss-2} in continuous time with the \gls{bptt} algorithm, we discretize the hardware observables assuming the same step size, see \cref{sec:bss-2}.
We allocate $\bar{N}^\text{i} = 10$ input neurons per sample of which only a subset is active, depending on the sample value, see \cref{fig:snn_results}.
We use the cross entropy on the max-over-time voltage values as the objective function.
The parameters of the \gls{snn} and input encoding are listed in \cref{tab:snn_params}.
In case of emulation on \gls{bss-2}, these parameters are used as calibration targets, resp.\ for \gls{itl} training (see \cref{sec:bss-2}).

\begin{table}[!t]
    \begin{center}
        \caption{Parameters of the Input Encoding and \Acrshort{snn}}
        \vspace{-5px}
        \label{tab:snn_params}
        \begin{tabular}{cc|cc}
            \hline \hline
            \multicolumn{2}{c|}{Input Encoding} & \multicolumn{2}{c}{SNN} \\ \hline
            $\alpha$ & \SI{8}{\micro\second} & $N^\text{h}$ & $40$ \\
            $o$ sim.\ & \SI{0}{\micro\second} & $N^\text{o}$ & $4$  \\
            $o$ \gls{bss-2} & \SI{1}{\micro\second} & $\tau_\text{m}$ & \SI{6}{\micro\second} \\
            $\bar{N}^\text{i}$ & $10$ & $\tau_\text{s}$ & \SI{6}{\micro\second} \\ [-0.3em]
            $\chi_i$ & $0.\bar{7} \cdot i \text{\, with \,} i \in \mathbb{N}_0^{<\bar{N}^\text{i}}$ & $v_\text{l}$ & \SI{0}{\au} \\
            $o$ & \SI{0}{\micro\second} in SW and \SI{1}{\micro\second} on \gls{bss-2} & $v_\text{r}$ & \SI{0}{\au} \\
            $t_\text{c}$ & \SI{15}{\micro\second} & $\vartheta$ & \SI{1}{\au} \\ [0.2em] \cline{1-4}
            \multicolumn{2}{c}{$\Delta t$} & \multicolumn{2}{c}{\SI{0.5}{\micro\second}} \\
            \multicolumn{2}{c}{$T$} & \multicolumn{2}{c}{\SI{30}{\micro\second}} \\
            \hline \hline
        \end{tabular}
    \end{center}
\end{table}

%% file: figures/lif_dynamic.pgf
\begingroup%
\makeatletter%
\begin{pgfpicture}%
\pgfpathrectangle{\pgfpointorigin}{\pgfqpoint{3.500000in}{1.300000in}}%
\pgfusepath{use as bounding box, clip}%
\begin{pgfscope}%
\pgfsetbuttcap%
\pgfsetmiterjoin%
\definecolor{currentfill}{rgb}{1.000000,1.000000,1.000000}%
\pgfsetfillcolor{currentfill}%
\pgfsetlinewidth{0.000000pt}%
\definecolor{currentstroke}{rgb}{1.000000,1.000000,1.000000}%
\pgfsetstrokecolor{currentstroke}%
\pgfsetdash{}{0pt}%
\pgfpathmoveto{\pgfqpoint{0.000000in}{0.000000in}}%
\pgfpathlineto{\pgfqpoint{3.500000in}{0.000000in}}%
\pgfpathlineto{\pgfqpoint{3.500000in}{1.300000in}}%
\pgfpathlineto{\pgfqpoint{0.000000in}{1.300000in}}%
\pgfpathclose%
\pgfusepath{fill}%
\end{pgfscope}%
\begin{pgfscope}%
\pgfsetbuttcap%
\pgfsetmiterjoin%
\definecolor{currentfill}{rgb}{1.000000,1.000000,1.000000}%
\pgfsetfillcolor{currentfill}%
\pgfsetlinewidth{0.000000pt}%
\definecolor{currentstroke}{rgb}{0.000000,0.000000,0.000000}%
\pgfsetstrokecolor{currentstroke}%
\pgfsetstrokeopacity{0.000000}%
\pgfsetdash{}{0pt}%
\pgfpathmoveto{\pgfqpoint{0.090000in}{0.255556in}}%
\pgfpathlineto{\pgfqpoint{3.410000in}{0.255556in}}%
\pgfpathlineto{\pgfqpoint{3.410000in}{1.210000in}}%
\pgfpathlineto{\pgfqpoint{0.090000in}{1.210000in}}%
\pgfpathclose%
\pgfusepath{fill}%
\end{pgfscope}%
\begin{pgfscope}%
\pgfpathrectangle{\pgfqpoint{0.090000in}{0.255556in}}{\pgfqpoint{3.320000in}{0.954444in}}%
\pgfusepath{clip}%
\pgfsetbuttcap%
\pgfsetmiterjoin%
\definecolor{currentfill}{rgb}{0.160784,0.423529,0.835294}%
\pgfsetfillcolor{currentfill}%
\pgfsetlinewidth{1.003750pt}%
\definecolor{currentstroke}{rgb}{0.160784,0.423529,0.835294}%
\pgfsetstrokecolor{currentstroke}%
\pgfsetdash{}{0pt}%
\pgfpathmoveto{\pgfqpoint{1.836320in}{1.160510in}}%
\pgfpathlineto{\pgfqpoint{1.846280in}{1.125160in}}%
\pgfpathlineto{\pgfqpoint{1.839640in}{1.125160in}}%
\pgfpathlineto{\pgfqpoint{1.839640in}{1.068601in}}%
\pgfpathlineto{\pgfqpoint{1.833000in}{1.068601in}}%
\pgfpathlineto{\pgfqpoint{1.833000in}{1.125160in}}%
\pgfpathlineto{\pgfqpoint{1.826360in}{1.125160in}}%
\pgfpathclose%
\pgfusepath{stroke,fill}%
\end{pgfscope}%
\begin{pgfscope}%
\pgfpathrectangle{\pgfqpoint{0.090000in}{0.255556in}}{\pgfqpoint{3.320000in}{0.954444in}}%
\pgfusepath{clip}%
\pgfsetbuttcap%
\pgfsetmiterjoin%
\definecolor{currentfill}{rgb}{0.949020,0.674510,0.113725}%
\pgfsetfillcolor{currentfill}%
\pgfsetlinewidth{1.003750pt}%
\definecolor{currentstroke}{rgb}{0.949020,0.674510,0.113725}%
\pgfsetstrokecolor{currentstroke}%
\pgfsetdash{}{0pt}%
\pgfpathmoveto{\pgfqpoint{0.222800in}{1.160510in}}%
\pgfpathlineto{\pgfqpoint{0.232760in}{1.125160in}}%
\pgfpathlineto{\pgfqpoint{0.226120in}{1.125160in}}%
\pgfpathlineto{\pgfqpoint{0.226120in}{1.068601in}}%
\pgfpathlineto{\pgfqpoint{0.219480in}{1.068601in}}%
\pgfpathlineto{\pgfqpoint{0.219480in}{1.125160in}}%
\pgfpathlineto{\pgfqpoint{0.212840in}{1.125160in}}%
\pgfpathclose%
\pgfusepath{stroke,fill}%
\end{pgfscope}%
\begin{pgfscope}%
\pgfpathrectangle{\pgfqpoint{0.090000in}{0.255556in}}{\pgfqpoint{3.320000in}{0.954444in}}%
\pgfusepath{clip}%
\pgfsetbuttcap%
\pgfsetmiterjoin%
\definecolor{currentfill}{rgb}{0.949020,0.674510,0.113725}%
\pgfsetfillcolor{currentfill}%
\pgfsetlinewidth{1.003750pt}%
\definecolor{currentstroke}{rgb}{0.949020,0.674510,0.113725}%
\pgfsetstrokecolor{currentstroke}%
\pgfsetdash{}{0pt}%
\pgfpathmoveto{\pgfqpoint{1.152400in}{1.160510in}}%
\pgfpathlineto{\pgfqpoint{1.162360in}{1.125160in}}%
\pgfpathlineto{\pgfqpoint{1.155720in}{1.125160in}}%
\pgfpathlineto{\pgfqpoint{1.155720in}{1.068601in}}%
\pgfpathlineto{\pgfqpoint{1.149080in}{1.068601in}}%
\pgfpathlineto{\pgfqpoint{1.149080in}{1.125160in}}%
\pgfpathlineto{\pgfqpoint{1.142440in}{1.125160in}}%
\pgfpathclose%
\pgfusepath{stroke,fill}%
\end{pgfscope}%
\begin{pgfscope}%
\pgfpathrectangle{\pgfqpoint{0.090000in}{0.255556in}}{\pgfqpoint{3.320000in}{0.954444in}}%
\pgfusepath{clip}%
\pgfsetbuttcap%
\pgfsetmiterjoin%
\definecolor{currentfill}{rgb}{0.949020,0.674510,0.113725}%
\pgfsetfillcolor{currentfill}%
\pgfsetlinewidth{1.003750pt}%
\definecolor{currentstroke}{rgb}{0.949020,0.674510,0.113725}%
\pgfsetstrokecolor{currentstroke}%
\pgfsetdash{}{0pt}%
\pgfpathmoveto{\pgfqpoint{1.484400in}{1.160510in}}%
\pgfpathlineto{\pgfqpoint{1.494360in}{1.125160in}}%
\pgfpathlineto{\pgfqpoint{1.487720in}{1.125160in}}%
\pgfpathlineto{\pgfqpoint{1.487720in}{1.068601in}}%
\pgfpathlineto{\pgfqpoint{1.481080in}{1.068601in}}%
\pgfpathlineto{\pgfqpoint{1.481080in}{1.125160in}}%
\pgfpathlineto{\pgfqpoint{1.474440in}{1.125160in}}%
\pgfpathclose%
\pgfusepath{stroke,fill}%
\end{pgfscope}%
\begin{pgfscope}%
\pgfpathrectangle{\pgfqpoint{0.090000in}{0.255556in}}{\pgfqpoint{3.320000in}{0.954444in}}%
\pgfusepath{clip}%
\pgfsetbuttcap%
\pgfsetmiterjoin%
\definecolor{currentfill}{rgb}{0.949020,0.674510,0.113725}%
\pgfsetfillcolor{currentfill}%
\pgfsetlinewidth{1.003750pt}%
\definecolor{currentstroke}{rgb}{0.949020,0.674510,0.113725}%
\pgfsetstrokecolor{currentstroke}%
\pgfsetdash{}{0pt}%
\pgfpathmoveto{\pgfqpoint{1.750000in}{1.160510in}}%
\pgfpathlineto{\pgfqpoint{1.759960in}{1.125160in}}%
\pgfpathlineto{\pgfqpoint{1.753320in}{1.125160in}}%
\pgfpathlineto{\pgfqpoint{1.753320in}{1.068601in}}%
\pgfpathlineto{\pgfqpoint{1.746680in}{1.068601in}}%
\pgfpathlineto{\pgfqpoint{1.746680in}{1.125160in}}%
\pgfpathlineto{\pgfqpoint{1.740040in}{1.125160in}}%
\pgfpathclose%
\pgfusepath{stroke,fill}%
\end{pgfscope}%
\begin{pgfscope}%
\pgfpathrectangle{\pgfqpoint{0.090000in}{0.255556in}}{\pgfqpoint{3.320000in}{0.954444in}}%
\pgfusepath{clip}%
\pgfsetbuttcap%
\pgfsetmiterjoin%
\definecolor{currentfill}{rgb}{0.949020,0.674510,0.113725}%
\pgfsetfillcolor{currentfill}%
\pgfsetlinewidth{1.003750pt}%
\definecolor{currentstroke}{rgb}{0.949020,0.674510,0.113725}%
\pgfsetstrokecolor{currentstroke}%
\pgfsetdash{}{0pt}%
\pgfpathmoveto{\pgfqpoint{2.746000in}{1.160510in}}%
\pgfpathlineto{\pgfqpoint{2.755960in}{1.125160in}}%
\pgfpathlineto{\pgfqpoint{2.749320in}{1.125160in}}%
\pgfpathlineto{\pgfqpoint{2.749320in}{1.068601in}}%
\pgfpathlineto{\pgfqpoint{2.742680in}{1.068601in}}%
\pgfpathlineto{\pgfqpoint{2.742680in}{1.125160in}}%
\pgfpathlineto{\pgfqpoint{2.736040in}{1.125160in}}%
\pgfpathclose%
\pgfusepath{stroke,fill}%
\end{pgfscope}%
\begin{pgfscope}%
\definecolor{textcolor}{rgb}{0.000000,0.000000,0.000000}%
\pgfsetstrokecolor{textcolor}%
\pgfsetfillcolor{textcolor}%
\pgftext[x=1.750000in,y=0.200000in,,top]{\color{textcolor}\sffamily\fontsize{6.000000}{7.200000}\selectfont \(\displaystyle t\)}%
\end{pgfscope}%
\begin{pgfscope}%
\pgfpathrectangle{\pgfqpoint{0.090000in}{0.255556in}}{\pgfqpoint{3.320000in}{0.954444in}}%
\pgfusepath{clip}%
\pgfsetbuttcap%
\pgfsetroundjoin%
\pgfsetlinewidth{1.003750pt}%
\definecolor{currentstroke}{rgb}{0.501961,0.501961,0.501961}%
\pgfsetstrokecolor{currentstroke}%
\pgfsetdash{{1.000000pt}{1.650000pt}}{0.000000pt}%
\pgfpathmoveto{\pgfqpoint{0.090000in}{0.997901in}}%
\pgfpathlineto{\pgfqpoint{3.410000in}{0.997901in}}%
\pgfusepath{stroke}%
\end{pgfscope}%
\begin{pgfscope}%
\pgfpathrectangle{\pgfqpoint{0.090000in}{0.255556in}}{\pgfqpoint{3.320000in}{0.954444in}}%
\pgfusepath{clip}%
\pgfsetrectcap%
\pgfsetroundjoin%
\pgfsetlinewidth{1.003750pt}%
\definecolor{currentstroke}{rgb}{0.788235,0.149020,0.509804}%
\pgfsetstrokecolor{currentstroke}%
\pgfsetdash{}{0pt}%
\pgfpathmoveto{\pgfqpoint{0.090000in}{0.290905in}}%
\pgfpathlineto{\pgfqpoint{0.216160in}{0.290905in}}%
\pgfpathlineto{\pgfqpoint{0.222800in}{0.630263in}}%
\pgfpathlineto{\pgfqpoint{0.236080in}{0.597176in}}%
\pgfpathlineto{\pgfqpoint{0.249360in}{0.567315in}}%
\pgfpathlineto{\pgfqpoint{0.262640in}{0.540365in}}%
\pgfpathlineto{\pgfqpoint{0.275920in}{0.516042in}}%
\pgfpathlineto{\pgfqpoint{0.289200in}{0.494092in}}%
\pgfpathlineto{\pgfqpoint{0.302480in}{0.474281in}}%
\pgfpathlineto{\pgfqpoint{0.315760in}{0.456402in}}%
\pgfpathlineto{\pgfqpoint{0.329040in}{0.440266in}}%
\pgfpathlineto{\pgfqpoint{0.342320in}{0.425703in}}%
\pgfpathlineto{\pgfqpoint{0.355600in}{0.412560in}}%
\pgfpathlineto{\pgfqpoint{0.368880in}{0.400699in}}%
\pgfpathlineto{\pgfqpoint{0.382160in}{0.389994in}}%
\pgfpathlineto{\pgfqpoint{0.395440in}{0.380333in}}%
\pgfpathlineto{\pgfqpoint{0.408720in}{0.371614in}}%
\pgfpathlineto{\pgfqpoint{0.422000in}{0.363745in}}%
\pgfpathlineto{\pgfqpoint{0.435280in}{0.356643in}}%
\pgfpathlineto{\pgfqpoint{0.448560in}{0.350233in}}%
\pgfpathlineto{\pgfqpoint{0.461840in}{0.344449in}}%
\pgfpathlineto{\pgfqpoint{0.481760in}{0.336812in}}%
\pgfpathlineto{\pgfqpoint{0.501680in}{0.330265in}}%
\pgfpathlineto{\pgfqpoint{0.521600in}{0.324651in}}%
\pgfpathlineto{\pgfqpoint{0.541520in}{0.319838in}}%
\pgfpathlineto{\pgfqpoint{0.561440in}{0.315712in}}%
\pgfpathlineto{\pgfqpoint{0.588000in}{0.311110in}}%
\pgfpathlineto{\pgfqpoint{0.614560in}{0.307362in}}%
\pgfpathlineto{\pgfqpoint{0.641120in}{0.304310in}}%
\pgfpathlineto{\pgfqpoint{0.674320in}{0.301277in}}%
\pgfpathlineto{\pgfqpoint{0.714160in}{0.298530in}}%
\pgfpathlineto{\pgfqpoint{0.760640in}{0.296230in}}%
\pgfpathlineto{\pgfqpoint{0.813760in}{0.294438in}}%
\pgfpathlineto{\pgfqpoint{0.880160in}{0.293020in}}%
\pgfpathlineto{\pgfqpoint{0.966480in}{0.291991in}}%
\pgfpathlineto{\pgfqpoint{1.092640in}{0.291315in}}%
\pgfpathlineto{\pgfqpoint{1.145760in}{0.291177in}}%
\pgfpathlineto{\pgfqpoint{1.152400in}{0.630522in}}%
\pgfpathlineto{\pgfqpoint{1.165680in}{0.597409in}}%
\pgfpathlineto{\pgfqpoint{1.178960in}{0.567525in}}%
\pgfpathlineto{\pgfqpoint{1.192240in}{0.540554in}}%
\pgfpathlineto{\pgfqpoint{1.205520in}{0.516214in}}%
\pgfpathlineto{\pgfqpoint{1.218800in}{0.494246in}}%
\pgfpathlineto{\pgfqpoint{1.232080in}{0.474420in}}%
\pgfpathlineto{\pgfqpoint{1.245360in}{0.456528in}}%
\pgfpathlineto{\pgfqpoint{1.258640in}{0.440379in}}%
\pgfpathlineto{\pgfqpoint{1.271920in}{0.425806in}}%
\pgfpathlineto{\pgfqpoint{1.285200in}{0.412653in}}%
\pgfpathlineto{\pgfqpoint{1.298480in}{0.400783in}}%
\pgfpathlineto{\pgfqpoint{1.311760in}{0.390070in}}%
\pgfpathlineto{\pgfqpoint{1.325040in}{0.380401in}}%
\pgfpathlineto{\pgfqpoint{1.338320in}{0.371675in}}%
\pgfpathlineto{\pgfqpoint{1.351600in}{0.363800in}}%
\pgfpathlineto{\pgfqpoint{1.364880in}{0.356693in}}%
\pgfpathlineto{\pgfqpoint{1.378160in}{0.350279in}}%
\pgfpathlineto{\pgfqpoint{1.391440in}{0.344490in}}%
\pgfpathlineto{\pgfqpoint{1.411360in}{0.336847in}}%
\pgfpathlineto{\pgfqpoint{1.431280in}{0.330295in}}%
\pgfpathlineto{\pgfqpoint{1.451200in}{0.324677in}}%
\pgfpathlineto{\pgfqpoint{1.471120in}{0.319860in}}%
\pgfpathlineto{\pgfqpoint{1.477760in}{0.318412in}}%
\pgfpathlineto{\pgfqpoint{1.484400in}{0.656395in}}%
\pgfpathlineto{\pgfqpoint{1.497680in}{0.620760in}}%
\pgfpathlineto{\pgfqpoint{1.510960in}{0.588599in}}%
\pgfpathlineto{\pgfqpoint{1.524240in}{0.559574in}}%
\pgfpathlineto{\pgfqpoint{1.537520in}{0.533379in}}%
\pgfpathlineto{\pgfqpoint{1.550800in}{0.509738in}}%
\pgfpathlineto{\pgfqpoint{1.564080in}{0.488401in}}%
\pgfpathlineto{\pgfqpoint{1.577360in}{0.469146in}}%
\pgfpathlineto{\pgfqpoint{1.590640in}{0.451767in}}%
\pgfpathlineto{\pgfqpoint{1.603920in}{0.436083in}}%
\pgfpathlineto{\pgfqpoint{1.617200in}{0.421928in}}%
\pgfpathlineto{\pgfqpoint{1.630480in}{0.409154in}}%
\pgfpathlineto{\pgfqpoint{1.643760in}{0.397624in}}%
\pgfpathlineto{\pgfqpoint{1.657040in}{0.387219in}}%
\pgfpathlineto{\pgfqpoint{1.670320in}{0.377829in}}%
\pgfpathlineto{\pgfqpoint{1.683600in}{0.369354in}}%
\pgfpathlineto{\pgfqpoint{1.696880in}{0.361705in}}%
\pgfpathlineto{\pgfqpoint{1.710160in}{0.354802in}}%
\pgfpathlineto{\pgfqpoint{1.723440in}{0.348572in}}%
\pgfpathlineto{\pgfqpoint{1.736720in}{0.342950in}}%
\pgfpathlineto{\pgfqpoint{1.743360in}{0.340347in}}%
\pgfpathlineto{\pgfqpoint{1.750000in}{0.677233in}}%
\pgfpathlineto{\pgfqpoint{1.763280in}{0.639566in}}%
\pgfpathlineto{\pgfqpoint{1.776560in}{0.605572in}}%
\pgfpathlineto{\pgfqpoint{1.789840in}{0.574892in}}%
\pgfpathlineto{\pgfqpoint{1.803120in}{0.547203in}}%
\pgfpathlineto{\pgfqpoint{1.816400in}{0.522214in}}%
\pgfpathlineto{\pgfqpoint{1.829680in}{0.499661in}}%
\pgfpathlineto{\pgfqpoint{1.842960in}{0.479308in}}%
\pgfpathlineto{\pgfqpoint{1.856240in}{0.460939in}}%
\pgfpathlineto{\pgfqpoint{1.869520in}{0.444360in}}%
\pgfpathlineto{\pgfqpoint{1.882800in}{0.429398in}}%
\pgfpathlineto{\pgfqpoint{1.896080in}{0.415895in}}%
\pgfpathlineto{\pgfqpoint{1.909360in}{0.403709in}}%
\pgfpathlineto{\pgfqpoint{1.922640in}{0.392710in}}%
\pgfpathlineto{\pgfqpoint{1.935920in}{0.382784in}}%
\pgfpathlineto{\pgfqpoint{1.949200in}{0.373826in}}%
\pgfpathlineto{\pgfqpoint{1.962480in}{0.365741in}}%
\pgfpathlineto{\pgfqpoint{1.975760in}{0.358445in}}%
\pgfpathlineto{\pgfqpoint{1.989040in}{0.351860in}}%
\pgfpathlineto{\pgfqpoint{2.002320in}{0.345917in}}%
\pgfpathlineto{\pgfqpoint{2.015600in}{0.340553in}}%
\pgfpathlineto{\pgfqpoint{2.035520in}{0.333472in}}%
\pgfpathlineto{\pgfqpoint{2.055440in}{0.327401in}}%
\pgfpathlineto{\pgfqpoint{2.075360in}{0.322196in}}%
\pgfpathlineto{\pgfqpoint{2.095280in}{0.317733in}}%
\pgfpathlineto{\pgfqpoint{2.115200in}{0.313907in}}%
\pgfpathlineto{\pgfqpoint{2.141760in}{0.309640in}}%
\pgfpathlineto{\pgfqpoint{2.168320in}{0.306165in}}%
\pgfpathlineto{\pgfqpoint{2.201520in}{0.302713in}}%
\pgfpathlineto{\pgfqpoint{2.234720in}{0.300042in}}%
\pgfpathlineto{\pgfqpoint{2.274560in}{0.297621in}}%
\pgfpathlineto{\pgfqpoint{2.321040in}{0.295595in}}%
\pgfpathlineto{\pgfqpoint{2.374160in}{0.294017in}}%
\pgfpathlineto{\pgfqpoint{2.447200in}{0.292675in}}%
\pgfpathlineto{\pgfqpoint{2.540160in}{0.291768in}}%
\pgfpathlineto{\pgfqpoint{2.686240in}{0.291185in}}%
\pgfpathlineto{\pgfqpoint{2.739360in}{0.291091in}}%
\pgfpathlineto{\pgfqpoint{2.746000in}{0.630439in}}%
\pgfpathlineto{\pgfqpoint{2.759280in}{0.597335in}}%
\pgfpathlineto{\pgfqpoint{2.772560in}{0.567458in}}%
\pgfpathlineto{\pgfqpoint{2.785840in}{0.540494in}}%
\pgfpathlineto{\pgfqpoint{2.799120in}{0.516159in}}%
\pgfpathlineto{\pgfqpoint{2.812400in}{0.494197in}}%
\pgfpathlineto{\pgfqpoint{2.825680in}{0.474376in}}%
\pgfpathlineto{\pgfqpoint{2.838960in}{0.456488in}}%
\pgfpathlineto{\pgfqpoint{2.852240in}{0.440343in}}%
\pgfpathlineto{\pgfqpoint{2.865520in}{0.425773in}}%
\pgfpathlineto{\pgfqpoint{2.878800in}{0.412623in}}%
\pgfpathlineto{\pgfqpoint{2.892080in}{0.400756in}}%
\pgfpathlineto{\pgfqpoint{2.905360in}{0.390046in}}%
\pgfpathlineto{\pgfqpoint{2.918640in}{0.380379in}}%
\pgfpathlineto{\pgfqpoint{2.931920in}{0.371656in}}%
\pgfpathlineto{\pgfqpoint{2.945200in}{0.363782in}}%
\pgfpathlineto{\pgfqpoint{2.958480in}{0.356677in}}%
\pgfpathlineto{\pgfqpoint{2.971760in}{0.350264in}}%
\pgfpathlineto{\pgfqpoint{2.985040in}{0.344477in}}%
\pgfpathlineto{\pgfqpoint{3.004960in}{0.336836in}}%
\pgfpathlineto{\pgfqpoint{3.024880in}{0.330285in}}%
\pgfpathlineto{\pgfqpoint{3.044800in}{0.324669in}}%
\pgfpathlineto{\pgfqpoint{3.064720in}{0.319853in}}%
\pgfpathlineto{\pgfqpoint{3.084640in}{0.315724in}}%
\pgfpathlineto{\pgfqpoint{3.111200in}{0.311121in}}%
\pgfpathlineto{\pgfqpoint{3.137760in}{0.307371in}}%
\pgfpathlineto{\pgfqpoint{3.164320in}{0.304317in}}%
\pgfpathlineto{\pgfqpoint{3.197520in}{0.301283in}}%
\pgfpathlineto{\pgfqpoint{3.237360in}{0.298534in}}%
\pgfpathlineto{\pgfqpoint{3.283840in}{0.296232in}}%
\pgfpathlineto{\pgfqpoint{3.336960in}{0.294439in}}%
\pgfpathlineto{\pgfqpoint{3.403360in}{0.293021in}}%
\pgfpathlineto{\pgfqpoint{3.403360in}{0.293021in}}%
\pgfusepath{stroke}%
\end{pgfscope}%
\begin{pgfscope}%
\pgfpathrectangle{\pgfqpoint{0.090000in}{0.255556in}}{\pgfqpoint{3.320000in}{0.954444in}}%
\pgfusepath{clip}%
\pgfsetrectcap%
\pgfsetroundjoin%
\pgfsetlinewidth{1.003750pt}%
\definecolor{currentstroke}{rgb}{0.160784,0.423529,0.835294}%
\pgfsetstrokecolor{currentstroke}%
\pgfsetdash{}{0pt}%
\pgfpathmoveto{\pgfqpoint{0.090000in}{0.290905in}}%
\pgfpathlineto{\pgfqpoint{0.216160in}{0.290905in}}%
\pgfpathlineto{\pgfqpoint{0.222800in}{0.319185in}}%
\pgfpathlineto{\pgfqpoint{0.229440in}{0.345580in}}%
\pgfpathlineto{\pgfqpoint{0.236080in}{0.370191in}}%
\pgfpathlineto{\pgfqpoint{0.242720in}{0.393116in}}%
\pgfpathlineto{\pgfqpoint{0.249360in}{0.414447in}}%
\pgfpathlineto{\pgfqpoint{0.256000in}{0.434270in}}%
\pgfpathlineto{\pgfqpoint{0.262640in}{0.452669in}}%
\pgfpathlineto{\pgfqpoint{0.269280in}{0.469722in}}%
\pgfpathlineto{\pgfqpoint{0.275920in}{0.485503in}}%
\pgfpathlineto{\pgfqpoint{0.282560in}{0.500083in}}%
\pgfpathlineto{\pgfqpoint{0.289200in}{0.513529in}}%
\pgfpathlineto{\pgfqpoint{0.295840in}{0.525904in}}%
\pgfpathlineto{\pgfqpoint{0.302480in}{0.537269in}}%
\pgfpathlineto{\pgfqpoint{0.309120in}{0.547680in}}%
\pgfpathlineto{\pgfqpoint{0.315760in}{0.557192in}}%
\pgfpathlineto{\pgfqpoint{0.322400in}{0.565855in}}%
\pgfpathlineto{\pgfqpoint{0.329040in}{0.573719in}}%
\pgfpathlineto{\pgfqpoint{0.335680in}{0.580830in}}%
\pgfpathlineto{\pgfqpoint{0.342320in}{0.587231in}}%
\pgfpathlineto{\pgfqpoint{0.348960in}{0.592964in}}%
\pgfpathlineto{\pgfqpoint{0.355600in}{0.598068in}}%
\pgfpathlineto{\pgfqpoint{0.362240in}{0.602579in}}%
\pgfpathlineto{\pgfqpoint{0.368880in}{0.606534in}}%
\pgfpathlineto{\pgfqpoint{0.375520in}{0.609966in}}%
\pgfpathlineto{\pgfqpoint{0.382160in}{0.612905in}}%
\pgfpathlineto{\pgfqpoint{0.388800in}{0.615383in}}%
\pgfpathlineto{\pgfqpoint{0.395440in}{0.617428in}}%
\pgfpathlineto{\pgfqpoint{0.408720in}{0.620322in}}%
\pgfpathlineto{\pgfqpoint{0.422000in}{0.621786in}}%
\pgfpathlineto{\pgfqpoint{0.435280in}{0.621997in}}%
\pgfpathlineto{\pgfqpoint{0.448560in}{0.621114in}}%
\pgfpathlineto{\pgfqpoint{0.461840in}{0.619279in}}%
\pgfpathlineto{\pgfqpoint{0.475120in}{0.616620in}}%
\pgfpathlineto{\pgfqpoint{0.488400in}{0.613249in}}%
\pgfpathlineto{\pgfqpoint{0.501680in}{0.609269in}}%
\pgfpathlineto{\pgfqpoint{0.521600in}{0.602350in}}%
\pgfpathlineto{\pgfqpoint{0.541520in}{0.594526in}}%
\pgfpathlineto{\pgfqpoint{0.568080in}{0.583064in}}%
\pgfpathlineto{\pgfqpoint{0.601280in}{0.567665in}}%
\pgfpathlineto{\pgfqpoint{0.661040in}{0.538748in}}%
\pgfpathlineto{\pgfqpoint{0.714160in}{0.513320in}}%
\pgfpathlineto{\pgfqpoint{0.754000in}{0.495045in}}%
\pgfpathlineto{\pgfqpoint{0.787200in}{0.480529in}}%
\pgfpathlineto{\pgfqpoint{0.820400in}{0.466745in}}%
\pgfpathlineto{\pgfqpoint{0.853600in}{0.453732in}}%
\pgfpathlineto{\pgfqpoint{0.886800in}{0.441506in}}%
\pgfpathlineto{\pgfqpoint{0.920000in}{0.430063in}}%
\pgfpathlineto{\pgfqpoint{0.953200in}{0.419384in}}%
\pgfpathlineto{\pgfqpoint{0.986400in}{0.409444in}}%
\pgfpathlineto{\pgfqpoint{1.019600in}{0.400212in}}%
\pgfpathlineto{\pgfqpoint{1.052800in}{0.391651in}}%
\pgfpathlineto{\pgfqpoint{1.092640in}{0.382210in}}%
\pgfpathlineto{\pgfqpoint{1.132480in}{0.373616in}}%
\pgfpathlineto{\pgfqpoint{1.145760in}{0.370928in}}%
\pgfpathlineto{\pgfqpoint{1.152400in}{0.397896in}}%
\pgfpathlineto{\pgfqpoint{1.159040in}{0.422999in}}%
\pgfpathlineto{\pgfqpoint{1.165680in}{0.446340in}}%
\pgfpathlineto{\pgfqpoint{1.172320in}{0.468014in}}%
\pgfpathlineto{\pgfqpoint{1.178960in}{0.488114in}}%
\pgfpathlineto{\pgfqpoint{1.185600in}{0.506726in}}%
\pgfpathlineto{\pgfqpoint{1.192240in}{0.523933in}}%
\pgfpathlineto{\pgfqpoint{1.198880in}{0.539813in}}%
\pgfpathlineto{\pgfqpoint{1.205520in}{0.554440in}}%
\pgfpathlineto{\pgfqpoint{1.212160in}{0.567885in}}%
\pgfpathlineto{\pgfqpoint{1.218800in}{0.580214in}}%
\pgfpathlineto{\pgfqpoint{1.225440in}{0.591490in}}%
\pgfpathlineto{\pgfqpoint{1.232080in}{0.601773in}}%
\pgfpathlineto{\pgfqpoint{1.238720in}{0.611120in}}%
\pgfpathlineto{\pgfqpoint{1.245360in}{0.619585in}}%
\pgfpathlineto{\pgfqpoint{1.252000in}{0.627219in}}%
\pgfpathlineto{\pgfqpoint{1.258640in}{0.634070in}}%
\pgfpathlineto{\pgfqpoint{1.265280in}{0.640184in}}%
\pgfpathlineto{\pgfqpoint{1.271920in}{0.645604in}}%
\pgfpathlineto{\pgfqpoint{1.278560in}{0.650372in}}%
\pgfpathlineto{\pgfqpoint{1.285200in}{0.654527in}}%
\pgfpathlineto{\pgfqpoint{1.291840in}{0.658105in}}%
\pgfpathlineto{\pgfqpoint{1.298480in}{0.661141in}}%
\pgfpathlineto{\pgfqpoint{1.305120in}{0.663669in}}%
\pgfpathlineto{\pgfqpoint{1.311760in}{0.665720in}}%
\pgfpathlineto{\pgfqpoint{1.318400in}{0.667324in}}%
\pgfpathlineto{\pgfqpoint{1.325040in}{0.668508in}}%
\pgfpathlineto{\pgfqpoint{1.338320in}{0.669724in}}%
\pgfpathlineto{\pgfqpoint{1.351600in}{0.669564in}}%
\pgfpathlineto{\pgfqpoint{1.364880in}{0.668204in}}%
\pgfpathlineto{\pgfqpoint{1.378160in}{0.665802in}}%
\pgfpathlineto{\pgfqpoint{1.391440in}{0.662497in}}%
\pgfpathlineto{\pgfqpoint{1.404720in}{0.658415in}}%
\pgfpathlineto{\pgfqpoint{1.418000in}{0.653668in}}%
\pgfpathlineto{\pgfqpoint{1.437920in}{0.645518in}}%
\pgfpathlineto{\pgfqpoint{1.457840in}{0.636387in}}%
\pgfpathlineto{\pgfqpoint{1.477760in}{0.626520in}}%
\pgfpathlineto{\pgfqpoint{1.484400in}{0.651384in}}%
\pgfpathlineto{\pgfqpoint{1.491040in}{0.674310in}}%
\pgfpathlineto{\pgfqpoint{1.497680in}{0.695408in}}%
\pgfpathlineto{\pgfqpoint{1.504320in}{0.714780in}}%
\pgfpathlineto{\pgfqpoint{1.510960in}{0.732523in}}%
\pgfpathlineto{\pgfqpoint{1.517600in}{0.748730in}}%
\pgfpathlineto{\pgfqpoint{1.524240in}{0.763489in}}%
\pgfpathlineto{\pgfqpoint{1.530880in}{0.776882in}}%
\pgfpathlineto{\pgfqpoint{1.537520in}{0.788989in}}%
\pgfpathlineto{\pgfqpoint{1.544160in}{0.799883in}}%
\pgfpathlineto{\pgfqpoint{1.550800in}{0.809636in}}%
\pgfpathlineto{\pgfqpoint{1.557440in}{0.818315in}}%
\pgfpathlineto{\pgfqpoint{1.564080in}{0.825983in}}%
\pgfpathlineto{\pgfqpoint{1.570720in}{0.832700in}}%
\pgfpathlineto{\pgfqpoint{1.577360in}{0.838523in}}%
\pgfpathlineto{\pgfqpoint{1.584000in}{0.843507in}}%
\pgfpathlineto{\pgfqpoint{1.590640in}{0.847702in}}%
\pgfpathlineto{\pgfqpoint{1.597280in}{0.851157in}}%
\pgfpathlineto{\pgfqpoint{1.603920in}{0.853918in}}%
\pgfpathlineto{\pgfqpoint{1.610560in}{0.856027in}}%
\pgfpathlineto{\pgfqpoint{1.617200in}{0.857527in}}%
\pgfpathlineto{\pgfqpoint{1.623840in}{0.858456in}}%
\pgfpathlineto{\pgfqpoint{1.630480in}{0.858851in}}%
\pgfpathlineto{\pgfqpoint{1.637120in}{0.858746in}}%
\pgfpathlineto{\pgfqpoint{1.643760in}{0.858176in}}%
\pgfpathlineto{\pgfqpoint{1.650400in}{0.857170in}}%
\pgfpathlineto{\pgfqpoint{1.663680in}{0.853969in}}%
\pgfpathlineto{\pgfqpoint{1.676960in}{0.849361in}}%
\pgfpathlineto{\pgfqpoint{1.690240in}{0.843540in}}%
\pgfpathlineto{\pgfqpoint{1.703520in}{0.836678in}}%
\pgfpathlineto{\pgfqpoint{1.716800in}{0.828932in}}%
\pgfpathlineto{\pgfqpoint{1.730080in}{0.820438in}}%
\pgfpathlineto{\pgfqpoint{1.743360in}{0.811319in}}%
\pgfpathlineto{\pgfqpoint{1.750000in}{0.834839in}}%
\pgfpathlineto{\pgfqpoint{1.756640in}{0.856358in}}%
\pgfpathlineto{\pgfqpoint{1.763280in}{0.875989in}}%
\pgfpathlineto{\pgfqpoint{1.769920in}{0.893840in}}%
\pgfpathlineto{\pgfqpoint{1.776560in}{0.910013in}}%
\pgfpathlineto{\pgfqpoint{1.783200in}{0.924606in}}%
\pgfpathlineto{\pgfqpoint{1.789840in}{0.937710in}}%
\pgfpathlineto{\pgfqpoint{1.796480in}{0.949412in}}%
\pgfpathlineto{\pgfqpoint{1.803120in}{0.959795in}}%
\pgfpathlineto{\pgfqpoint{1.809760in}{0.968937in}}%
\pgfpathlineto{\pgfqpoint{1.816400in}{0.976912in}}%
\pgfpathlineto{\pgfqpoint{1.823040in}{0.983791in}}%
\pgfpathlineto{\pgfqpoint{1.829680in}{0.989639in}}%
\pgfpathlineto{\pgfqpoint{1.836320in}{0.994520in}}%
\pgfpathlineto{\pgfqpoint{1.842960in}{0.290905in}}%
\pgfpathlineto{\pgfqpoint{1.849600in}{0.305821in}}%
\pgfpathlineto{\pgfqpoint{1.856240in}{0.319741in}}%
\pgfpathlineto{\pgfqpoint{1.862880in}{0.332722in}}%
\pgfpathlineto{\pgfqpoint{1.869520in}{0.344813in}}%
\pgfpathlineto{\pgfqpoint{1.876160in}{0.356063in}}%
\pgfpathlineto{\pgfqpoint{1.882800in}{0.366518in}}%
\pgfpathlineto{\pgfqpoint{1.889440in}{0.376222in}}%
\pgfpathlineto{\pgfqpoint{1.896080in}{0.385216in}}%
\pgfpathlineto{\pgfqpoint{1.902720in}{0.393539in}}%
\pgfpathlineto{\pgfqpoint{1.909360in}{0.401229in}}%
\pgfpathlineto{\pgfqpoint{1.916000in}{0.408320in}}%
\pgfpathlineto{\pgfqpoint{1.922640in}{0.414847in}}%
\pgfpathlineto{\pgfqpoint{1.929280in}{0.420841in}}%
\pgfpathlineto{\pgfqpoint{1.942560in}{0.431348in}}%
\pgfpathlineto{\pgfqpoint{1.955840in}{0.440066in}}%
\pgfpathlineto{\pgfqpoint{1.969120in}{0.447192in}}%
\pgfpathlineto{\pgfqpoint{1.982400in}{0.452907in}}%
\pgfpathlineto{\pgfqpoint{1.995680in}{0.457373in}}%
\pgfpathlineto{\pgfqpoint{2.008960in}{0.460733in}}%
\pgfpathlineto{\pgfqpoint{2.022240in}{0.463118in}}%
\pgfpathlineto{\pgfqpoint{2.035520in}{0.464644in}}%
\pgfpathlineto{\pgfqpoint{2.048800in}{0.465416in}}%
\pgfpathlineto{\pgfqpoint{2.062080in}{0.465528in}}%
\pgfpathlineto{\pgfqpoint{2.075360in}{0.465062in}}%
\pgfpathlineto{\pgfqpoint{2.095280in}{0.463443in}}%
\pgfpathlineto{\pgfqpoint{2.115200in}{0.460914in}}%
\pgfpathlineto{\pgfqpoint{2.135120in}{0.457660in}}%
\pgfpathlineto{\pgfqpoint{2.161680in}{0.452460in}}%
\pgfpathlineto{\pgfqpoint{2.194880in}{0.444994in}}%
\pgfpathlineto{\pgfqpoint{2.234720in}{0.435201in}}%
\pgfpathlineto{\pgfqpoint{2.380800in}{0.398571in}}%
\pgfpathlineto{\pgfqpoint{2.427280in}{0.387960in}}%
\pgfpathlineto{\pgfqpoint{2.473760in}{0.378122in}}%
\pgfpathlineto{\pgfqpoint{2.520240in}{0.369095in}}%
\pgfpathlineto{\pgfqpoint{2.566720in}{0.360872in}}%
\pgfpathlineto{\pgfqpoint{2.613200in}{0.353425in}}%
\pgfpathlineto{\pgfqpoint{2.659680in}{0.346708in}}%
\pgfpathlineto{\pgfqpoint{2.706160in}{0.340669in}}%
\pgfpathlineto{\pgfqpoint{2.739360in}{0.336741in}}%
\pgfpathlineto{\pgfqpoint{2.746000in}{0.364272in}}%
\pgfpathlineto{\pgfqpoint{2.752640in}{0.389929in}}%
\pgfpathlineto{\pgfqpoint{2.759280in}{0.413814in}}%
\pgfpathlineto{\pgfqpoint{2.765920in}{0.436025in}}%
\pgfpathlineto{\pgfqpoint{2.772560in}{0.456652in}}%
\pgfpathlineto{\pgfqpoint{2.779200in}{0.475783in}}%
\pgfpathlineto{\pgfqpoint{2.785840in}{0.493501in}}%
\pgfpathlineto{\pgfqpoint{2.792480in}{0.509883in}}%
\pgfpathlineto{\pgfqpoint{2.799120in}{0.525005in}}%
\pgfpathlineto{\pgfqpoint{2.805760in}{0.538936in}}%
\pgfpathlineto{\pgfqpoint{2.812400in}{0.551743in}}%
\pgfpathlineto{\pgfqpoint{2.819040in}{0.563490in}}%
\pgfpathlineto{\pgfqpoint{2.825680in}{0.574236in}}%
\pgfpathlineto{\pgfqpoint{2.832320in}{0.584038in}}%
\pgfpathlineto{\pgfqpoint{2.838960in}{0.592951in}}%
\pgfpathlineto{\pgfqpoint{2.845600in}{0.601026in}}%
\pgfpathlineto{\pgfqpoint{2.852240in}{0.608310in}}%
\pgfpathlineto{\pgfqpoint{2.858880in}{0.614851in}}%
\pgfpathlineto{\pgfqpoint{2.865520in}{0.620691in}}%
\pgfpathlineto{\pgfqpoint{2.872160in}{0.625871in}}%
\pgfpathlineto{\pgfqpoint{2.878800in}{0.630432in}}%
\pgfpathlineto{\pgfqpoint{2.885440in}{0.634409in}}%
\pgfpathlineto{\pgfqpoint{2.892080in}{0.637838in}}%
\pgfpathlineto{\pgfqpoint{2.898720in}{0.640752in}}%
\pgfpathlineto{\pgfqpoint{2.905360in}{0.643183in}}%
\pgfpathlineto{\pgfqpoint{2.912000in}{0.645161in}}%
\pgfpathlineto{\pgfqpoint{2.918640in}{0.646712in}}%
\pgfpathlineto{\pgfqpoint{2.931920in}{0.648646in}}%
\pgfpathlineto{\pgfqpoint{2.945200in}{0.649180in}}%
\pgfpathlineto{\pgfqpoint{2.958480in}{0.648491in}}%
\pgfpathlineto{\pgfqpoint{2.971760in}{0.646737in}}%
\pgfpathlineto{\pgfqpoint{2.985040in}{0.644060in}}%
\pgfpathlineto{\pgfqpoint{2.998320in}{0.640586in}}%
\pgfpathlineto{\pgfqpoint{3.011600in}{0.636427in}}%
\pgfpathlineto{\pgfqpoint{3.024880in}{0.631684in}}%
\pgfpathlineto{\pgfqpoint{3.044800in}{0.623668in}}%
\pgfpathlineto{\pgfqpoint{3.064720in}{0.614800in}}%
\pgfpathlineto{\pgfqpoint{3.091280in}{0.602023in}}%
\pgfpathlineto{\pgfqpoint{3.124480in}{0.585101in}}%
\pgfpathlineto{\pgfqpoint{3.244000in}{0.523108in}}%
\pgfpathlineto{\pgfqpoint{3.283840in}{0.503742in}}%
\pgfpathlineto{\pgfqpoint{3.317040in}{0.488430in}}%
\pgfpathlineto{\pgfqpoint{3.350240in}{0.473935in}}%
\pgfpathlineto{\pgfqpoint{3.383440in}{0.460285in}}%
\pgfpathlineto{\pgfqpoint{3.403360in}{0.452504in}}%
\pgfpathlineto{\pgfqpoint{3.403360in}{0.452504in}}%
\pgfusepath{stroke}%
\end{pgfscope}%
\begin{pgfscope}%
\pgfsetrectcap%
\pgfsetmiterjoin%
\pgfsetlinewidth{0.803000pt}%
\definecolor{currentstroke}{rgb}{0.000000,0.000000,0.000000}%
\pgfsetstrokecolor{currentstroke}%
\pgfsetdash{}{0pt}%
\pgfpathmoveto{\pgfqpoint{0.090000in}{0.255556in}}%
\pgfpathlineto{\pgfqpoint{3.410000in}{0.255556in}}%
\pgfusepath{stroke}%
\end{pgfscope}%
\begin{pgfscope}%
\definecolor{textcolor}{rgb}{0.501961,0.501961,0.501961}%
\pgfsetstrokecolor{textcolor}%
\pgfsetfillcolor{textcolor}%
\pgftext[x=0.090000in,y=0.891852in,left,base]{\color{textcolor}\sffamily\fontsize{6.000000}{7.200000}\selectfont \(\displaystyle \vartheta\)}%
\end{pgfscope}%
\begin{pgfscope}%
\definecolor{textcolor}{rgb}{0.160784,0.423529,0.835294}%
\pgfsetstrokecolor{textcolor}%
\pgfsetfillcolor{textcolor}%
\pgftext[x=0.090000in,y=0.821152in,left,base]{\color{textcolor}\sffamily\fontsize{6.000000}{7.200000}\selectfont \(\displaystyle v_\mathrm{m}(t)\)}%
\end{pgfscope}%
\begin{pgfscope}%
\definecolor{textcolor}{rgb}{0.788235,0.149020,0.509804}%
\pgfsetstrokecolor{textcolor}%
\pgfsetfillcolor{textcolor}%
\pgftext[x=0.090000in,y=0.715103in,left,base]{\color{textcolor}\sffamily\fontsize{6.000000}{7.200000}\selectfont \(\displaystyle I_\mathrm{syn}(t)\)}%
\end{pgfscope}%
\begin{pgfscope}%
\definecolor{textcolor}{rgb}{0.949020,0.674510,0.113725}%
\pgfsetstrokecolor{textcolor}%
\pgfsetfillcolor{textcolor}%
\pgftext[x=0.422000in,y=1.096881in,left,base]{\color{textcolor}\sffamily\fontsize{6.000000}{7.200000}\selectfont \(\displaystyle z_i (t)\)}%
\end{pgfscope}%
\begin{pgfscope}%
\definecolor{textcolor}{rgb}{0.160784,0.423529,0.835294}%
\pgfsetstrokecolor{textcolor}%
\pgfsetfillcolor{textcolor}%
\pgftext[x=1.882800in,y=1.096881in,left,base]{\color{textcolor}\sffamily\fontsize{6.000000}{7.200000}\selectfont \(\displaystyle z_j (t)\)}%
\end{pgfscope}%
\end{pgfpicture}%
\makeatother%
\endgroup%

%% file: figures/network_setup.tex
\begin{tikzpicture}
    \tikzset{
        neuron/.style={
            circle,
            inner sep=0pt,
            outer sep=3pt,
            align=center,
            thick,
            minimum size=7pt},
        labelneuron/.style={
            circle,
            inner sep=1pt,
            align=center,
            minimum size=7pt,
            rounded corners=1pt},
        spike/.style={
            rectangle,
            fill=black!50,
            minimum width=1pt,
            minimum height=3pt,
            inner sep=0pt,
            draw=black!50},
        sample_outer/.style={
            rectangle,
            align=center,
            draw=black,
            minimum height=1.8cm,
            minimum width=0.5cm,
            rounded corners=1pt},
        input_outer/.style={
            rectangle,
            align=center,
            minimum height=1.8cm,
            minimum width=1.3cm,
            rounded corners=1pt},
        hidden_outer/.style={
            rectangle,
            align=center,
            minimum height=1.8cm,
            minimum width=1.3cm,
            rounded corners=1pt},
        output_outer/.style={
            rectangle,
            align=center,
            minimum height=1.8cm,
            minimum width=1.5cm,
            rounded corners=1pt},
        sample/.style={
            circle,
            very thick,
            inner sep=0pt,
            align=center,
            minimum size=4pt},
        outer_argmax/.style={
            rectangle,
            inner sep=4pt,
            align=center,
            rounded corners=1pt,
            minimum height=1.8cm,
            font=\scriptsize,
            draw=black},
        outer_demapper/.style={
            rectangle,
            inner sep=4pt,
            rounded corners=1pt,
            minimum height=1.8cm,
            draw=black,
            align=center}
    }

    \def\dt{0.05}
    \def\spacing{0.15}
    \def\inoff{1.5}
    \def\outoff{1.2}

    \node[color=black] (c_n) at (-0.3, 0.9+\inoff) {$\scriptstyle \Tilde{y}_{n}$};

    \node (n_begin) at (0.3, 4.6) {};
    \node[sample, fill=black!20] (s_n-4) at (0.3, 2.6+\inoff) {};
    \node[sample, fill=black!40] (s_n-3) at (0.3, 2.3+\inoff) {};
    \node[sample, fill=black!60] (s_n-2) at (0.3, 2.0+\inoff) {};
    \node[sample, fill=black, outer sep=3pt] (s_n-1) at (0.3, 1.5+\inoff) {};
    \node[sample, fill=black, outer sep=3pt] (s_n) at (0.3, 0.9+\inoff) {};
    \node[sample, fill=black, outer sep=3pt] (s_n+1) at (0.3, 0.3+\inoff) {};
    \node (n_end) at (0.3, 0.8) {};
    \node[sample_outer] (samples) at (0.3, 0.9+\inoff) {};

    \draw[thick, draw=black, dotted] (s_n-1.south) -- (s_n.north);
    \draw[thick, draw=black, dotted] (s_n.south) -- (s_n+1.north);
    \draw[thick, draw=black, -latex] (samples.south) -- node[left]{$\scriptstyle n$}(n_end.north);

    \node[draw=black!50, thick, dotted, rounded corners=1pt, minimum width=5.2cm, minimum height=2.6cm] (bss2) at (3.7, 2.4) {};
    \node[font=\scriptsize] (bss2_) at (3.7, 4.0) {On BSS-2};

    \node[neuron, draw=black] (n_i_0) at (1.4, 1.75) {};
    \node[neuron, draw=black] (n_i_N) at (1.4, 3.05) {};
    \node[input_outer] (inputs) at (1.9, 0.9+\inoff) {};
    \draw[thick, draw=black, dotted] (n_i_0.north) -- (n_i_N.south);
    \draw[thick, -latex] (samples.east) -- node[above]{$\scriptstyle \Tilde{\boldsymbol{y}}_n$}(bss2.west);

    \def\lineoffset{0.06}
    \draw[thick, draw=black] (1.7, 1.75 - \lineoffset) -- (1.7, 1.75 + \lineoffset);
    \draw[thick, draw=black] (2.0, 3.05 - \lineoffset) -- (2.0, 3.05 + \lineoffset);
    \draw[thick, draw=black] (1.8, 2.7 - \lineoffset) -- (1.8, 2.7 + \lineoffset);
    \draw[thick, draw=black] (2.1, 2.4 - \lineoffset) -- (2.1, 2.4 + \lineoffset);
    \draw[thick, draw=black] (2.2, 2.9 - \lineoffset) -- (2.2, 2.9 + \lineoffset);
    \draw[thick, draw=black] (2.0, 2.0 - \lineoffset) -- (2.0, 2.0 + \lineoffset);

    \node[neuron, draw=black] (n_h_0) at (3.3, 1.75) {};
    \node[neuron, draw=black] (n_h_N) at (3.3, 3.05) {};
    \node[hidden_outer] (hiddens) at (3.7, 0.9+\inoff) {};
    \draw[thick, draw=black, dotted] (n_h_0.north) -- (n_h_N.south);

    \draw[thick, draw=black] (4.0, 1.75 - \lineoffset) -- (4.0, 1.75 + \lineoffset);
    \draw[thick, draw=black] (4.1, 3.05 - \lineoffset) -- (4.1, 3.05 + \lineoffset);
    \draw[thick, draw=black] (3.7, 2.7 - \lineoffset) -- (3.7, 2.7 + \lineoffset);
    \draw[thick, draw=black] (3.9, 2.4 - \lineoffset) -- (3.9, 2.4 + \lineoffset);
    \draw[thick, draw=black] (3.9, 2.9 - \lineoffset) -- (3.9, 2.9 + \lineoffset);
    \draw[thick, draw=black] (3.6, 2.0 - \lineoffset) -- (3.6, 2.0 + \lineoffset);

    \node[neuron, draw=red, thick] (no_0) at (5.1, \outoff + 0.75 ) {};
    \node[neuron, draw=blue, fill=blue!30, thick] (no_1) at (5.1, \outoff + 0.75 + 2.0*\spacing*1) {};
    \node[neuron, draw=yellow, thick] (no_2) at (5.1, \outoff + 0.75 + 2.0*\spacing*2) {};
    \node[neuron, draw=green, thick] (no_3) at (5.1, \outoff + 0.75 + 2.0*\spacing*3) {};
    
    \node[output_outer] (outputs) at (5.6, \outoff + 1.2) {};

    \draw [blue, very thick] plot [smooth, tension=1] coordinates { (5.3, 2.4) (5.6, 2.8) (5.9, 2.4) (6.2, 2.2) };
    \draw [yellow!50, thick] plot [smooth, tension=1] coordinates { (5.3, 2.4) (5.6, 2.3) (5.9, 2.2) (6.2, 2.3) };
    \draw [green!50, thick] plot [smooth, tension=1] coordinates { (5.3, 2.4) (5.6, 2.1) (5.9, 1.9) (6.2, 2.4) };
    \draw [red!50, thick] plot [smooth, tension=1] coordinates { (5.3, 2.4) (5.6, 2.0) (5.9, 2.1) (6.2, 2.4) };
    
    \draw [-latex, thick] (inputs.east) -- node[above]{$\scriptstyle w^\text{ih}$}(hiddens.west);

    \draw [-latex, thick] (hiddens.east) -- node[above]{$\scriptstyle w^\text{ho}$}(outputs.west);

    \node[outer_argmax] (max) at (7.4, \outoff + 1.2) {$ \scriptstyle \max_{t}(v_k)$\\and\\$\scriptstyle \argmax_k$};
    \draw [-latex, thick] (bss2.east) -- (max.west);

    \node[outer_demapper] (map) at (9.0, \outoff + 1.2) {
        \color{green!50}{$\scriptstyle 3 \mapsto \texttt{10}$} \\
        \color{yellow!50}{$\scriptstyle 2 \mapsto \texttt{11}$}\\
        \color{blue}{$\scriptstyle \boldsymbol{1} \mapsto \texttt{01}$}\\
        \color{red!50}{$\scriptstyle \; 0 \mapsto \texttt{00}$}
    };
    \draw [-latex, thick] (max.east) -- (map.west);

    \node (out) at (10.6, \outoff + 1.2) {$ \scriptstyle \left[\hat{b}_1\hat{b}_2\right]_n$};
     \draw [-latex, thick] (map.east) -- (out.west);

    \draw [-latex, draw=black] (1.5, 1.50) -- node[below]{$\scriptstyle t$} (2.1, 1.50);
    \draw [-latex, draw=black] (3.4, 1.50) -- node[below]{$\scriptstyle t$} (4.0, 1.50);
    \draw [-latex, draw=black] (5.2, 1.50) -- node[below]{$\scriptstyle t$} (5.8, 1.50);

    \node (out) at (5.1, 3.1) {$ \scriptstyle v_k$};
    \node[font=\scriptsize] (out) at (3.8, 3.5) {LIF Neurons};
    \node[font=\scriptsize] (out) at (2.0, 3.5) {Input Neurons};
    \node[font=\scriptsize] (out) at (5.5, 3.5) {LI Neurons};

\end{tikzpicture}

%% file: figures/linear_encoding.pgf
\begingroup%
\makeatletter%
\begin{pgfpicture}%
\pgfpathrectangle{\pgfpointorigin}{\pgfqpoint{1.500000in}{2.800000in}}%
\pgfusepath{use as bounding box, clip}%
\begin{pgfscope}%
\pgfsetbuttcap%
\pgfsetmiterjoin%
\definecolor{currentfill}{rgb}{1.000000,1.000000,1.000000}%
\pgfsetfillcolor{currentfill}%
\pgfsetlinewidth{0.000000pt}%
\definecolor{currentstroke}{rgb}{1.000000,1.000000,1.000000}%
\pgfsetstrokecolor{currentstroke}%
\pgfsetdash{}{0pt}%
\pgfpathmoveto{\pgfqpoint{0.000000in}{0.000000in}}%
\pgfpathlineto{\pgfqpoint{1.500000in}{0.000000in}}%
\pgfpathlineto{\pgfqpoint{1.500000in}{2.800000in}}%
\pgfpathlineto{\pgfqpoint{0.000000in}{2.800000in}}%
\pgfpathclose%
\pgfusepath{fill}%
\end{pgfscope}%
\begin{pgfscope}%
\pgfsetbuttcap%
\pgfsetmiterjoin%
\definecolor{currentfill}{rgb}{1.000000,1.000000,1.000000}%
\pgfsetfillcolor{currentfill}%
\pgfsetlinewidth{0.000000pt}%
\definecolor{currentstroke}{rgb}{0.000000,0.000000,0.000000}%
\pgfsetstrokecolor{currentstroke}%
\pgfsetstrokeopacity{0.000000}%
\pgfsetdash{}{0pt}%
\pgfpathmoveto{\pgfqpoint{0.539023in}{1.045974in}}%
\pgfpathlineto{\pgfqpoint{1.377678in}{1.045974in}}%
\pgfpathlineto{\pgfqpoint{1.377678in}{2.675000in}}%
\pgfpathlineto{\pgfqpoint{0.539023in}{2.675000in}}%
\pgfpathclose%
\pgfusepath{fill}%
\end{pgfscope}%
\begin{pgfscope}%
\pgfsetbuttcap%
\pgfsetroundjoin%
\definecolor{currentfill}{rgb}{0.000000,0.000000,0.000000}%
\pgfsetfillcolor{currentfill}%
\pgfsetlinewidth{0.803000pt}%
\definecolor{currentstroke}{rgb}{0.000000,0.000000,0.000000}%
\pgfsetstrokecolor{currentstroke}%
\pgfsetdash{}{0pt}%
\pgfsys@defobject{currentmarker}{\pgfqpoint{-0.048611in}{0.000000in}}{\pgfqpoint{-0.000000in}{0.000000in}}{%
\pgfpathmoveto{\pgfqpoint{-0.000000in}{0.000000in}}%
\pgfpathlineto{\pgfqpoint{-0.048611in}{0.000000in}}%
\pgfusepath{stroke,fill}%
}%
\begin{pgfscope}%
\pgfsys@transformshift{0.539023in}{1.045974in}%
\pgfsys@useobject{currentmarker}{}%
\end{pgfscope}%
\end{pgfscope}%
\begin{pgfscope}%
\definecolor{textcolor}{rgb}{0.000000,0.000000,0.000000}%
\pgfsetstrokecolor{textcolor}%
\pgfsetfillcolor{textcolor}%
\pgftext[x=0.388782in, y=1.014317in, left, base]{\color{textcolor}\sffamily\fontsize{6.000000}{7.200000}\selectfont 0}%
\end{pgfscope}%
\begin{pgfscope}%
\pgfsetbuttcap%
\pgfsetroundjoin%
\definecolor{currentfill}{rgb}{0.000000,0.000000,0.000000}%
\pgfsetfillcolor{currentfill}%
\pgfsetlinewidth{0.803000pt}%
\definecolor{currentstroke}{rgb}{0.000000,0.000000,0.000000}%
\pgfsetstrokecolor{currentstroke}%
\pgfsetdash{}{0pt}%
\pgfsys@defobject{currentmarker}{\pgfqpoint{-0.048611in}{0.000000in}}{\pgfqpoint{-0.000000in}{0.000000in}}{%
\pgfpathmoveto{\pgfqpoint{-0.000000in}{0.000000in}}%
\pgfpathlineto{\pgfqpoint{-0.048611in}{0.000000in}}%
\pgfusepath{stroke,fill}%
}%
\begin{pgfscope}%
\pgfsys@transformshift{0.539023in}{1.278692in}%
\pgfsys@useobject{currentmarker}{}%
\end{pgfscope}%
\end{pgfscope}%
\begin{pgfscope}%
\definecolor{textcolor}{rgb}{0.000000,0.000000,0.000000}%
\pgfsetstrokecolor{textcolor}%
\pgfsetfillcolor{textcolor}%
\pgftext[x=0.388782in, y=1.247035in, left, base]{\color{textcolor}\sffamily\fontsize{6.000000}{7.200000}\selectfont 1}%
\end{pgfscope}%
\begin{pgfscope}%
\pgfsetbuttcap%
\pgfsetroundjoin%
\definecolor{currentfill}{rgb}{0.000000,0.000000,0.000000}%
\pgfsetfillcolor{currentfill}%
\pgfsetlinewidth{0.803000pt}%
\definecolor{currentstroke}{rgb}{0.000000,0.000000,0.000000}%
\pgfsetstrokecolor{currentstroke}%
\pgfsetdash{}{0pt}%
\pgfsys@defobject{currentmarker}{\pgfqpoint{-0.048611in}{0.000000in}}{\pgfqpoint{-0.000000in}{0.000000in}}{%
\pgfpathmoveto{\pgfqpoint{-0.000000in}{0.000000in}}%
\pgfpathlineto{\pgfqpoint{-0.048611in}{0.000000in}}%
\pgfusepath{stroke,fill}%
}%
\begin{pgfscope}%
\pgfsys@transformshift{0.539023in}{1.511410in}%
\pgfsys@useobject{currentmarker}{}%
\end{pgfscope}%
\end{pgfscope}%
\begin{pgfscope}%
\definecolor{textcolor}{rgb}{0.000000,0.000000,0.000000}%
\pgfsetstrokecolor{textcolor}%
\pgfsetfillcolor{textcolor}%
\pgftext[x=0.388782in, y=1.479753in, left, base]{\color{textcolor}\sffamily\fontsize{6.000000}{7.200000}\selectfont 2}%
\end{pgfscope}%
\begin{pgfscope}%
\pgfsetbuttcap%
\pgfsetroundjoin%
\definecolor{currentfill}{rgb}{0.000000,0.000000,0.000000}%
\pgfsetfillcolor{currentfill}%
\pgfsetlinewidth{0.803000pt}%
\definecolor{currentstroke}{rgb}{0.000000,0.000000,0.000000}%
\pgfsetstrokecolor{currentstroke}%
\pgfsetdash{}{0pt}%
\pgfsys@defobject{currentmarker}{\pgfqpoint{-0.048611in}{0.000000in}}{\pgfqpoint{-0.000000in}{0.000000in}}{%
\pgfpathmoveto{\pgfqpoint{-0.000000in}{0.000000in}}%
\pgfpathlineto{\pgfqpoint{-0.048611in}{0.000000in}}%
\pgfusepath{stroke,fill}%
}%
\begin{pgfscope}%
\pgfsys@transformshift{0.539023in}{1.744128in}%
\pgfsys@useobject{currentmarker}{}%
\end{pgfscope}%
\end{pgfscope}%
\begin{pgfscope}%
\definecolor{textcolor}{rgb}{0.000000,0.000000,0.000000}%
\pgfsetstrokecolor{textcolor}%
\pgfsetfillcolor{textcolor}%
\pgftext[x=0.388782in, y=1.712471in, left, base]{\color{textcolor}\sffamily\fontsize{6.000000}{7.200000}\selectfont 3}%
\end{pgfscope}%
\begin{pgfscope}%
\pgfsetbuttcap%
\pgfsetroundjoin%
\definecolor{currentfill}{rgb}{0.000000,0.000000,0.000000}%
\pgfsetfillcolor{currentfill}%
\pgfsetlinewidth{0.803000pt}%
\definecolor{currentstroke}{rgb}{0.000000,0.000000,0.000000}%
\pgfsetstrokecolor{currentstroke}%
\pgfsetdash{}{0pt}%
\pgfsys@defobject{currentmarker}{\pgfqpoint{-0.048611in}{0.000000in}}{\pgfqpoint{-0.000000in}{0.000000in}}{%
\pgfpathmoveto{\pgfqpoint{-0.000000in}{0.000000in}}%
\pgfpathlineto{\pgfqpoint{-0.048611in}{0.000000in}}%
\pgfusepath{stroke,fill}%
}%
\begin{pgfscope}%
\pgfsys@transformshift{0.539023in}{1.976846in}%
\pgfsys@useobject{currentmarker}{}%
\end{pgfscope}%
\end{pgfscope}%
\begin{pgfscope}%
\definecolor{textcolor}{rgb}{0.000000,0.000000,0.000000}%
\pgfsetstrokecolor{textcolor}%
\pgfsetfillcolor{textcolor}%
\pgftext[x=0.388782in, y=1.945189in, left, base]{\color{textcolor}\sffamily\fontsize{6.000000}{7.200000}\selectfont 4}%
\end{pgfscope}%
\begin{pgfscope}%
\pgfsetbuttcap%
\pgfsetroundjoin%
\definecolor{currentfill}{rgb}{0.000000,0.000000,0.000000}%
\pgfsetfillcolor{currentfill}%
\pgfsetlinewidth{0.803000pt}%
\definecolor{currentstroke}{rgb}{0.000000,0.000000,0.000000}%
\pgfsetstrokecolor{currentstroke}%
\pgfsetdash{}{0pt}%
\pgfsys@defobject{currentmarker}{\pgfqpoint{-0.048611in}{0.000000in}}{\pgfqpoint{-0.000000in}{0.000000in}}{%
\pgfpathmoveto{\pgfqpoint{-0.000000in}{0.000000in}}%
\pgfpathlineto{\pgfqpoint{-0.048611in}{0.000000in}}%
\pgfusepath{stroke,fill}%
}%
\begin{pgfscope}%
\pgfsys@transformshift{0.539023in}{2.209564in}%
\pgfsys@useobject{currentmarker}{}%
\end{pgfscope}%
\end{pgfscope}%
\begin{pgfscope}%
\definecolor{textcolor}{rgb}{0.000000,0.000000,0.000000}%
\pgfsetstrokecolor{textcolor}%
\pgfsetfillcolor{textcolor}%
\pgftext[x=0.388782in, y=2.177907in, left, base]{\color{textcolor}\sffamily\fontsize{6.000000}{7.200000}\selectfont 5}%
\end{pgfscope}%
\begin{pgfscope}%
\pgfsetbuttcap%
\pgfsetroundjoin%
\definecolor{currentfill}{rgb}{0.000000,0.000000,0.000000}%
\pgfsetfillcolor{currentfill}%
\pgfsetlinewidth{0.803000pt}%
\definecolor{currentstroke}{rgb}{0.000000,0.000000,0.000000}%
\pgfsetstrokecolor{currentstroke}%
\pgfsetdash{}{0pt}%
\pgfsys@defobject{currentmarker}{\pgfqpoint{-0.048611in}{0.000000in}}{\pgfqpoint{-0.000000in}{0.000000in}}{%
\pgfpathmoveto{\pgfqpoint{-0.000000in}{0.000000in}}%
\pgfpathlineto{\pgfqpoint{-0.048611in}{0.000000in}}%
\pgfusepath{stroke,fill}%
}%
\begin{pgfscope}%
\pgfsys@transformshift{0.539023in}{1.395051in}%
\pgfsys@useobject{currentmarker}{}%
\end{pgfscope}%
\end{pgfscope}%
\begin{pgfscope}%
\definecolor{textcolor}{rgb}{0.000000,0.000000,0.000000}%
\pgfsetstrokecolor{textcolor}%
\pgfsetfillcolor{textcolor}%
\pgftext[x=0.247629in, y=1.366915in, left, base]{\color{textcolor}\sffamily\fontsize{6.000000}{7.200000}\selectfont \(\displaystyle \tilde{y}_{n,\ell}\)}%
\end{pgfscope}%
\begin{pgfscope}%
\pgfsetbuttcap%
\pgfsetroundjoin%
\definecolor{currentfill}{rgb}{0.000000,0.000000,0.000000}%
\pgfsetfillcolor{currentfill}%
\pgfsetlinewidth{0.803000pt}%
\definecolor{currentstroke}{rgb}{0.000000,0.000000,0.000000}%
\pgfsetstrokecolor{currentstroke}%
\pgfsetdash{}{0pt}%
\pgfsys@defobject{currentmarker}{\pgfqpoint{-0.048611in}{0.000000in}}{\pgfqpoint{-0.000000in}{0.000000in}}{%
\pgfpathmoveto{\pgfqpoint{-0.000000in}{0.000000in}}%
\pgfpathlineto{\pgfqpoint{-0.048611in}{0.000000in}}%
\pgfusepath{stroke,fill}%
}%
\begin{pgfscope}%
\pgfsys@transformshift{0.539023in}{2.442282in}%
\pgfsys@useobject{currentmarker}{}%
\end{pgfscope}%
\end{pgfscope}%
\begin{pgfscope}%
\definecolor{textcolor}{rgb}{0.000000,0.000000,0.000000}%
\pgfsetstrokecolor{textcolor}%
\pgfsetfillcolor{textcolor}%
\pgftext[x=0.388782in, y=2.410625in, left, base]{\color{textcolor}\sffamily\fontsize{6.000000}{7.200000}\selectfont 6}%
\end{pgfscope}%
\begin{pgfscope}%
\pgfsetbuttcap%
\pgfsetroundjoin%
\definecolor{currentfill}{rgb}{0.000000,0.000000,0.000000}%
\pgfsetfillcolor{currentfill}%
\pgfsetlinewidth{0.803000pt}%
\definecolor{currentstroke}{rgb}{0.000000,0.000000,0.000000}%
\pgfsetstrokecolor{currentstroke}%
\pgfsetdash{}{0pt}%
\pgfsys@defobject{currentmarker}{\pgfqpoint{-0.048611in}{0.000000in}}{\pgfqpoint{-0.000000in}{0.000000in}}{%
\pgfpathmoveto{\pgfqpoint{-0.000000in}{0.000000in}}%
\pgfpathlineto{\pgfqpoint{-0.048611in}{0.000000in}}%
\pgfusepath{stroke,fill}%
}%
\begin{pgfscope}%
\pgfsys@transformshift{0.539023in}{2.675000in}%
\pgfsys@useobject{currentmarker}{}%
\end{pgfscope}%
\end{pgfscope}%
\begin{pgfscope}%
\definecolor{textcolor}{rgb}{0.000000,0.000000,0.000000}%
\pgfsetstrokecolor{textcolor}%
\pgfsetfillcolor{textcolor}%
\pgftext[x=0.388782in, y=2.643343in, left, base]{\color{textcolor}\sffamily\fontsize{6.000000}{7.200000}\selectfont 7}%
\end{pgfscope}%
\begin{pgfscope}%
\definecolor{textcolor}{rgb}{0.000000,0.000000,0.000000}%
\pgfsetstrokecolor{textcolor}%
\pgfsetfillcolor{textcolor}%
\pgftext[x=0.192073in,y=1.860487in,,bottom,rotate=90.000000]{\color{textcolor}\sffamily\fontsize{6.000000}{7.200000}\selectfont \(\displaystyle \tilde{y}\) [a.u.]}%
\end{pgfscope}%
\begin{pgfscope}%
\pgfpathrectangle{\pgfqpoint{0.539023in}{1.045974in}}{\pgfqpoint{0.838655in}{1.629026in}}%
\pgfusepath{clip}%
\pgfsetbuttcap%
\pgfsetroundjoin%
\pgfsetlinewidth{1.003750pt}%
\definecolor{currentstroke}{rgb}{0.788235,0.149020,0.509804}%
\pgfsetstrokecolor{currentstroke}%
\pgfsetdash{{1.000000pt}{1.650000pt}}{0.000000pt}%
\pgfpathmoveto{\pgfqpoint{0.566076in}{1.395051in}}%
\pgfpathlineto{\pgfqpoint{1.377678in}{1.395051in}}%
\pgfusepath{stroke}%
\end{pgfscope}%
\begin{pgfscope}%
\pgfpathrectangle{\pgfqpoint{0.539023in}{1.045974in}}{\pgfqpoint{0.838655in}{1.629026in}}%
\pgfusepath{clip}%
\pgfsetbuttcap%
\pgfsetroundjoin%
\pgfsetlinewidth{1.003750pt}%
\definecolor{currentstroke}{rgb}{0.160784,0.423529,0.835294}%
\pgfsetstrokecolor{currentstroke}%
\pgfsetdash{{1.000000pt}{1.650000pt}}{0.000000pt}%
\pgfpathmoveto{\pgfqpoint{0.566076in}{1.769986in}}%
\pgfpathlineto{\pgfqpoint{1.025984in}{1.769986in}}%
\pgfusepath{stroke}%
\end{pgfscope}%
\begin{pgfscope}%
\pgfpathrectangle{\pgfqpoint{0.539023in}{1.045974in}}{\pgfqpoint{0.838655in}{1.629026in}}%
\pgfusepath{clip}%
\pgfsetbuttcap%
\pgfsetroundjoin%
\pgfsetlinewidth{1.003750pt}%
\definecolor{currentstroke}{rgb}{0.949020,0.674510,0.113725}%
\pgfsetstrokecolor{currentstroke}%
\pgfsetdash{{1.000000pt}{1.650000pt}}{0.000000pt}%
\pgfpathmoveto{\pgfqpoint{0.566076in}{2.493997in}}%
\pgfpathlineto{\pgfqpoint{1.025984in}{2.493997in}}%
\pgfusepath{stroke}%
\end{pgfscope}%
\begin{pgfscope}%
\pgfpathrectangle{\pgfqpoint{0.539023in}{1.045974in}}{\pgfqpoint{0.838655in}{1.629026in}}%
\pgfusepath{clip}%
\pgfsetrectcap%
\pgfsetroundjoin%
\pgfsetlinewidth{0.501875pt}%
\definecolor{currentstroke}{rgb}{0.501961,0.501961,0.501961}%
\pgfsetstrokecolor{currentstroke}%
\pgfsetdash{}{0pt}%
\pgfpathmoveto{\pgfqpoint{0.566076in}{1.045974in}}%
\pgfpathlineto{\pgfqpoint{0.970984in}{1.481359in}}%
\pgfpathlineto{\pgfqpoint{0.970984in}{1.481359in}}%
\pgfusepath{stroke}%
\end{pgfscope}%
\begin{pgfscope}%
\pgfpathrectangle{\pgfqpoint{0.539023in}{1.045974in}}{\pgfqpoint{0.838655in}{1.629026in}}%
\pgfusepath{clip}%
\pgfsetrectcap%
\pgfsetroundjoin%
\pgfsetlinewidth{0.501875pt}%
\definecolor{currentstroke}{rgb}{0.501961,0.501961,0.501961}%
\pgfsetstrokecolor{currentstroke}%
\pgfsetdash{}{0pt}%
\pgfpathmoveto{\pgfqpoint{0.734409in}{1.045974in}}%
\pgfpathlineto{\pgfqpoint{0.566076in}{1.226977in}}%
\pgfpathlineto{\pgfqpoint{0.970984in}{1.662362in}}%
\pgfpathlineto{\pgfqpoint{0.970984in}{1.662362in}}%
\pgfusepath{stroke}%
\end{pgfscope}%
\begin{pgfscope}%
\pgfpathrectangle{\pgfqpoint{0.539023in}{1.045974in}}{\pgfqpoint{0.838655in}{1.629026in}}%
\pgfusepath{clip}%
\pgfsetrectcap%
\pgfsetroundjoin%
\pgfsetlinewidth{0.501875pt}%
\definecolor{currentstroke}{rgb}{0.501961,0.501961,0.501961}%
\pgfsetstrokecolor{currentstroke}%
\pgfsetdash{}{0pt}%
\pgfpathmoveto{\pgfqpoint{0.902741in}{1.045974in}}%
\pgfpathlineto{\pgfqpoint{0.566076in}{1.407980in}}%
\pgfpathlineto{\pgfqpoint{0.970984in}{1.843365in}}%
\pgfpathlineto{\pgfqpoint{0.970984in}{1.843365in}}%
\pgfusepath{stroke}%
\end{pgfscope}%
\begin{pgfscope}%
\pgfpathrectangle{\pgfqpoint{0.539023in}{1.045974in}}{\pgfqpoint{0.838655in}{1.629026in}}%
\pgfusepath{clip}%
\pgfsetrectcap%
\pgfsetroundjoin%
\pgfsetlinewidth{0.501875pt}%
\definecolor{currentstroke}{rgb}{0.501961,0.501961,0.501961}%
\pgfsetstrokecolor{currentstroke}%
\pgfsetdash{}{0pt}%
\pgfpathmoveto{\pgfqpoint{0.970984in}{1.153597in}}%
\pgfpathlineto{\pgfqpoint{0.566076in}{1.588983in}}%
\pgfpathlineto{\pgfqpoint{0.970984in}{2.024368in}}%
\pgfpathlineto{\pgfqpoint{0.970984in}{2.024368in}}%
\pgfusepath{stroke}%
\end{pgfscope}%
\begin{pgfscope}%
\pgfpathrectangle{\pgfqpoint{0.539023in}{1.045974in}}{\pgfqpoint{0.838655in}{1.629026in}}%
\pgfusepath{clip}%
\pgfsetrectcap%
\pgfsetroundjoin%
\pgfsetlinewidth{1.003750pt}%
\definecolor{currentstroke}{rgb}{0.160784,0.423529,0.835294}%
\pgfsetstrokecolor{currentstroke}%
\pgfsetdash{}{0pt}%
\pgfpathmoveto{\pgfqpoint{0.970984in}{1.334600in}}%
\pgfpathlineto{\pgfqpoint{0.566076in}{1.769986in}}%
\pgfpathlineto{\pgfqpoint{0.970984in}{2.205371in}}%
\pgfpathlineto{\pgfqpoint{0.970984in}{2.205371in}}%
\pgfusepath{stroke}%
\end{pgfscope}%
\begin{pgfscope}%
\pgfpathrectangle{\pgfqpoint{0.539023in}{1.045974in}}{\pgfqpoint{0.838655in}{1.629026in}}%
\pgfusepath{clip}%
\pgfsetrectcap%
\pgfsetroundjoin%
\pgfsetlinewidth{0.501875pt}%
\definecolor{currentstroke}{rgb}{0.501961,0.501961,0.501961}%
\pgfsetstrokecolor{currentstroke}%
\pgfsetdash{}{0pt}%
\pgfpathmoveto{\pgfqpoint{0.970984in}{1.515603in}}%
\pgfpathlineto{\pgfqpoint{0.566076in}{1.950988in}}%
\pgfpathlineto{\pgfqpoint{0.970984in}{2.386374in}}%
\pgfpathlineto{\pgfqpoint{0.970984in}{2.386374in}}%
\pgfusepath{stroke}%
\end{pgfscope}%
\begin{pgfscope}%
\pgfpathrectangle{\pgfqpoint{0.539023in}{1.045974in}}{\pgfqpoint{0.838655in}{1.629026in}}%
\pgfusepath{clip}%
\pgfsetrectcap%
\pgfsetroundjoin%
\pgfsetlinewidth{0.501875pt}%
\definecolor{currentstroke}{rgb}{0.501961,0.501961,0.501961}%
\pgfsetstrokecolor{currentstroke}%
\pgfsetdash{}{0pt}%
\pgfpathmoveto{\pgfqpoint{0.970984in}{1.696606in}}%
\pgfpathlineto{\pgfqpoint{0.566076in}{2.131991in}}%
\pgfpathlineto{\pgfqpoint{0.970984in}{2.567377in}}%
\pgfpathlineto{\pgfqpoint{0.970984in}{2.567377in}}%
\pgfusepath{stroke}%
\end{pgfscope}%
\begin{pgfscope}%
\pgfpathrectangle{\pgfqpoint{0.539023in}{1.045974in}}{\pgfqpoint{0.838655in}{1.629026in}}%
\pgfusepath{clip}%
\pgfsetrectcap%
\pgfsetroundjoin%
\pgfsetlinewidth{0.501875pt}%
\definecolor{currentstroke}{rgb}{0.501961,0.501961,0.501961}%
\pgfsetstrokecolor{currentstroke}%
\pgfsetdash{}{0pt}%
\pgfpathmoveto{\pgfqpoint{0.970984in}{1.877609in}}%
\pgfpathlineto{\pgfqpoint{0.566076in}{2.312994in}}%
\pgfpathlineto{\pgfqpoint{0.902741in}{2.675000in}}%
\pgfpathlineto{\pgfqpoint{0.902741in}{2.675000in}}%
\pgfusepath{stroke}%
\end{pgfscope}%
\begin{pgfscope}%
\pgfpathrectangle{\pgfqpoint{0.539023in}{1.045974in}}{\pgfqpoint{0.838655in}{1.629026in}}%
\pgfusepath{clip}%
\pgfsetrectcap%
\pgfsetroundjoin%
\pgfsetlinewidth{0.501875pt}%
\definecolor{currentstroke}{rgb}{0.501961,0.501961,0.501961}%
\pgfsetstrokecolor{currentstroke}%
\pgfsetdash{}{0pt}%
\pgfpathmoveto{\pgfqpoint{0.970984in}{2.058612in}}%
\pgfpathlineto{\pgfqpoint{0.566076in}{2.493997in}}%
\pgfpathlineto{\pgfqpoint{0.734409in}{2.675000in}}%
\pgfpathlineto{\pgfqpoint{0.734409in}{2.675000in}}%
\pgfusepath{stroke}%
\end{pgfscope}%
\begin{pgfscope}%
\pgfpathrectangle{\pgfqpoint{0.539023in}{1.045974in}}{\pgfqpoint{0.838655in}{1.629026in}}%
\pgfusepath{clip}%
\pgfsetrectcap%
\pgfsetroundjoin%
\pgfsetlinewidth{0.501875pt}%
\definecolor{currentstroke}{rgb}{0.501961,0.501961,0.501961}%
\pgfsetstrokecolor{currentstroke}%
\pgfsetdash{}{0pt}%
\pgfpathmoveto{\pgfqpoint{0.970984in}{2.239615in}}%
\pgfpathlineto{\pgfqpoint{0.566076in}{2.675000in}}%
\pgfpathlineto{\pgfqpoint{0.566076in}{2.675000in}}%
\pgfusepath{stroke}%
\end{pgfscope}%
\begin{pgfscope}%
\pgfsetrectcap%
\pgfsetmiterjoin%
\pgfsetlinewidth{0.803000pt}%
\definecolor{currentstroke}{rgb}{0.000000,0.000000,0.000000}%
\pgfsetstrokecolor{currentstroke}%
\pgfsetdash{}{0pt}%
\pgfpathmoveto{\pgfqpoint{0.539023in}{1.045974in}}%
\pgfpathlineto{\pgfqpoint{0.539023in}{2.675000in}}%
\pgfusepath{stroke}%
\end{pgfscope}%
\begin{pgfscope}%
\pgfsetrectcap%
\pgfsetmiterjoin%
\pgfsetlinewidth{0.803000pt}%
\definecolor{currentstroke}{rgb}{0.000000,0.000000,0.000000}%
\pgfsetstrokecolor{currentstroke}%
\pgfsetdash{}{0pt}%
\pgfpathmoveto{\pgfqpoint{0.539023in}{1.045974in}}%
\pgfpathlineto{\pgfqpoint{1.377678in}{1.045974in}}%
\pgfusepath{stroke}%
\end{pgfscope}%
\begin{pgfscope}%
\definecolor{textcolor}{rgb}{0.000000,0.000000,0.000000}%
\pgfsetstrokecolor{textcolor}%
\pgfsetfillcolor{textcolor}%
\pgftext[x=1.053038in,y=1.744128in,left,base]{\color{textcolor}\sffamily\fontsize{6.000000}{7.200000}\selectfont \(\displaystyle \chi_4 = {3.89}\)}%
\end{pgfscope}%
\begin{pgfscope}%
\definecolor{textcolor}{rgb}{0.000000,0.000000,0.000000}%
\pgfsetstrokecolor{textcolor}%
\pgfsetfillcolor{textcolor}%
\pgftext[x=1.053038in,y=2.465554in,left,base]{\color{textcolor}\sffamily\fontsize{6.000000}{7.200000}\selectfont \(\displaystyle \chi_8 = {6.22}\)}%
\end{pgfscope}%
\begin{pgfscope}%
\pgfsetbuttcap%
\pgfsetmiterjoin%
\definecolor{currentfill}{rgb}{1.000000,1.000000,1.000000}%
\pgfsetfillcolor{currentfill}%
\pgfsetlinewidth{0.000000pt}%
\definecolor{currentstroke}{rgb}{0.000000,0.000000,0.000000}%
\pgfsetstrokecolor{currentstroke}%
\pgfsetstrokeopacity{0.000000}%
\pgfsetdash{}{0pt}%
\pgfpathmoveto{\pgfqpoint{0.539023in}{0.502965in}}%
\pgfpathlineto{\pgfqpoint{1.377678in}{0.502965in}}%
\pgfpathlineto{\pgfqpoint{1.377678in}{1.045974in}}%
\pgfpathlineto{\pgfqpoint{0.539023in}{1.045974in}}%
\pgfpathclose%
\pgfusepath{fill}%
\end{pgfscope}%
\begin{pgfscope}%
\pgfpathrectangle{\pgfqpoint{0.539023in}{0.502965in}}{\pgfqpoint{0.838655in}{0.543009in}}%
\pgfusepath{clip}%
\pgfsetbuttcap%
\pgfsetroundjoin%
\definecolor{currentfill}{rgb}{0.000000,0.000000,0.000000}%
\pgfsetfillcolor{currentfill}%
\pgfsetlinewidth{1.505625pt}%
\definecolor{currentstroke}{rgb}{0.000000,0.000000,0.000000}%
\pgfsetstrokecolor{currentstroke}%
\pgfsetdash{}{0pt}%
\pgfsys@defobject{currentmarker}{\pgfqpoint{0.000000in}{-0.019642in}}{\pgfqpoint{0.000000in}{0.019642in}}{%
\pgfpathmoveto{\pgfqpoint{0.000000in}{-0.019642in}}%
\pgfpathlineto{\pgfqpoint{0.000000in}{0.019642in}}%
\pgfusepath{stroke,fill}%
}%
\begin{pgfscope}%
\pgfsys@transformshift{0.890717in}{0.530116in}%
\pgfsys@useobject{currentmarker}{}%
\end{pgfscope}%
\end{pgfscope}%
\begin{pgfscope}%
\pgfpathrectangle{\pgfqpoint{0.539023in}{0.502965in}}{\pgfqpoint{0.838655in}{0.543009in}}%
\pgfusepath{clip}%
\pgfsetbuttcap%
\pgfsetroundjoin%
\definecolor{currentfill}{rgb}{0.000000,0.000000,0.000000}%
\pgfsetfillcolor{currentfill}%
\pgfsetlinewidth{1.505625pt}%
\definecolor{currentstroke}{rgb}{0.000000,0.000000,0.000000}%
\pgfsetstrokecolor{currentstroke}%
\pgfsetdash{}{0pt}%
\pgfsys@defobject{currentmarker}{\pgfqpoint{0.000000in}{-0.019642in}}{\pgfqpoint{0.000000in}{0.019642in}}{%
\pgfpathmoveto{\pgfqpoint{0.000000in}{-0.019642in}}%
\pgfpathlineto{\pgfqpoint{0.000000in}{0.019642in}}%
\pgfusepath{stroke,fill}%
}%
\begin{pgfscope}%
\pgfsys@transformshift{0.722385in}{0.584417in}%
\pgfsys@useobject{currentmarker}{}%
\end{pgfscope}%
\end{pgfscope}%
\begin{pgfscope}%
\pgfpathrectangle{\pgfqpoint{0.539023in}{0.502965in}}{\pgfqpoint{0.838655in}{0.543009in}}%
\pgfusepath{clip}%
\pgfsetbuttcap%
\pgfsetroundjoin%
\definecolor{currentfill}{rgb}{0.000000,0.000000,0.000000}%
\pgfsetfillcolor{currentfill}%
\pgfsetlinewidth{1.505625pt}%
\definecolor{currentstroke}{rgb}{0.000000,0.000000,0.000000}%
\pgfsetstrokecolor{currentstroke}%
\pgfsetdash{}{0pt}%
\pgfsys@defobject{currentmarker}{\pgfqpoint{0.000000in}{-0.019642in}}{\pgfqpoint{0.000000in}{0.019642in}}{%
\pgfpathmoveto{\pgfqpoint{0.000000in}{-0.019642in}}%
\pgfpathlineto{\pgfqpoint{0.000000in}{0.019642in}}%
\pgfusepath{stroke,fill}%
}%
\begin{pgfscope}%
\pgfsys@transformshift{0.578100in}{0.638717in}%
\pgfsys@useobject{currentmarker}{}%
\end{pgfscope}%
\end{pgfscope}%
\begin{pgfscope}%
\pgfpathrectangle{\pgfqpoint{0.539023in}{0.502965in}}{\pgfqpoint{0.838655in}{0.543009in}}%
\pgfusepath{clip}%
\pgfsetbuttcap%
\pgfsetroundjoin%
\definecolor{currentfill}{rgb}{0.000000,0.000000,0.000000}%
\pgfsetfillcolor{currentfill}%
\pgfsetlinewidth{1.505625pt}%
\definecolor{currentstroke}{rgb}{0.000000,0.000000,0.000000}%
\pgfsetstrokecolor{currentstroke}%
\pgfsetdash{}{0pt}%
\pgfsys@defobject{currentmarker}{\pgfqpoint{0.000000in}{-0.019642in}}{\pgfqpoint{0.000000in}{0.019642in}}{%
\pgfpathmoveto{\pgfqpoint{0.000000in}{-0.019642in}}%
\pgfpathlineto{\pgfqpoint{0.000000in}{0.019642in}}%
\pgfusepath{stroke,fill}%
}%
\begin{pgfscope}%
\pgfsys@transformshift{0.746432in}{0.693018in}%
\pgfsys@useobject{currentmarker}{}%
\end{pgfscope}%
\end{pgfscope}%
\begin{pgfscope}%
\pgfpathrectangle{\pgfqpoint{0.539023in}{0.502965in}}{\pgfqpoint{0.838655in}{0.543009in}}%
\pgfusepath{clip}%
\pgfsetbuttcap%
\pgfsetroundjoin%
\definecolor{currentfill}{rgb}{0.160784,0.423529,0.835294}%
\pgfsetfillcolor{currentfill}%
\pgfsetlinewidth{1.505625pt}%
\definecolor{currentstroke}{rgb}{0.160784,0.423529,0.835294}%
\pgfsetstrokecolor{currentstroke}%
\pgfsetdash{}{0pt}%
\pgfsys@defobject{currentmarker}{\pgfqpoint{0.000000in}{-0.019642in}}{\pgfqpoint{0.000000in}{0.019642in}}{%
\pgfpathmoveto{\pgfqpoint{0.000000in}{-0.019642in}}%
\pgfpathlineto{\pgfqpoint{0.000000in}{0.019642in}}%
\pgfusepath{stroke,fill}%
}%
\begin{pgfscope}%
\pgfsys@transformshift{0.914765in}{0.747319in}%
\pgfsys@useobject{currentmarker}{}%
\end{pgfscope}%
\end{pgfscope}%
\begin{pgfscope}%
\pgfpathrectangle{\pgfqpoint{0.539023in}{0.502965in}}{\pgfqpoint{0.838655in}{0.543009in}}%
\pgfusepath{clip}%
\pgfsetbuttcap%
\pgfsetroundjoin%
\definecolor{currentfill}{rgb}{0.000000,0.000000,0.000000}%
\pgfsetfillcolor{currentfill}%
\pgfsetlinewidth{1.505625pt}%
\definecolor{currentstroke}{rgb}{0.000000,0.000000,0.000000}%
\pgfsetstrokecolor{currentstroke}%
\pgfsetdash{}{0pt}%
\pgfsys@defobject{currentmarker}{\pgfqpoint{0.000000in}{-0.019642in}}{\pgfqpoint{0.000000in}{0.019642in}}{%
\pgfpathmoveto{\pgfqpoint{0.000000in}{-0.019642in}}%
\pgfpathlineto{\pgfqpoint{0.000000in}{0.019642in}}%
\pgfusepath{stroke,fill}%
}%
\begin{pgfscope}%
\pgfsys@transformshift{27.592418in}{0.801620in}%
\pgfsys@useobject{currentmarker}{}%
\end{pgfscope}%
\end{pgfscope}%
\begin{pgfscope}%
\pgfpathrectangle{\pgfqpoint{0.539023in}{0.502965in}}{\pgfqpoint{0.838655in}{0.543009in}}%
\pgfusepath{clip}%
\pgfsetbuttcap%
\pgfsetroundjoin%
\definecolor{currentfill}{rgb}{0.000000,0.000000,0.000000}%
\pgfsetfillcolor{currentfill}%
\pgfsetlinewidth{1.505625pt}%
\definecolor{currentstroke}{rgb}{0.000000,0.000000,0.000000}%
\pgfsetstrokecolor{currentstroke}%
\pgfsetdash{}{0pt}%
\pgfsys@defobject{currentmarker}{\pgfqpoint{0.000000in}{-0.019642in}}{\pgfqpoint{0.000000in}{0.019642in}}{%
\pgfpathmoveto{\pgfqpoint{0.000000in}{-0.019642in}}%
\pgfpathlineto{\pgfqpoint{0.000000in}{0.019642in}}%
\pgfusepath{stroke,fill}%
}%
\begin{pgfscope}%
\pgfsys@transformshift{27.592418in}{0.855921in}%
\pgfsys@useobject{currentmarker}{}%
\end{pgfscope}%
\end{pgfscope}%
\begin{pgfscope}%
\pgfpathrectangle{\pgfqpoint{0.539023in}{0.502965in}}{\pgfqpoint{0.838655in}{0.543009in}}%
\pgfusepath{clip}%
\pgfsetbuttcap%
\pgfsetroundjoin%
\definecolor{currentfill}{rgb}{0.000000,0.000000,0.000000}%
\pgfsetfillcolor{currentfill}%
\pgfsetlinewidth{1.505625pt}%
\definecolor{currentstroke}{rgb}{0.000000,0.000000,0.000000}%
\pgfsetstrokecolor{currentstroke}%
\pgfsetdash{}{0pt}%
\pgfsys@defobject{currentmarker}{\pgfqpoint{0.000000in}{-0.019642in}}{\pgfqpoint{0.000000in}{0.019642in}}{%
\pgfpathmoveto{\pgfqpoint{0.000000in}{-0.019642in}}%
\pgfpathlineto{\pgfqpoint{0.000000in}{0.019642in}}%
\pgfusepath{stroke,fill}%
}%
\begin{pgfscope}%
\pgfsys@transformshift{27.592418in}{0.910222in}%
\pgfsys@useobject{currentmarker}{}%
\end{pgfscope}%
\end{pgfscope}%
\begin{pgfscope}%
\pgfpathrectangle{\pgfqpoint{0.539023in}{0.502965in}}{\pgfqpoint{0.838655in}{0.543009in}}%
\pgfusepath{clip}%
\pgfsetbuttcap%
\pgfsetroundjoin%
\definecolor{currentfill}{rgb}{0.000000,0.000000,0.000000}%
\pgfsetfillcolor{currentfill}%
\pgfsetlinewidth{1.505625pt}%
\definecolor{currentstroke}{rgb}{0.000000,0.000000,0.000000}%
\pgfsetstrokecolor{currentstroke}%
\pgfsetdash{}{0pt}%
\pgfsys@defobject{currentmarker}{\pgfqpoint{0.000000in}{-0.019642in}}{\pgfqpoint{0.000000in}{0.019642in}}{%
\pgfpathmoveto{\pgfqpoint{0.000000in}{-0.019642in}}%
\pgfpathlineto{\pgfqpoint{0.000000in}{0.019642in}}%
\pgfusepath{stroke,fill}%
}%
\begin{pgfscope}%
\pgfsys@transformshift{27.592418in}{0.964523in}%
\pgfsys@useobject{currentmarker}{}%
\end{pgfscope}%
\end{pgfscope}%
\begin{pgfscope}%
\pgfpathrectangle{\pgfqpoint{0.539023in}{0.502965in}}{\pgfqpoint{0.838655in}{0.543009in}}%
\pgfusepath{clip}%
\pgfsetbuttcap%
\pgfsetroundjoin%
\definecolor{currentfill}{rgb}{0.000000,0.000000,0.000000}%
\pgfsetfillcolor{currentfill}%
\pgfsetlinewidth{1.505625pt}%
\definecolor{currentstroke}{rgb}{0.000000,0.000000,0.000000}%
\pgfsetstrokecolor{currentstroke}%
\pgfsetdash{}{0pt}%
\pgfsys@defobject{currentmarker}{\pgfqpoint{0.000000in}{-0.019642in}}{\pgfqpoint{0.000000in}{0.019642in}}{%
\pgfpathmoveto{\pgfqpoint{0.000000in}{-0.019642in}}%
\pgfpathlineto{\pgfqpoint{0.000000in}{0.019642in}}%
\pgfusepath{stroke,fill}%
}%
\begin{pgfscope}%
\pgfsys@transformshift{27.592418in}{1.018824in}%
\pgfsys@useobject{currentmarker}{}%
\end{pgfscope}%
\end{pgfscope}%
\begin{pgfscope}%
\pgfsetroundcap%
\pgfsetroundjoin%
\pgfsetlinewidth{0.501875pt}%
\definecolor{currentstroke}{rgb}{0.160784,0.423529,0.835294}%
\pgfsetstrokecolor{currentstroke}%
\pgfsetdash{{0.500000pt}{0.825000pt}}{0.000000pt}%
\pgfpathmoveto{\pgfqpoint{0.914765in}{1.395051in}}%
\pgfpathquadraticcurveto{\pgfqpoint{0.914765in}{1.079330in}}{\pgfqpoint{0.914765in}{0.763609in}}%
\pgfusepath{stroke}%
\end{pgfscope}%
\begin{pgfscope}%
\pgfsetbuttcap%
\pgfsetroundjoin%
\definecolor{currentfill}{rgb}{0.000000,0.000000,0.000000}%
\pgfsetfillcolor{currentfill}%
\pgfsetlinewidth{0.803000pt}%
\definecolor{currentstroke}{rgb}{0.000000,0.000000,0.000000}%
\pgfsetstrokecolor{currentstroke}%
\pgfsetdash{}{0pt}%
\pgfsys@defobject{currentmarker}{\pgfqpoint{0.000000in}{-0.048611in}}{\pgfqpoint{0.000000in}{0.000000in}}{%
\pgfpathmoveto{\pgfqpoint{0.000000in}{0.000000in}}%
\pgfpathlineto{\pgfqpoint{0.000000in}{-0.048611in}}%
\pgfusepath{stroke,fill}%
}%
\begin{pgfscope}%
\pgfsys@transformshift{0.566076in}{0.502965in}%
\pgfsys@useobject{currentmarker}{}%
\end{pgfscope}%
\end{pgfscope}%
\begin{pgfscope}%
\definecolor{textcolor}{rgb}{0.000000,0.000000,0.000000}%
\pgfsetstrokecolor{textcolor}%
\pgfsetfillcolor{textcolor}%
\pgftext[x=0.566076in,y=0.405743in,,top]{\color{textcolor}\sffamily\fontsize{6.000000}{7.200000}\selectfont 0}%
\end{pgfscope}%
\begin{pgfscope}%
\pgfsetbuttcap%
\pgfsetroundjoin%
\definecolor{currentfill}{rgb}{0.000000,0.000000,0.000000}%
\pgfsetfillcolor{currentfill}%
\pgfsetlinewidth{0.803000pt}%
\definecolor{currentstroke}{rgb}{0.000000,0.000000,0.000000}%
\pgfsetstrokecolor{currentstroke}%
\pgfsetdash{}{0pt}%
\pgfsys@defobject{currentmarker}{\pgfqpoint{0.000000in}{-0.048611in}}{\pgfqpoint{0.000000in}{0.000000in}}{%
\pgfpathmoveto{\pgfqpoint{0.000000in}{0.000000in}}%
\pgfpathlineto{\pgfqpoint{0.000000in}{-0.048611in}}%
\pgfusepath{stroke,fill}%
}%
\begin{pgfscope}%
\pgfsys@transformshift{0.971877in}{0.502965in}%
\pgfsys@useobject{currentmarker}{}%
\end{pgfscope}%
\end{pgfscope}%
\begin{pgfscope}%
\definecolor{textcolor}{rgb}{0.000000,0.000000,0.000000}%
\pgfsetstrokecolor{textcolor}%
\pgfsetfillcolor{textcolor}%
\pgftext[x=0.971877in,y=0.405743in,,top]{\color{textcolor}\sffamily\fontsize{6.000000}{7.200000}\selectfont \(\displaystyle t_\mathrm{c}=\)\SI{15}{\micro\second}}%
\end{pgfscope}%
\begin{pgfscope}%
\definecolor{textcolor}{rgb}{0.000000,0.000000,0.000000}%
\pgfsetstrokecolor{textcolor}%
\pgfsetfillcolor{textcolor}%
\pgftext[x=0.958351in,y=0.269540in,,top]{\color{textcolor}\sffamily\fontsize{6.000000}{7.200000}\selectfont \(\displaystyle t^\mathrm{s}_{i, \ell}\) [\si{\micro\second}]}%
\end{pgfscope}%
\begin{pgfscope}%
\pgfsetbuttcap%
\pgfsetroundjoin%
\definecolor{currentfill}{rgb}{0.000000,0.000000,0.000000}%
\pgfsetfillcolor{currentfill}%
\pgfsetlinewidth{0.803000pt}%
\definecolor{currentstroke}{rgb}{0.000000,0.000000,0.000000}%
\pgfsetstrokecolor{currentstroke}%
\pgfsetdash{}{0pt}%
\pgfsys@defobject{currentmarker}{\pgfqpoint{-0.048611in}{0.000000in}}{\pgfqpoint{-0.000000in}{0.000000in}}{%
\pgfpathmoveto{\pgfqpoint{-0.000000in}{0.000000in}}%
\pgfpathlineto{\pgfqpoint{-0.048611in}{0.000000in}}%
\pgfusepath{stroke,fill}%
}%
\begin{pgfscope}%
\pgfsys@transformshift{0.539023in}{0.530116in}%
\pgfsys@useobject{currentmarker}{}%
\end{pgfscope}%
\end{pgfscope}%
\begin{pgfscope}%
\definecolor{textcolor}{rgb}{0.000000,0.000000,0.000000}%
\pgfsetstrokecolor{textcolor}%
\pgfsetfillcolor{textcolor}%
\pgftext[x=0.388782in, y=0.498459in, left, base]{\color{textcolor}\sffamily\fontsize{6.000000}{7.200000}\selectfont 0}%
\end{pgfscope}%
\begin{pgfscope}%
\pgfsetbuttcap%
\pgfsetroundjoin%
\definecolor{currentfill}{rgb}{0.000000,0.000000,0.000000}%
\pgfsetfillcolor{currentfill}%
\pgfsetlinewidth{0.803000pt}%
\definecolor{currentstroke}{rgb}{0.000000,0.000000,0.000000}%
\pgfsetstrokecolor{currentstroke}%
\pgfsetdash{}{0pt}%
\pgfsys@defobject{currentmarker}{\pgfqpoint{-0.048611in}{0.000000in}}{\pgfqpoint{-0.000000in}{0.000000in}}{%
\pgfpathmoveto{\pgfqpoint{-0.000000in}{0.000000in}}%
\pgfpathlineto{\pgfqpoint{-0.048611in}{0.000000in}}%
\pgfusepath{stroke,fill}%
}%
\begin{pgfscope}%
\pgfsys@transformshift{0.539023in}{0.855921in}%
\pgfsys@useobject{currentmarker}{}%
\end{pgfscope}%
\end{pgfscope}%
\begin{pgfscope}%
\definecolor{textcolor}{rgb}{0.000000,0.000000,0.000000}%
\pgfsetstrokecolor{textcolor}%
\pgfsetfillcolor{textcolor}%
\pgftext[x=0.388782in, y=0.824264in, left, base]{\color{textcolor}\sffamily\fontsize{6.000000}{7.200000}\selectfont 6}%
\end{pgfscope}%
\begin{pgfscope}%
\definecolor{textcolor}{rgb}{0.000000,0.000000,0.000000}%
\pgfsetstrokecolor{textcolor}%
\pgfsetfillcolor{textcolor}%
\pgftext[x=0.333226in,y=0.774470in,,bottom,rotate=90.000000]{\color{textcolor}\sffamily\fontsize{6.000000}{7.200000}\selectfont Neuron \(\displaystyle i\)}%
\end{pgfscope}%
\begin{pgfscope}%
\pgfsetrectcap%
\pgfsetmiterjoin%
\pgfsetlinewidth{0.803000pt}%
\definecolor{currentstroke}{rgb}{0.000000,0.000000,0.000000}%
\pgfsetstrokecolor{currentstroke}%
\pgfsetdash{}{0pt}%
\pgfpathmoveto{\pgfqpoint{0.539023in}{0.502965in}}%
\pgfpathlineto{\pgfqpoint{0.539023in}{1.045974in}}%
\pgfusepath{stroke}%
\end{pgfscope}%
\begin{pgfscope}%
\pgfsetrectcap%
\pgfsetmiterjoin%
\pgfsetlinewidth{0.803000pt}%
\definecolor{currentstroke}{rgb}{0.000000,0.000000,0.000000}%
\pgfsetstrokecolor{currentstroke}%
\pgfsetdash{}{0pt}%
\pgfpathmoveto{\pgfqpoint{0.539023in}{0.502965in}}%
\pgfpathlineto{\pgfqpoint{1.377678in}{0.502965in}}%
\pgfusepath{stroke}%
\end{pgfscope}%
\end{pgfpicture}%
\makeatother%
\endgroup%

%% file: tex/bss2.tex
\section{\acrlong{bss-2} Neuromorphic System}\label{sec:bss2}
\label{sec:bss-2}

\begin{figure}[!t]
    \centering
    \tikzset{
        panel/.style={
            inner sep=0pt,
            outer sep=0pt,
            execute at begin node={\tikzset{anchor=center, inner sep=.33333em}}},
        label/.style={
            anchor=north west,
            inner sep=0,
            outer sep=0}}

    \begin{tikzpicture}
        \node[panel, anchor=north west] (a) at (0.6, 0) {
            \includegraphics[width=128px]{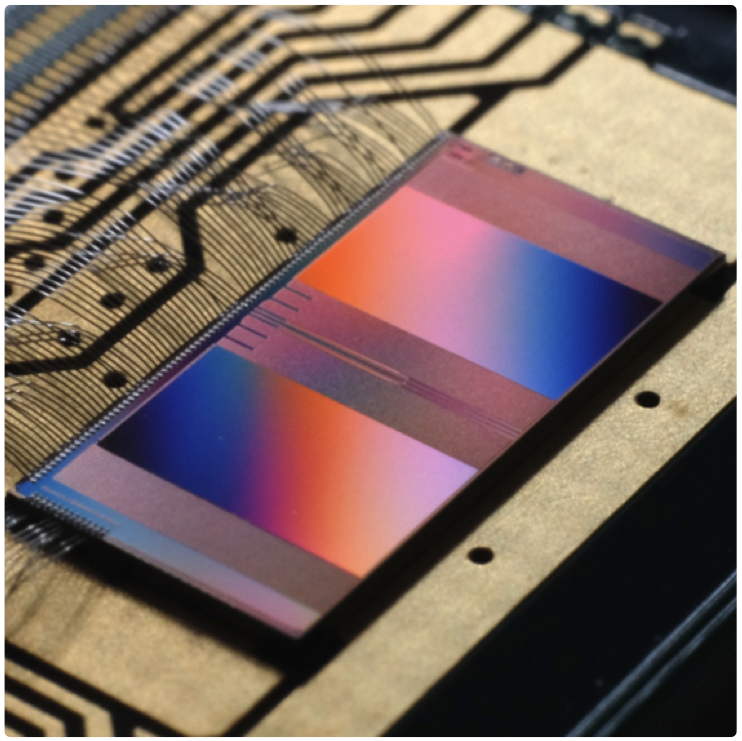}};
        \node[label] at (0.3, 0) {\textbf{A}};

        \node[panel, anchor=north west] (b) at (5.0, -0.1) {
            \input{./figures/chip_schematic.tex}};
        \node[label] at (5.2, 0) {\textbf{B}};
    \end{tikzpicture}

    \caption{%
    \textbf{(A)} The \acrlong{bss-2} neuromorphic \gls{asic} bonded onto its carrier board:
    the \gls{asic} is about \SI{4}{\milli\meter} \texttimes{} \SI{8}{\milli\meter} in size.
    It is organized in two hemispheres, each containing 256 spiking neurons in analog circuits.
    \textbf{(B)} The schematic of \gls{bss-2}:
    one parallel \acrlong{cadc} and one general-purpose \acrfull{simd} processor are available per hemisphere.}
    \label{fig:bss2}
\end{figure}
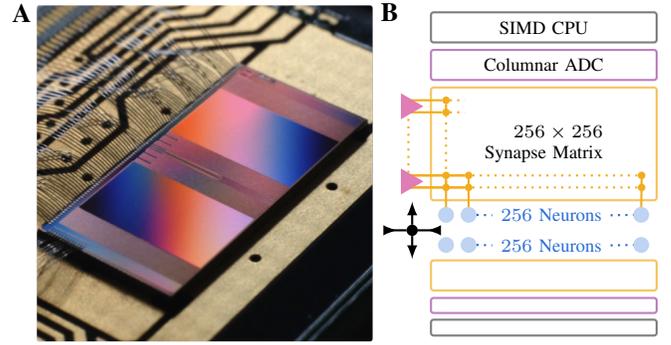

We now discuss the emulation of the \gls{snn} demappers on the \acrfull{bss-2} system~\cite{pehle2022brainscales2}.

\gls{bss-2}, depicted in \cref{fig:bss2}, is an accelerated neuromorphic mixed-signal hardware platform developed at Heidelberg University.
Its \acrlong{anncore} features 512 \gls{adex}~\cite{brette2005adaptive} neuron compartments on two chip hemispheres, each implemented as an analog circuit and emulated in continuous time.

Each neuron compartment receives stimuli from a column of 256 synapses with 6-bit weights.
Increased fan-in can be achieved by building larger `logical' neurons from multiple compartments.
Synapses can be configured to be inhibitory (negative sign) or excitatory (positive sign) in a row-wise fashion.
To realize a specific network topology, on-chip spike routing connects neurons to target synapses.
In addition, off-chip spikes are injected as an input stimulus for the network.

Neuron circuits can be configured to emulate \gls{lif} neurons as well as \gls{li} neurons by disabling the spiking mechanism.
Each hemisphere on \gls{bss-2} has one \gls{cadc} to measure and digitize in parallel the membrane potentials of the neurons of each column in the synapse matrix with an effective sampling period of about \SI{2}{\micro\second} including time stamping and writing to memory.
\Gls{cadc} measurements and spike events can be recorded on an \acrshort{fpga}-managed \gls{dram} and read out by the host computer.

Recording hardware observables allows for hardware \acrshort{itl} training.
In the case of our \gls{snn} demapper on \gls{bss-2},
the forward pass is performed on \gls{bss-2} and the hardware gradient is estimated on the host computer by utilizing the network's recorded hardware observables to calculate the weight updates~\cite{schmitt2017neuromorphic,cramer2022surrogate}.

To obtain an equivalent experiment configuration on \gls{bss-2}, our software stack
translates the high-level \gls{snn} experiment description to a data flow graph representation, places and routes neurons and synapses on the hardware substrate, and compiles stimulus inputs, recording settings and other runtime dynamics into an experiment program~\cite{mueller2022scalable}.

The analog circuits on \gls{bss-2} are subject to device variations (fixed-pattern noise) that can be compensated for by calibration.
Therefore, one part of the system configuration consists of a calibration data set that is loaded to obtain a chip operating point, which most closely resembles the desired target dynamics with minimal variation, e.g., with respect to model parameters such as neuron membrane time constants or synaptic efficacy.

To represent one signed software weight $w_\text{sw}$ on \gls{bss-2}, two hardware synapses, with the respectively excitatory and inhibitory weights
\begin{equation}
    w^\text{inh}_\text{hw} = \max \left(0, -w_\text{sw}\right) \quad \text{and} \quad w^\text{exc}_\text{hw} = \max \left(0, w_\text{sw}\right),
\end{equation}
are allocated and constitute one signed hardware weight $w_\text{hw} \in \left[ -63, 63 \right]$.
We scale each weight $w_\text{sw}$ linearly into a hardware-compatible range and round it to the nearest value representable on \gls{bss-2}.
The batched input spikes are injected into \gls{bss-2} and the \gls{snn} is emulated for $T = \SI{30}{\micro\second}$ per batch entry, i.e., for demapping a single sample.
During emulation, spike events are recorded and the \gls{cadc} samples membrane voltages of the hidden neurons and the readout neurons.
After the emulation, the host computer reads back and post-processes the recorded data.
The post-processing step includes a linear interpolation to convert event-based \gls{cadc} recordings to a \texttt{torch::Tensor} expressed on a fixed time grid.
To facilitate hardware-\gls{itl} training on \gls{bss-2}, we utilized \texttt{hxtorch.snn}~\cite{mueller2022scalable}, a \texttt{PyTorch}-based~\cite{paszke2017automatic} library that automates and abstracts away hardware-specific procedures and provides data conversions from and to \texttt{PyTorch}.

%% file: figures/chip_schematic.tex
\begin{tikzpicture}
    \usetikzlibrary{shapes}
    \tikzset{
    neuron/.style={
        anchor=south,
        align=center,
        fill=blue!30,
        circle,
        inner sep=0,
        outer sep=0,
        minimum size=6pt},
    synapse_matrix/.style={
        thick,
        font=\scriptsize,
        draw=yellow!70,
        anchor=south,
        rectangle,
        inner sep=4pt,
        outer sep=0,
        rounded corners=1pt,
        minimum height=1.8cm,
        align=left},
    ppus/.style={
        font=\scriptsize,
        thick,
        anchor=south,
        draw=black!50,
        rectangle,
        inner sep=3pt,
        outer sep=0,
        rounded corners=1pt,
        minimum height=0.4cm,
        align=left},
    cadcs/.style={
        thick,
        font=\scriptsize,
        anchor=south,
        rectangle,
        draw=violet!50,
        inner sep=1pt,
        outer sep=0,
        rounded corners=1pt,
        minimum height=0.4cm,
        align=left},
    drivers/.style={
        regular polygon,
        regular polygon sides=3,
        inner sep=0,
        fill=red!60,
        shape border rotate=30,
        minimum size=0.4cm},
    syn/.style={
        circle,
        fill=yellow,
        minimum size=1.mm,
        inner sep=0}
    }

    \def\stackwidth{3.0cm};
    \def\columnstep{0.3};

    \node[ppus, minimum height=0.2cm, minimum width=\stackwidth] (ppu1) at (0, 0) {};
    \node[cadcs, minimum height=0.2cm, minimum width=\stackwidth] (cadc1) at (0, 0.3) {};
    \node[synapse_matrix, minimum height=0.4cm, minimum width=\stackwidth] (syn_mat_0) at (0, 0.6) {};

    \def\neuronbase{1.1}
    \node[neuron] (n_0_l) at (-1.3 + 0 * \columnstep, \neuronbase) {};
    \node[neuron] (n_1_l) at (-1.3 + 1 * \columnstep, \neuronbase) {};
    \node[font=\scriptsize, color=blue] (n_t_l) at (-1.3 + 1.4, 0.1 + \neuronbase) {$256$ Neurons};
    \node[neuron] (n_N_l) at (-1.3 + 8.65 * \columnstep, \neuronbase) {};
    \draw[thick, draw=blue, dotted] (n_1_l.east) -- (n_t_l.west);
    \draw[thick, draw=blue, dotted] (n_t_l.east) -- (n_N_l.west);

    \def\neuronbaseupper{1.5}
    \node[neuron] (n_0_u) at (-1.3 + 0 * \columnstep, \neuronbaseupper) {};
    \node[neuron] (n_1_u) at (-1.3 + 1 * \columnstep, \neuronbaseupper) {};
    \node[font=\scriptsize, color=blue] (n_t_u) at (-1.3 + 1.4, 0.1 + \neuronbaseupper) {$256$ Neurons};
    \node[neuron] (n_N_u) at (-1.3 + 8.65 * \columnstep, \neuronbaseupper) {};
    \draw[thick, draw=blue, dotted] (n_1_u.east) -- (n_t_u.west);
    \draw[thick, draw=blue, dotted] (n_t_u.east) -- (n_N_u.west);

    \node[synapse_matrix, align=right, minimum height=1.5cm, minimum width=\stackwidth] (syn_mat_1) at (0, 1.8) {$256\times 256$ \\ Synapse Matrix};
    \node[cadcs, minimum width=\stackwidth] (cadc2) at (0, 3.4) {Columnar ADC};
    \node[ppus, minimum width=\stackwidth] (ppu2) at (0, 3.9) {SIMD CPU};

    \node[syn] (syn_1) at (-1.30, 1.97) {};
    \node[syn] (syn_2) at (-1.30, 2.13) {};
    \node[syn] (syn_3) at (-1.30 + \columnstep, 1.97) {};
    \node[syn] (syn_4) at (-1.30 + \columnstep, 2.13) {};
    \node[syn, fill=none] (syn_1_f) at (-1.80, 1.97) {};
    \node[syn, fill=none] (syn_2_f) at (-1.80, 2.13) {};
    \node[syn] (syn_5) at (1.3, 1.97) {};
    \node[syn] (syn_6) at (1.3, 2.13) {};
    
    \node[syn, fill=none] (syn_7_f) at (-1.80, 2.97) {};
    \node[syn, fill=none] (syn_8_f) at (-1.80, 3.13) {};
    \node[syn] (syn_7) at (-1.30, 2.97) {};
    \node[syn] (syn_8) at (-1.30, 3.13) {};
    \node[syn, fill=none] (syn_9) at (-1.30 + \columnstep, 2.97) {};
    \node[syn, fill=none] (syn_10) at (-1.30 + \columnstep, 3.13) {};

    \draw[thick, draw=yellow] (syn_1_f.west) -- (syn_3.west);
    \draw[thick, draw=yellow] (syn_2_f.west) -- (syn_4.west);
    \draw[thick, draw=yellow, dotted] (syn_3.east) -- (syn_5.west);
    \draw[thick, draw=yellow, dotted] (syn_4.east) -- (syn_6.west);
    \draw[thick, draw=yellow] (syn_1_f.west) -- (syn_3.west);
    \draw[thick, draw=yellow] (syn_2_f.west) -- (syn_4.west);
    \draw[thick, draw=yellow] (syn_7_f.west) -- (syn_7.west);
    \draw[thick, draw=yellow] (syn_8_f.west) -- (syn_8.west);
    \draw[thick, draw=yellow, dotted] (syn_7.east) -- (syn_9.west);
    \draw[thick, draw=yellow, dotted] (syn_8.east) -- (syn_10.west);
    \draw[thick, draw=yellow, dotted] (syn_1.north) -- (syn_7.south);
    \draw[thick, draw=yellow] (syn_7.north) -- (syn_8.south);

    \node[drivers] (drv1) at (-1.8, 2.05) {};
    \node[drivers] (drv2) at (-1.8, 3.05) {};
    \draw[thick, draw=yellow, dotted] (drv1.north) -- (drv2.south);

    \draw[thick, draw=yellow] (syn_2.south) -- (n_0_u.north);
    \draw[thick, draw=yellow] (syn_4.south) -- (n_1_u.north);
    \draw[thick, draw=yellow] (syn_6.south) -- (n_N_u.north);

    \node[circle, fill=black, minimum size=1.5mm, inner sep=0] (cb) at (-1.75, 1.4) {};
    \node (cb_up) at (-1.75, 1.9) {};
    \node (cb_down) at (-1.75, 0.9) {};
    \node (cb_right) at (-1.25, 1.4) {};
    \node (cb_left) at (-2.25, 1.4) {};

    \draw[thick, draw=black, -latex] (cb.north) -- (cb_up.south);
    \draw[thick, draw=black, -latex] (cb.south) -- (cb_down.north);
    \draw[thick, draw=black, -latex reversed] (cb.east) -- (cb_right.west);
    \draw[thick, draw=black, -latex reversed] (cb.west) -- (cb_left.east);

\end{tikzpicture}

%% file: tex/training.tex
\section{Training and Testing}
\label{sec:training}
To measure the \gls{ber} of the demappers against the noise-level in the \gls{imdd} link, we train our models with successively increasing noise-levels $\sigma^2$. At each noise-level, we perform validation runs on independent data and store the model parameters of the best performing demapper. At the next noise-level, we restore the best model from the previous noise-level and continue training. This procedure is repeated for five different random seeds, affecting model initialization, \gls{imdd}-data generation and sampling permutations. We select the best-performing demappers for each noise-level over the seeds according to their respective validation runs and benchmark the models on independent test data. The tests are run until a minimum of $2000$ bit error events are encountered.

%% file: tex/results.tex
\section{Results}
\label{sec:results}

\begin{figure}[!t]
    \centering
    \tikzset{
        panel/.style={
            inner sep=0pt,
            outer sep=0pt,
            execute at begin node={\tikzset{anchor=center, inner sep=.33333em}}},
        label/.style={
            anchor=north west,
            inner sep=0,
            outer sep=0}}
    \begin{tikzpicture}
        \node[panel, anchor=north west] (a) at (0,  0) {
            \input{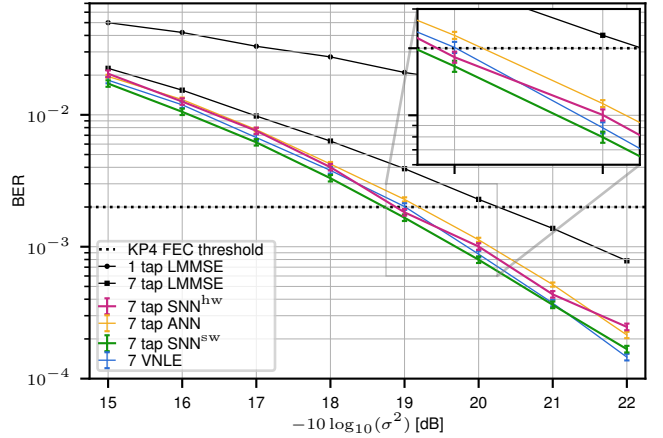}};
    \end{tikzpicture}
    \caption{The \acrshort{ber} of \acrshort{snn} equalizers in simulation and on the \acrshort{bss-2} system over the noise-levels $\sigma^2$ in the \acrshort{imdd} link compared to \acrshort{ann}, \acrshort{vnle}, and \acrshort{lmmse} reference equalizers. The error bars denote the 99\% credibility intervals.}
    \label{fig:ber_snr_sweep}
\end{figure}

In \Cref{fig:ber_snr_sweep}, we compare our $7$-tap \gls{snn} demapper emulated on the analog neuromorphic \gls{bss-2} system (SNN$^\text{hw}$) to a $7$-tap \gls{snn} demapper simulated in software (SNN$^\text{sw}$) in terms of \gls{ber} versus noise-level $\sigma^2$ in the link.
We benchmark our \gls{snn} performances against the \gls{lmmse}, with 7 taps (LE7) and without memory (LE1).
As additional nonlinear references, we consider a 7-tap \gls{ann} demapper with two hidden layers (see \cref{sec:ann}) and a 7-tap \gls{vnle}.
All demapper configurations are specified in \cref{tab:bers}.

\begin{table}[!t]
    \caption{Demapper Definitions}
    \vspace{-10px}
    \label{tab:bers}
    \begin{center}
        \begin{tabular}{c|ccc}
        \hline \hline
        Name & Type & Layers & $n_\text{tap}$ \\ \hline
        LE1 & \acrshort{lmmse} & $1 - 1$ & $1$ \\
        LE7 & \acrshort{lmmse} & $7 - 1$ & $7$ \\ \hline
        ANN & \acrshort{ann} & $7$ -- $40$ -- $20$ -- $4$ & $7$ \\ \hline
        VNLE & \acrshort{vnle} & - & $7$ \\ \hline
        SNN$^\text{sw}$ & \acrshort{snn} in sim.\ & $70$ -- $40$ -- $4$ & $7$ \\
        SNN$^\text{hw}$ & \acrshort{snn} on \acrshort{bss-2} & $70$ -- $40$ -- $4$ & $7$ \\
        \hline \hline
    \end{tabular}
    \end{center}
\end{table}

Both the simulated SNN$^\text{sw}$ demapper and the SNN$^\text{hw}$ demapper on \gls{bss-2} outperform the \gls{lmmse} demapper.
At a pre-\gls{fec} \gls{ber} of \num[exponent-product=\cdot]{2e-3}, we observe a gain of about \SI{1.5}{dB} of the SNN$^\text{sw}$ demapper to the LE7 demapper and a gain of \SI{0.5}{dB} to the nonlinear \gls{ann} demapper.
Compared to the \gls{vnle} demapper, the SNN$^\text{sw}$ demapper shows superior performance for noise levels higher than \SI{-21}{dB}, in particular at the considered pre-\gls{fec} \gls{ber} threshold, it shows a \SI{0.3}{dB} improvement, however, for noise levels lower than \SI{-21}{dB} the \gls{vnle} demapper achieves a lower \gls{ber}.

The SNN$^\text{hw}$ demapper on \gls{bss-2} approaches the performance observed with the simulated \gls{snn} and only suffers from a small hardware penalty with respect to the SNN$^\text{sw}$ of about \SI{0.2}{dB} at a \gls{ber} of \num[exponent-product=\cdot]{2e-3}, outperforming all reference strategies.

\begin{figure*}[!t]
    \centering
    \tikzset{
        panel/.style={
            inner sep=0pt,
            outer sep=0pt,
            execute at begin node={\tikzset{anchor=center, inner sep=.33333em}}},
        label/.style={
            anchor=north west,
            inner sep=0,
            outer sep=0}}

    \begin{tikzpicture}
        \node[panel, anchor=north west] (a) at (0,  0) {
            \input{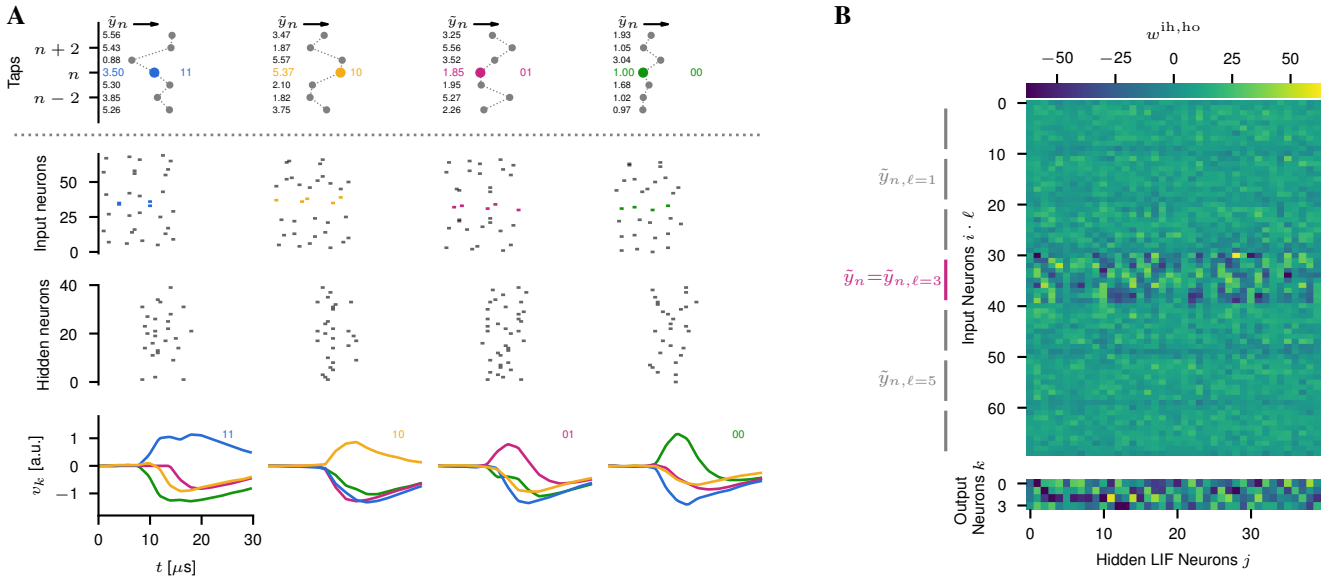}};
        \node[label] at (a.north west) {\textbf{A}};

        \node[panel, anchor=north west] (b) at (11,  0) {
            \input{./figures/weights.tex}};
        \node[label] at (b.north west) {\textbf{B}};
    \end{tikzpicture}

    \vspace{-10px}
    \caption{\textbf{(A)} Four different examples visualizing the bit inference process on \gls{bss-2}. The upper row depicts the chunk of samples used for demapping the innermost sample (colored). This chunk is translated to the spike events of the input neurons (10 per sample) as shown in the the second row. The third rows depicted the corresponding spike events of the hidden \acrshort{lif} neurons. In the last row, the membrane voltage traces of the four \acrshort{li} neurons are plotted. The output neuron corresponding to the estimated symbol produces the maximum voltage value over time. A Gray demapper provides bit decisions.
    \textbf{(B)} The \gls{snn} demapper's learned synapse matrices on \acrshort{bss-2}. The upper matrix shows the weights from the input to the hidden layer. Note, each of the $n_\text{tap}=7$ samples has $10$ input neurons assigned, resulting in $70$ input rows. The rows corresponding to the middle sample $\Tilde{y}_n$ contribute most to the decision. The lower matrix shows the weights from the hidden to the output layer.
    }
    \label{fig:snn_results}
\end{figure*}

In \cref{fig:snn_results}A, we visualize the process of joint equalization and demapping on \gls{bss-2} on four different samples.
The upper row indicates the sample set $\Tilde{\boldsymbol{y}}_n$, with the sample of interest $\Tilde{y}_n$ highlighted.
Each sample in this set is translated to spike times of $10$ input neurons, depicted in the second row.
For $n_\text{tap} = 7$, the hidden \gls{lif} layer receives spike events from $70$ input neurons, of which the majority are silent due to a cutoff time of \SI{15}{\micro\second} (see \cref{sec:ssn_eq}).
These input spike events activate the $40$ \gls{lif} neurons in the hidden layer, exciting them to emit spike events themselves as shown in the third row.
These spikes events constitute a meaningful pattern, driving the membrane voltage of the correct \gls{li} output neuron to the maximum voltage value over time, from which the bits are inferred via an \gls{hd}.
This behavior is observed in the analog membrane traces in the lowermost row.
The membrane voltage of the readout neuron corresponding to the estimated symbol is deflected upwards while the others drop below zero and hence do not intervene in the decision.
Note that the dynamics visualized in each column from the second to fourth row all happen simultaneously in \gls{bss-2}'s analog circuits.

The weight matrices learned on \gls{bss-2} are shown in \cref{fig:snn_results}B.
The input-to-hidden weight matrix $w^\text{ih}_{ij}$ shows a greater weight magnitude for rows with indices $i \in [30, 39]$. This is expected as these rows receive the input spike events encoding the most significant sample to demap $\Tilde{y}_n$ in the innermost tap.
For the outer rows, one can observe a pattern repeating with the number of input rows per sample, $\bar{N}^\text{i}=10$.
The lower plot depicts the hidden-to-output weight matrix $w^\text{ho}_{jk}$.

%% file: figures/weights.tex
\begin{tikzpicture}
    \tikzset{
        neuron/.style={
            circle,
            inner sep=0pt,
            outer sep=3pt,
            align=center,
            thick,
            minimum size=7pt}
    }

    \def\spacing{0.67}

    \node (plot) at (0,  0) {\input{./figures/weights_snr_22.pgf}};

    \draw[very thick, draw=red] (-2.6, 0.01) -- (-2.6, 0.55);
    \node[color=red] (sample) at (-3.3, 0.3) {$\scriptstyle \Tilde{y}_{n} =  \Tilde{y}_{n, \ell = 3}$};
    \node[color=black!50] (sample) at (-3.1, 0.3 + 2 * \spacing) {$\scriptstyle \Tilde{y}_{n, \ell = 1}$};
    \node[color=black!50] (sample) at (-3.1, 0.3 - 2 * \spacing) {$\scriptstyle \Tilde{y}_{n, \ell = 5}$};

    \draw[very thick, draw=black!50] (-2.6, 0.01 + 1 * \spacing) -- (-2.6, 0.55 + 1 * \spacing);
    \draw[very thick, draw=black!50] (-2.6, 0.01 + 2 * \spacing) -- (-2.6, 0.55 + 2 * \spacing);
    \draw[very thick, draw=black!50] (-2.6, 0.01 + 3 * \spacing) -- (-2.6, 0.55 + 3 * \spacing);

    \draw[very thick, draw=black!50] (-2.6, 0.01 - 1 * \spacing) -- (-2.6, 0.55 - 1 * \spacing);
    \draw[very thick, draw=black!50] (-2.6, 0.01 - 2 * \spacing) -- (-2.6, 0.55 - 2 * \spacing);
    \draw[very thick, draw=black!50] (-2.6, 0.01 - 3 * \spacing) -- (-2.6, 0.55 - 3 * \spacing);

\end{tikzpicture}

%% file: figures/weights_snr_22.pgf
\begingroup%
\makeatletter%
\begin{pgfpicture}%
\pgfpathrectangle{\pgfpointorigin}{\pgfqpoint{2.000000in}{3.000000in}}%
\pgfusepath{use as bounding box, clip}%
\begin{pgfscope}%
\pgfsetbuttcap%
\pgfsetmiterjoin%
\definecolor{currentfill}{rgb}{1.000000,1.000000,1.000000}%
\pgfsetfillcolor{currentfill}%
\pgfsetlinewidth{0.000000pt}%
\definecolor{currentstroke}{rgb}{1.000000,1.000000,1.000000}%
\pgfsetstrokecolor{currentstroke}%
\pgfsetdash{}{0pt}%
\pgfpathmoveto{\pgfqpoint{0.000000in}{0.000000in}}%
\pgfpathlineto{\pgfqpoint{2.000000in}{0.000000in}}%
\pgfpathlineto{\pgfqpoint{2.000000in}{3.000000in}}%
\pgfpathlineto{\pgfqpoint{0.000000in}{3.000000in}}%
\pgfpathclose%
\pgfusepath{fill}%
\end{pgfscope}%
\begin{pgfscope}%
\pgfsetbuttcap%
\pgfsetmiterjoin%
\definecolor{currentfill}{rgb}{1.000000,1.000000,1.000000}%
\pgfsetfillcolor{currentfill}%
\pgfsetlinewidth{0.000000pt}%
\definecolor{currentstroke}{rgb}{0.000000,0.000000,0.000000}%
\pgfsetstrokecolor{currentstroke}%
\pgfsetstrokeopacity{0.000000}%
\pgfsetdash{}{0pt}%
\pgfpathmoveto{\pgfqpoint{0.400000in}{2.565509in}}%
\pgfpathlineto{\pgfqpoint{1.940000in}{2.565509in}}%
\pgfpathlineto{\pgfqpoint{1.940000in}{2.640000in}}%
\pgfpathlineto{\pgfqpoint{0.400000in}{2.640000in}}%
\pgfpathclose%
\pgfusepath{fill}%
\end{pgfscope}%
\begin{pgfscope}%
\pgfpathrectangle{\pgfqpoint{0.400000in}{2.565509in}}{\pgfqpoint{1.540000in}{0.074491in}}%
\pgfusepath{clip}%
\pgfsetbuttcap%
\pgfsetmiterjoin%
\definecolor{currentfill}{rgb}{1.000000,1.000000,1.000000}%
\pgfsetfillcolor{currentfill}%
\pgfsetlinewidth{0.010037pt}%
\definecolor{currentstroke}{rgb}{1.000000,1.000000,1.000000}%
\pgfsetstrokecolor{currentstroke}%
\pgfsetdash{}{0pt}%
\pgfpathmoveto{\pgfqpoint{0.400000in}{2.565509in}}%
\pgfpathlineto{\pgfqpoint{0.406016in}{2.565509in}}%
\pgfpathlineto{\pgfqpoint{1.933984in}{2.565509in}}%
\pgfpathlineto{\pgfqpoint{1.940000in}{2.565509in}}%
\pgfpathlineto{\pgfqpoint{1.940000in}{2.640000in}}%
\pgfpathlineto{\pgfqpoint{1.933984in}{2.640000in}}%
\pgfpathlineto{\pgfqpoint{0.406016in}{2.640000in}}%
\pgfpathlineto{\pgfqpoint{0.400000in}{2.640000in}}%
\pgfpathlineto{\pgfqpoint{0.400000in}{2.640000in}}%
\pgfpathclose%
\pgfusepath{stroke,fill}%
\end{pgfscope}%
\begin{pgfscope}%
\pgfsys@transformshift{0.400000in}{2.566667in}%
\pgftext[left,bottom]{\includegraphics[interpolate=true,width=1.540000in,height=0.073333in]{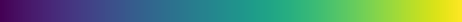}}%
\end{pgfscope}%
\begin{pgfscope}%
\pgfsetbuttcap%
\pgfsetroundjoin%
\definecolor{currentfill}{rgb}{0.000000,0.000000,0.000000}%
\pgfsetfillcolor{currentfill}%
\pgfsetlinewidth{0.803000pt}%
\definecolor{currentstroke}{rgb}{0.000000,0.000000,0.000000}%
\pgfsetstrokecolor{currentstroke}%
\pgfsetdash{}{0pt}%
\pgfsys@defobject{currentmarker}{\pgfqpoint{0.000000in}{0.000000in}}{\pgfqpoint{0.000000in}{0.048611in}}{%
\pgfpathmoveto{\pgfqpoint{0.000000in}{0.000000in}}%
\pgfpathlineto{\pgfqpoint{0.000000in}{0.048611in}}%
\pgfusepath{stroke,fill}%
}%
\begin{pgfscope}%
\pgfsys@transformshift{0.558889in}{2.640000in}%
\pgfsys@useobject{currentmarker}{}%
\end{pgfscope}%
\end{pgfscope}%
\begin{pgfscope}%
\definecolor{textcolor}{rgb}{0.000000,0.000000,0.000000}%
\pgfsetstrokecolor{textcolor}%
\pgfsetfillcolor{textcolor}%
\pgftext[x=0.558889in,y=2.737222in,,bottom]{\color{textcolor}\sffamily\fontsize{6.000000}{7.200000}\selectfont \ensuremath{-}50}%
\end{pgfscope}%
\begin{pgfscope}%
\pgfsetbuttcap%
\pgfsetroundjoin%
\definecolor{currentfill}{rgb}{0.000000,0.000000,0.000000}%
\pgfsetfillcolor{currentfill}%
\pgfsetlinewidth{0.803000pt}%
\definecolor{currentstroke}{rgb}{0.000000,0.000000,0.000000}%
\pgfsetstrokecolor{currentstroke}%
\pgfsetdash{}{0pt}%
\pgfsys@defobject{currentmarker}{\pgfqpoint{0.000000in}{0.000000in}}{\pgfqpoint{0.000000in}{0.048611in}}{%
\pgfpathmoveto{\pgfqpoint{0.000000in}{0.000000in}}%
\pgfpathlineto{\pgfqpoint{0.000000in}{0.048611in}}%
\pgfusepath{stroke,fill}%
}%
\begin{pgfscope}%
\pgfsys@transformshift{0.864444in}{2.640000in}%
\pgfsys@useobject{currentmarker}{}%
\end{pgfscope}%
\end{pgfscope}%
\begin{pgfscope}%
\definecolor{textcolor}{rgb}{0.000000,0.000000,0.000000}%
\pgfsetstrokecolor{textcolor}%
\pgfsetfillcolor{textcolor}%
\pgftext[x=0.864444in,y=2.737222in,,bottom]{\color{textcolor}\sffamily\fontsize{6.000000}{7.200000}\selectfont \ensuremath{-}25}%
\end{pgfscope}%
\begin{pgfscope}%
\pgfsetbuttcap%
\pgfsetroundjoin%
\definecolor{currentfill}{rgb}{0.000000,0.000000,0.000000}%
\pgfsetfillcolor{currentfill}%
\pgfsetlinewidth{0.803000pt}%
\definecolor{currentstroke}{rgb}{0.000000,0.000000,0.000000}%
\pgfsetstrokecolor{currentstroke}%
\pgfsetdash{}{0pt}%
\pgfsys@defobject{currentmarker}{\pgfqpoint{0.000000in}{0.000000in}}{\pgfqpoint{0.000000in}{0.048611in}}{%
\pgfpathmoveto{\pgfqpoint{0.000000in}{0.000000in}}%
\pgfpathlineto{\pgfqpoint{0.000000in}{0.048611in}}%
\pgfusepath{stroke,fill}%
}%
\begin{pgfscope}%
\pgfsys@transformshift{1.170000in}{2.640000in}%
\pgfsys@useobject{currentmarker}{}%
\end{pgfscope}%
\end{pgfscope}%
\begin{pgfscope}%
\definecolor{textcolor}{rgb}{0.000000,0.000000,0.000000}%
\pgfsetstrokecolor{textcolor}%
\pgfsetfillcolor{textcolor}%
\pgftext[x=1.170000in,y=2.737222in,,bottom]{\color{textcolor}\sffamily\fontsize{6.000000}{7.200000}\selectfont 0}%
\end{pgfscope}%
\begin{pgfscope}%
\pgfsetbuttcap%
\pgfsetroundjoin%
\definecolor{currentfill}{rgb}{0.000000,0.000000,0.000000}%
\pgfsetfillcolor{currentfill}%
\pgfsetlinewidth{0.803000pt}%
\definecolor{currentstroke}{rgb}{0.000000,0.000000,0.000000}%
\pgfsetstrokecolor{currentstroke}%
\pgfsetdash{}{0pt}%
\pgfsys@defobject{currentmarker}{\pgfqpoint{0.000000in}{0.000000in}}{\pgfqpoint{0.000000in}{0.048611in}}{%
\pgfpathmoveto{\pgfqpoint{0.000000in}{0.000000in}}%
\pgfpathlineto{\pgfqpoint{0.000000in}{0.048611in}}%
\pgfusepath{stroke,fill}%
}%
\begin{pgfscope}%
\pgfsys@transformshift{1.475556in}{2.640000in}%
\pgfsys@useobject{currentmarker}{}%
\end{pgfscope}%
\end{pgfscope}%
\begin{pgfscope}%
\definecolor{textcolor}{rgb}{0.000000,0.000000,0.000000}%
\pgfsetstrokecolor{textcolor}%
\pgfsetfillcolor{textcolor}%
\pgftext[x=1.475556in,y=2.737222in,,bottom]{\color{textcolor}\sffamily\fontsize{6.000000}{7.200000}\selectfont 25}%
\end{pgfscope}%
\begin{pgfscope}%
\pgfsetbuttcap%
\pgfsetroundjoin%
\definecolor{currentfill}{rgb}{0.000000,0.000000,0.000000}%
\pgfsetfillcolor{currentfill}%
\pgfsetlinewidth{0.803000pt}%
\definecolor{currentstroke}{rgb}{0.000000,0.000000,0.000000}%
\pgfsetstrokecolor{currentstroke}%
\pgfsetdash{}{0pt}%
\pgfsys@defobject{currentmarker}{\pgfqpoint{0.000000in}{0.000000in}}{\pgfqpoint{0.000000in}{0.048611in}}{%
\pgfpathmoveto{\pgfqpoint{0.000000in}{0.000000in}}%
\pgfpathlineto{\pgfqpoint{0.000000in}{0.048611in}}%
\pgfusepath{stroke,fill}%
}%
\begin{pgfscope}%
\pgfsys@transformshift{1.781111in}{2.640000in}%
\pgfsys@useobject{currentmarker}{}%
\end{pgfscope}%
\end{pgfscope}%
\begin{pgfscope}%
\definecolor{textcolor}{rgb}{0.000000,0.000000,0.000000}%
\pgfsetstrokecolor{textcolor}%
\pgfsetfillcolor{textcolor}%
\pgftext[x=1.781111in,y=2.737222in,,bottom]{\color{textcolor}\sffamily\fontsize{6.000000}{7.200000}\selectfont 50}%
\end{pgfscope}%
\begin{pgfscope}%
\definecolor{textcolor}{rgb}{0.000000,0.000000,0.000000}%
\pgfsetstrokecolor{textcolor}%
\pgfsetfillcolor{textcolor}%
\pgftext[x=1.170000in,y=2.873425in,,base]{\color{textcolor}\sffamily\fontsize{6.000000}{7.200000}\selectfont \(\displaystyle w^\mathrm{ih, ho}\)}%
\end{pgfscope}%
\begin{pgfscope}%
\pgfsetbuttcap%
\pgfsetmiterjoin%
\definecolor{currentfill}{rgb}{1.000000,1.000000,1.000000}%
\pgfsetfillcolor{currentfill}%
\pgfsetlinewidth{0.000000pt}%
\definecolor{currentstroke}{rgb}{0.000000,0.000000,0.000000}%
\pgfsetstrokecolor{currentstroke}%
\pgfsetstrokeopacity{0.000000}%
\pgfsetdash{}{0pt}%
\pgfpathmoveto{\pgfqpoint{0.400000in}{0.687848in}}%
\pgfpathlineto{\pgfqpoint{1.940000in}{0.687848in}}%
\pgfpathlineto{\pgfqpoint{1.940000in}{2.550115in}}%
\pgfpathlineto{\pgfqpoint{0.400000in}{2.550115in}}%
\pgfpathclose%
\pgfusepath{fill}%
\end{pgfscope}%
\begin{pgfscope}%
\pgfpathrectangle{\pgfqpoint{0.400000in}{0.687848in}}{\pgfqpoint{1.540000in}{1.862267in}}%
\pgfusepath{clip}%
\pgfsys@transformshift{0.400000in}{0.687848in}%
\pgftext[left,bottom]{\includegraphics[interpolate=true,width=1.540000in,height=1.863333in]{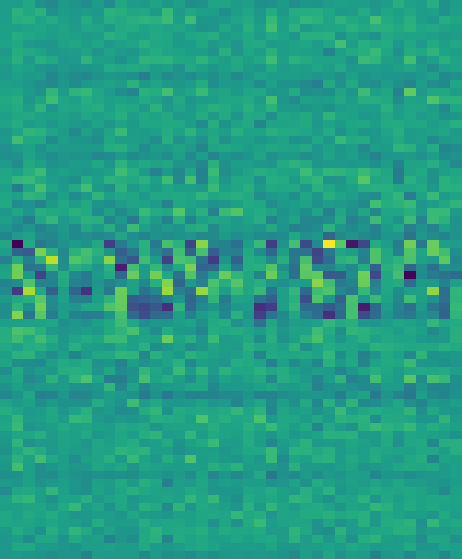}}%
\end{pgfscope}%
\begin{pgfscope}%
\pgfsetbuttcap%
\pgfsetroundjoin%
\definecolor{currentfill}{rgb}{0.000000,0.000000,0.000000}%
\pgfsetfillcolor{currentfill}%
\pgfsetlinewidth{0.803000pt}%
\definecolor{currentstroke}{rgb}{0.000000,0.000000,0.000000}%
\pgfsetstrokecolor{currentstroke}%
\pgfsetdash{}{0pt}%
\pgfsys@defobject{currentmarker}{\pgfqpoint{-0.048611in}{0.000000in}}{\pgfqpoint{-0.000000in}{0.000000in}}{%
\pgfpathmoveto{\pgfqpoint{-0.000000in}{0.000000in}}%
\pgfpathlineto{\pgfqpoint{-0.048611in}{0.000000in}}%
\pgfusepath{stroke,fill}%
}%
\begin{pgfscope}%
\pgfsys@transformshift{0.400000in}{2.536813in}%
\pgfsys@useobject{currentmarker}{}%
\end{pgfscope}%
\end{pgfscope}%
\begin{pgfscope}%
\definecolor{textcolor}{rgb}{0.000000,0.000000,0.000000}%
\pgfsetstrokecolor{textcolor}%
\pgfsetfillcolor{textcolor}%
\pgftext[x=0.249759in, y=2.505156in, left, base]{\color{textcolor}\sffamily\fontsize{6.000000}{7.200000}\selectfont 0}%
\end{pgfscope}%
\begin{pgfscope}%
\pgfsetbuttcap%
\pgfsetroundjoin%
\definecolor{currentfill}{rgb}{0.000000,0.000000,0.000000}%
\pgfsetfillcolor{currentfill}%
\pgfsetlinewidth{0.803000pt}%
\definecolor{currentstroke}{rgb}{0.000000,0.000000,0.000000}%
\pgfsetstrokecolor{currentstroke}%
\pgfsetdash{}{0pt}%
\pgfsys@defobject{currentmarker}{\pgfqpoint{-0.048611in}{0.000000in}}{\pgfqpoint{-0.000000in}{0.000000in}}{%
\pgfpathmoveto{\pgfqpoint{-0.000000in}{0.000000in}}%
\pgfpathlineto{\pgfqpoint{-0.048611in}{0.000000in}}%
\pgfusepath{stroke,fill}%
}%
\begin{pgfscope}%
\pgfsys@transformshift{0.400000in}{2.270775in}%
\pgfsys@useobject{currentmarker}{}%
\end{pgfscope}%
\end{pgfscope}%
\begin{pgfscope}%
\definecolor{textcolor}{rgb}{0.000000,0.000000,0.000000}%
\pgfsetstrokecolor{textcolor}%
\pgfsetfillcolor{textcolor}%
\pgftext[x=0.196739in, y=2.239118in, left, base]{\color{textcolor}\sffamily\fontsize{6.000000}{7.200000}\selectfont 10}%
\end{pgfscope}%
\begin{pgfscope}%
\pgfsetbuttcap%
\pgfsetroundjoin%
\definecolor{currentfill}{rgb}{0.000000,0.000000,0.000000}%
\pgfsetfillcolor{currentfill}%
\pgfsetlinewidth{0.803000pt}%
\definecolor{currentstroke}{rgb}{0.000000,0.000000,0.000000}%
\pgfsetstrokecolor{currentstroke}%
\pgfsetdash{}{0pt}%
\pgfsys@defobject{currentmarker}{\pgfqpoint{-0.048611in}{0.000000in}}{\pgfqpoint{-0.000000in}{0.000000in}}{%
\pgfpathmoveto{\pgfqpoint{-0.000000in}{0.000000in}}%
\pgfpathlineto{\pgfqpoint{-0.048611in}{0.000000in}}%
\pgfusepath{stroke,fill}%
}%
\begin{pgfscope}%
\pgfsys@transformshift{0.400000in}{2.004737in}%
\pgfsys@useobject{currentmarker}{}%
\end{pgfscope}%
\end{pgfscope}%
\begin{pgfscope}%
\definecolor{textcolor}{rgb}{0.000000,0.000000,0.000000}%
\pgfsetstrokecolor{textcolor}%
\pgfsetfillcolor{textcolor}%
\pgftext[x=0.196739in, y=1.973080in, left, base]{\color{textcolor}\sffamily\fontsize{6.000000}{7.200000}\selectfont 20}%
\end{pgfscope}%
\begin{pgfscope}%
\pgfsetbuttcap%
\pgfsetroundjoin%
\definecolor{currentfill}{rgb}{0.000000,0.000000,0.000000}%
\pgfsetfillcolor{currentfill}%
\pgfsetlinewidth{0.803000pt}%
\definecolor{currentstroke}{rgb}{0.000000,0.000000,0.000000}%
\pgfsetstrokecolor{currentstroke}%
\pgfsetdash{}{0pt}%
\pgfsys@defobject{currentmarker}{\pgfqpoint{-0.048611in}{0.000000in}}{\pgfqpoint{-0.000000in}{0.000000in}}{%
\pgfpathmoveto{\pgfqpoint{-0.000000in}{0.000000in}}%
\pgfpathlineto{\pgfqpoint{-0.048611in}{0.000000in}}%
\pgfusepath{stroke,fill}%
}%
\begin{pgfscope}%
\pgfsys@transformshift{0.400000in}{1.738698in}%
\pgfsys@useobject{currentmarker}{}%
\end{pgfscope}%
\end{pgfscope}%
\begin{pgfscope}%
\definecolor{textcolor}{rgb}{0.000000,0.000000,0.000000}%
\pgfsetstrokecolor{textcolor}%
\pgfsetfillcolor{textcolor}%
\pgftext[x=0.196739in, y=1.707042in, left, base]{\color{textcolor}\sffamily\fontsize{6.000000}{7.200000}\selectfont 30}%
\end{pgfscope}%
\begin{pgfscope}%
\pgfsetbuttcap%
\pgfsetroundjoin%
\definecolor{currentfill}{rgb}{0.000000,0.000000,0.000000}%
\pgfsetfillcolor{currentfill}%
\pgfsetlinewidth{0.803000pt}%
\definecolor{currentstroke}{rgb}{0.000000,0.000000,0.000000}%
\pgfsetstrokecolor{currentstroke}%
\pgfsetdash{}{0pt}%
\pgfsys@defobject{currentmarker}{\pgfqpoint{-0.048611in}{0.000000in}}{\pgfqpoint{-0.000000in}{0.000000in}}{%
\pgfpathmoveto{\pgfqpoint{-0.000000in}{0.000000in}}%
\pgfpathlineto{\pgfqpoint{-0.048611in}{0.000000in}}%
\pgfusepath{stroke,fill}%
}%
\begin{pgfscope}%
\pgfsys@transformshift{0.400000in}{1.472660in}%
\pgfsys@useobject{currentmarker}{}%
\end{pgfscope}%
\end{pgfscope}%
\begin{pgfscope}%
\definecolor{textcolor}{rgb}{0.000000,0.000000,0.000000}%
\pgfsetstrokecolor{textcolor}%
\pgfsetfillcolor{textcolor}%
\pgftext[x=0.196739in, y=1.441003in, left, base]{\color{textcolor}\sffamily\fontsize{6.000000}{7.200000}\selectfont 40}%
\end{pgfscope}%
\begin{pgfscope}%
\pgfsetbuttcap%
\pgfsetroundjoin%
\definecolor{currentfill}{rgb}{0.000000,0.000000,0.000000}%
\pgfsetfillcolor{currentfill}%
\pgfsetlinewidth{0.803000pt}%
\definecolor{currentstroke}{rgb}{0.000000,0.000000,0.000000}%
\pgfsetstrokecolor{currentstroke}%
\pgfsetdash{}{0pt}%
\pgfsys@defobject{currentmarker}{\pgfqpoint{-0.048611in}{0.000000in}}{\pgfqpoint{-0.000000in}{0.000000in}}{%
\pgfpathmoveto{\pgfqpoint{-0.000000in}{0.000000in}}%
\pgfpathlineto{\pgfqpoint{-0.048611in}{0.000000in}}%
\pgfusepath{stroke,fill}%
}%
\begin{pgfscope}%
\pgfsys@transformshift{0.400000in}{1.206622in}%
\pgfsys@useobject{currentmarker}{}%
\end{pgfscope}%
\end{pgfscope}%
\begin{pgfscope}%
\definecolor{textcolor}{rgb}{0.000000,0.000000,0.000000}%
\pgfsetstrokecolor{textcolor}%
\pgfsetfillcolor{textcolor}%
\pgftext[x=0.196739in, y=1.174965in, left, base]{\color{textcolor}\sffamily\fontsize{6.000000}{7.200000}\selectfont 50}%
\end{pgfscope}%
\begin{pgfscope}%
\pgfsetbuttcap%
\pgfsetroundjoin%
\definecolor{currentfill}{rgb}{0.000000,0.000000,0.000000}%
\pgfsetfillcolor{currentfill}%
\pgfsetlinewidth{0.803000pt}%
\definecolor{currentstroke}{rgb}{0.000000,0.000000,0.000000}%
\pgfsetstrokecolor{currentstroke}%
\pgfsetdash{}{0pt}%
\pgfsys@defobject{currentmarker}{\pgfqpoint{-0.048611in}{0.000000in}}{\pgfqpoint{-0.000000in}{0.000000in}}{%
\pgfpathmoveto{\pgfqpoint{-0.000000in}{0.000000in}}%
\pgfpathlineto{\pgfqpoint{-0.048611in}{0.000000in}}%
\pgfusepath{stroke,fill}%
}%
\begin{pgfscope}%
\pgfsys@transformshift{0.400000in}{0.940584in}%
\pgfsys@useobject{currentmarker}{}%
\end{pgfscope}%
\end{pgfscope}%
\begin{pgfscope}%
\definecolor{textcolor}{rgb}{0.000000,0.000000,0.000000}%
\pgfsetstrokecolor{textcolor}%
\pgfsetfillcolor{textcolor}%
\pgftext[x=0.196739in, y=0.908927in, left, base]{\color{textcolor}\sffamily\fontsize{6.000000}{7.200000}\selectfont 60}%
\end{pgfscope}%
\begin{pgfscope}%
\definecolor{textcolor}{rgb}{0.000000,0.000000,0.000000}%
\pgfsetstrokecolor{textcolor}%
\pgfsetfillcolor{textcolor}%
\pgftext[x=0.141184in,y=1.618981in,,bottom,rotate=90.000000]{\color{textcolor}\sffamily\fontsize{6.000000}{7.200000}\selectfont Input Neurons \(\displaystyle i \cdot \ell\)}%
\end{pgfscope}%
\begin{pgfscope}%
\pgfsetbuttcap%
\pgfsetmiterjoin%
\definecolor{currentfill}{rgb}{1.000000,1.000000,1.000000}%
\pgfsetfillcolor{currentfill}%
\pgfsetlinewidth{0.000000pt}%
\definecolor{currentstroke}{rgb}{0.000000,0.000000,0.000000}%
\pgfsetstrokecolor{currentstroke}%
\pgfsetstrokeopacity{0.000000}%
\pgfsetdash{}{0pt}%
\pgfpathmoveto{\pgfqpoint{0.400000in}{0.409227in}}%
\pgfpathlineto{\pgfqpoint{1.940000in}{0.409227in}}%
\pgfpathlineto{\pgfqpoint{1.940000in}{0.563227in}}%
\pgfpathlineto{\pgfqpoint{0.400000in}{0.563227in}}%
\pgfpathclose%
\pgfusepath{fill}%
\end{pgfscope}%
\begin{pgfscope}%
\pgfpathrectangle{\pgfqpoint{0.400000in}{0.409227in}}{\pgfqpoint{1.540000in}{0.154000in}}%
\pgfusepath{clip}%
\pgfsys@transformshift{0.400000in}{0.409227in}%
\pgftext[left,bottom]{\includegraphics[interpolate=true,width=1.540000in,height=0.156667in]{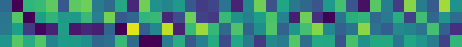}}%
\end{pgfscope}%
\begin{pgfscope}%
\pgfsetbuttcap%
\pgfsetroundjoin%
\definecolor{currentfill}{rgb}{0.000000,0.000000,0.000000}%
\pgfsetfillcolor{currentfill}%
\pgfsetlinewidth{0.803000pt}%
\definecolor{currentstroke}{rgb}{0.000000,0.000000,0.000000}%
\pgfsetstrokecolor{currentstroke}%
\pgfsetdash{}{0pt}%
\pgfsys@defobject{currentmarker}{\pgfqpoint{0.000000in}{-0.048611in}}{\pgfqpoint{0.000000in}{0.000000in}}{%
\pgfpathmoveto{\pgfqpoint{0.000000in}{0.000000in}}%
\pgfpathlineto{\pgfqpoint{0.000000in}{-0.048611in}}%
\pgfusepath{stroke,fill}%
}%
\begin{pgfscope}%
\pgfsys@transformshift{0.419250in}{0.409227in}%
\pgfsys@useobject{currentmarker}{}%
\end{pgfscope}%
\end{pgfscope}%
\begin{pgfscope}%
\definecolor{textcolor}{rgb}{0.000000,0.000000,0.000000}%
\pgfsetstrokecolor{textcolor}%
\pgfsetfillcolor{textcolor}%
\pgftext[x=0.419250in,y=0.312004in,,top]{\color{textcolor}\sffamily\fontsize{6.000000}{7.200000}\selectfont 0}%
\end{pgfscope}%
\begin{pgfscope}%
\pgfsetbuttcap%
\pgfsetroundjoin%
\definecolor{currentfill}{rgb}{0.000000,0.000000,0.000000}%
\pgfsetfillcolor{currentfill}%
\pgfsetlinewidth{0.803000pt}%
\definecolor{currentstroke}{rgb}{0.000000,0.000000,0.000000}%
\pgfsetstrokecolor{currentstroke}%
\pgfsetdash{}{0pt}%
\pgfsys@defobject{currentmarker}{\pgfqpoint{0.000000in}{-0.048611in}}{\pgfqpoint{0.000000in}{0.000000in}}{%
\pgfpathmoveto{\pgfqpoint{0.000000in}{0.000000in}}%
\pgfpathlineto{\pgfqpoint{0.000000in}{-0.048611in}}%
\pgfusepath{stroke,fill}%
}%
\begin{pgfscope}%
\pgfsys@transformshift{0.804250in}{0.409227in}%
\pgfsys@useobject{currentmarker}{}%
\end{pgfscope}%
\end{pgfscope}%
\begin{pgfscope}%
\definecolor{textcolor}{rgb}{0.000000,0.000000,0.000000}%
\pgfsetstrokecolor{textcolor}%
\pgfsetfillcolor{textcolor}%
\pgftext[x=0.804250in,y=0.312004in,,top]{\color{textcolor}\sffamily\fontsize{6.000000}{7.200000}\selectfont 10}%
\end{pgfscope}%
\begin{pgfscope}%
\pgfsetbuttcap%
\pgfsetroundjoin%
\definecolor{currentfill}{rgb}{0.000000,0.000000,0.000000}%
\pgfsetfillcolor{currentfill}%
\pgfsetlinewidth{0.803000pt}%
\definecolor{currentstroke}{rgb}{0.000000,0.000000,0.000000}%
\pgfsetstrokecolor{currentstroke}%
\pgfsetdash{}{0pt}%
\pgfsys@defobject{currentmarker}{\pgfqpoint{0.000000in}{-0.048611in}}{\pgfqpoint{0.000000in}{0.000000in}}{%
\pgfpathmoveto{\pgfqpoint{0.000000in}{0.000000in}}%
\pgfpathlineto{\pgfqpoint{0.000000in}{-0.048611in}}%
\pgfusepath{stroke,fill}%
}%
\begin{pgfscope}%
\pgfsys@transformshift{1.189250in}{0.409227in}%
\pgfsys@useobject{currentmarker}{}%
\end{pgfscope}%
\end{pgfscope}%
\begin{pgfscope}%
\definecolor{textcolor}{rgb}{0.000000,0.000000,0.000000}%
\pgfsetstrokecolor{textcolor}%
\pgfsetfillcolor{textcolor}%
\pgftext[x=1.189250in,y=0.312004in,,top]{\color{textcolor}\sffamily\fontsize{6.000000}{7.200000}\selectfont 20}%
\end{pgfscope}%
\begin{pgfscope}%
\pgfsetbuttcap%
\pgfsetroundjoin%
\definecolor{currentfill}{rgb}{0.000000,0.000000,0.000000}%
\pgfsetfillcolor{currentfill}%
\pgfsetlinewidth{0.803000pt}%
\definecolor{currentstroke}{rgb}{0.000000,0.000000,0.000000}%
\pgfsetstrokecolor{currentstroke}%
\pgfsetdash{}{0pt}%
\pgfsys@defobject{currentmarker}{\pgfqpoint{0.000000in}{-0.048611in}}{\pgfqpoint{0.000000in}{0.000000in}}{%
\pgfpathmoveto{\pgfqpoint{0.000000in}{0.000000in}}%
\pgfpathlineto{\pgfqpoint{0.000000in}{-0.048611in}}%
\pgfusepath{stroke,fill}%
}%
\begin{pgfscope}%
\pgfsys@transformshift{1.574250in}{0.409227in}%
\pgfsys@useobject{currentmarker}{}%
\end{pgfscope}%
\end{pgfscope}%
\begin{pgfscope}%
\definecolor{textcolor}{rgb}{0.000000,0.000000,0.000000}%
\pgfsetstrokecolor{textcolor}%
\pgfsetfillcolor{textcolor}%
\pgftext[x=1.574250in,y=0.312004in,,top]{\color{textcolor}\sffamily\fontsize{6.000000}{7.200000}\selectfont 30}%
\end{pgfscope}%
\begin{pgfscope}%
\definecolor{textcolor}{rgb}{0.000000,0.000000,0.000000}%
\pgfsetstrokecolor{textcolor}%
\pgfsetfillcolor{textcolor}%
\pgftext[x=1.170000in,y=0.175801in,,top]{\color{textcolor}\sffamily\fontsize{6.000000}{7.200000}\selectfont Hidden LIF Neurons \(\displaystyle j\)}%
\end{pgfscope}%
\begin{pgfscope}%
\pgfsetbuttcap%
\pgfsetroundjoin%
\definecolor{currentfill}{rgb}{0.000000,0.000000,0.000000}%
\pgfsetfillcolor{currentfill}%
\pgfsetlinewidth{0.803000pt}%
\definecolor{currentstroke}{rgb}{0.000000,0.000000,0.000000}%
\pgfsetstrokecolor{currentstroke}%
\pgfsetdash{}{0pt}%
\pgfsys@defobject{currentmarker}{\pgfqpoint{-0.048611in}{0.000000in}}{\pgfqpoint{-0.000000in}{0.000000in}}{%
\pgfpathmoveto{\pgfqpoint{-0.000000in}{0.000000in}}%
\pgfpathlineto{\pgfqpoint{-0.048611in}{0.000000in}}%
\pgfusepath{stroke,fill}%
}%
\begin{pgfscope}%
\pgfsys@transformshift{0.400000in}{0.543977in}%
\pgfsys@useobject{currentmarker}{}%
\end{pgfscope}%
\end{pgfscope}%
\begin{pgfscope}%
\definecolor{textcolor}{rgb}{0.000000,0.000000,0.000000}%
\pgfsetstrokecolor{textcolor}%
\pgfsetfillcolor{textcolor}%
\pgftext[x=0.249759in, y=0.512320in, left, base]{\color{textcolor}\sffamily\fontsize{6.000000}{7.200000}\selectfont 0}%
\end{pgfscope}%
\begin{pgfscope}%
\pgfsetbuttcap%
\pgfsetroundjoin%
\definecolor{currentfill}{rgb}{0.000000,0.000000,0.000000}%
\pgfsetfillcolor{currentfill}%
\pgfsetlinewidth{0.803000pt}%
\definecolor{currentstroke}{rgb}{0.000000,0.000000,0.000000}%
\pgfsetstrokecolor{currentstroke}%
\pgfsetdash{}{0pt}%
\pgfsys@defobject{currentmarker}{\pgfqpoint{-0.048611in}{0.000000in}}{\pgfqpoint{-0.000000in}{0.000000in}}{%
\pgfpathmoveto{\pgfqpoint{-0.000000in}{0.000000in}}%
\pgfpathlineto{\pgfqpoint{-0.048611in}{0.000000in}}%
\pgfusepath{stroke,fill}%
}%
\begin{pgfscope}%
\pgfsys@transformshift{0.400000in}{0.428477in}%
\pgfsys@useobject{currentmarker}{}%
\end{pgfscope}%
\end{pgfscope}%
\begin{pgfscope}%
\definecolor{textcolor}{rgb}{0.000000,0.000000,0.000000}%
\pgfsetstrokecolor{textcolor}%
\pgfsetfillcolor{textcolor}%
\pgftext[x=0.249759in, y=0.396820in, left, base]{\color{textcolor}\sffamily\fontsize{6.000000}{7.200000}\selectfont 3}%
\end{pgfscope}%
\begin{pgfscope}%
\definecolor{textcolor}{rgb}{0.000000,0.000000,0.000000}%
\pgfsetstrokecolor{textcolor}%
\pgfsetfillcolor{textcolor}%
\pgftext[x=0.083559in, y=0.328247in, left, base,rotate=90.000000]{\color{textcolor}\sffamily\fontsize{6.000000}{7.200000}\selectfont Output }%
\end{pgfscope}%
\begin{pgfscope}%
\definecolor{textcolor}{rgb}{0.000000,0.000000,0.000000}%
\pgfsetstrokecolor{textcolor}%
\pgfsetfillcolor{textcolor}%
\pgftext[x=0.176869in, y=0.271543in, left, base,rotate=90.000000]{\color{textcolor}\sffamily\fontsize{6.000000}{7.200000}\selectfont  Neurons \(\displaystyle k\)}%
\end{pgfscope}%
\end{pgfpicture}%
\makeatother%
\endgroup%

%% file: tex/conclusion.tex
\section{Conclusion}
\label{sec:conclusion}
This work successfully showcases the implementation of \gls{snn}-based joint equalization and demapping emulated on the accelerated analog neuromorphic hardware system \gls{bss-2}.
Our demapper on \gls{bss-2} approaches the performance of an \gls{snn} demapper simulated in software while outperforming an \gls{lmmse} equalizer and performing better than a nonlinear \gls{ann} reference demapper, both with the same number of taps.
A gain of \SI{1.5}{dB} at a \gls{ber} of \num[exponent-product=\cdot]{2e-3} of the simulated \gls{snn} over the \gls{lmmse} clearly demonstrates the nonlinear processing capability of the \gls{snn} demapper.
A small hardware penalty of about \SI{0.2}{dB} at the same \gls{ber} with respect to the \gls{snn} simulated in software is observed and is attributed to hardware imperfections like noise in the physical substrate, fixed-pattern noise artifacts of the production process, and potentially a sub-optimal hardware operation point.
Typically, the fixed-pattern noise effects are widely absorbed by gradient-based training.
An additional cause might be the limited precision of 6-bit hardware weights and the 8-bit \gls{cadc}.
Despite having multiple sources of noise and loss of information owing to limited precision, the \gls{snn} demapper on \gls{bss-2} shows an excellent performance and resilience to hardware impairments.
We conjecture that the chosen size of the \gls{snn} with $40$ hidden neurons ensures a robust behavior by encoding information redundantly.
Accordingly, we expect to observe a larger hardware penalty as the number of hidden neurons decreases.
An interesting direction for future research is to investigate how the complexity of the \gls{snn} on \gls{bss-2} can be reduced while maintaining its performance.

With the implementation at hand, the equalization and demapping of a single sample take about $T=\SI{30}{\micro\second}$.
Therefore, the \gls{bss-2} platform supports a maximum symbol rate of \SI{30}{\kilo Bd}.
However, this upper bound is due to the specific design target of \gls{bss-2} as a general purpose experimentation platform and does not follow from an intrinsic limitation of the underlying \gls{cmos} technology itself.
Significantly faster inference, and thus throughput, might be achieved by accelerating the emulation of the \gls{lif} dynamics.
\cite{schemmel_ijcnn06} presented a neuromorphic \gls{asic} exhibiting an acceleration of up to two additional \glspl{oom} with respect to \gls{bss-2}.
Given the fact that the cited implementation was fabricated in a \SI{180}{\nano\meter} \gls{cmos} process, it is reasonable to assume that a modern FinFET process could potentially gain at least another 2 \glspl{oom}.
This would result in a processing time in the order of nanoseconds per sample.
The throughput can further be increased by parallelization.
Several spiking network cores could be deployed in parallel, of which each could process multiple samples on the same physical substrate at once. To get nanoseconds per sample to 200 Gbit/s, a parallel processing factor of a few hundreds is enough, which is similar to the time-interleaving of multiple \glspl{adc} used in standard optical \gls{dsp} solutions \cite{Laperle}.

The spatio-temporal sparsity of \glspl{snn} promises an intrinsically favorable energy footprint when contrasted to traditional \gls{ann}-based solutions -- especially when combined with analog \gls{imc}~\cite{burr2017neuromorphic}.
Currently, the power consumption is dominated by I/O as well as the clock distribution and biasing of the individual subsystems -- a fact largely attributed to the flexible general-purpose approach of \gls{bss-2}.
Optimizing or omitting these subsystems in future, more specialized \glspl{asic} could dramatically reduce the overall energy footprint.

Future research aims to increase hardware resource efficiency by decreasing the architectural \gls{snn} complexity and investigate feature sharing in order to increase the throughput by parallelization.
Importantly, the power consumption of neuromorphic signal processing shall be analyzed, compared to a digital implementation, and optimized by minimization of the firing activity of the neurons and efficient design of the input and output interfaces of the \gls{snn}.

The presented results demonstrate that electrical neuromorphic hardware can implement signal processing with the accuracy required in optical transceivers. To successfully integrate SNN equalization in optical transceivers, efficient conversion of received signals into input spikes must be researched.

%% file: tex/acknowledgment.tex
\section*{Acknowledgment}
We thank L. Blessing, B. Cramer, and C. Pehle for insightful discussions, C. Mauch for keeping the \gls{bss-2} system on track, and all members of the Electronic Vision(s) research group who contributed to the \gls{bss-2} system.

\section*{Funding}
The contributions of the Electronic Vision(s) group have been supported by the EC Horizon 2020 Framework Programme under grant agreements 785907 (HBP SGA2) and 945539 (HBP SGA3),
\foreignlanguage{ngerman}{Deutsche Forschungsgemeinschaft} (DFG, German Research Foundation) under Germany's Excellence Strategy EXC 2181/1-390900948 (the Heidelberg STRUCTURES Excellence Cluster),
the Helmholtz Association Initiative and Networking Fund [Advanced Computing Architectures (ACA) under Project SO-092.